\documentclass[a4paper,11pt,titlepage=true]{scrartcl} 

\usepackage[english]{babel}
\usepackage[latin1]{inputenc}
\usepackage[T1]{fontenc}
\usepackage{fixltx2e}
\usepackage{lmodern}
\usepackage{authblk}
 
\usepackage{enumitem}
\usepackage{url}
\usepackage{stfloats}
\usepackage{graphicx}
\usepackage{tikz}
 \usetikzlibrary{arrows,shapes,fit,calc,positioning,spy,decorations.pathreplacing,decorations.pathmorphing,patterns,backgrounds,arrows.new}
 \tikzset{gene/.style={draw,circle,fill=JUlblue,minimum size=1.5em}, 
	  protein/.style={draw,regular polygon,regular polygon sides=3,fill=JUlblue,minimum size=1em,yshift=-0.2em}, 
	  compound/.style={draw,regular polygon,regular polygon sides=4,fill=JUlblue,minimum size=2em}, 
	  reaction/.style={draw,regular polygon,regular polygon sides=6,fill=JUlblue,minimum size=1.75em}, 
	  encodes/.style={-latex new,arrow head=1.75mm},
	  activates/.style={-latex new,arrow head=1.75mm,dashed,shorten >=0.15em},
	  inhibites/.style={-| new,arrow head=2.5mm,dashed,shorten >=0.15em},
	  box/.style={draw=black!60,rounded corners=2mm,inner sep=2mm,thick,dashed,fill=black!10},
	  node/.style={draw,black!70,fill,circle,inner sep=0.2em},
	  edge/.style={-latex,thick,mygray},
	  line/.style={dotted,thick,black!70}
	 }
\usepackage{tikzscale}
\usepackage{verbatim}
\usepackage{pifont}
\usepackage{xcolor,colortbl}
 \definecolor{JUdblue}{HTML}{143264}
 \definecolor{JUlblue}{HTML}{AAB4C9}
 \definecolor{JUbblue}{HTML}{CDD9F2}
 \definecolor{Tmagenta}{HTML}{AA00AA}
 \definecolor{Tcyan}{HTML}{09ADAD}
 \definecolor{Torange}{HTML}{FF8000}
 \definecolor{Tbrown}{HTML}{C0A080}
 \definecolor{Tgray}{HTML}{AAAAAA}
 \definecolor{myred}{HTML}{8B0000}
 \definecolor{mygreen}{HTML}{006400}
 \definecolor{myblue}{HTML}{00008B}
 \definecolor{myyellow}{HTML}{FFD700}
 \definecolor{myindigo}{HTML}{4B0082}
 \definecolor{myorange}{HTML}{D2691E}
 \definecolor{mygray}{HTML}{808080}
 \definecolor{mylightgray}{HTML}{C0C0C0}
 \definecolor{mycol}{HTML}{B8860B}
 \definecolor{mycol1}{HTML}{A0522D}
 \definecolor{dred}{HTML}{B22222}
 \definecolor{dgreen}{HTML}{008000}
 \definecolor{dblue}{HTML}{0000B0}
 \definecolor{gold}{HTML}{FFD700}
 \definecolor{chocolate}{HTML}{D2691E}
 \definecolor{purple}{HTML}{800080}
 \definecolor{teal}{HTML}{5F9EA0}
 \definecolor{steelblue}{HTML}{4682B4}
 \definecolor{violet}{HTML}{DB7093}
 \definecolor{what}{HTML}{808000}
 \definecolor{goldenrod}{HTML}{DAA520}
\usepackage{amsmath,amssymb,bbm,amsfonts,nicefrac}
\usepackage{MnSymbol} 
\usepackage{textcomp,wasysym}
\usepackage[normalem]{ulem} 
\usepackage{lscape}
\usepackage{tabularx,booktabs,longtable,multirow,dcolumn}
 \newcolumntype{L}[1]{>{\raggedright\arraybackslash}m{#1}} 
 \newcolumntype{C}[1]{>{\centering\arraybackslash}m{#1}} 
\usepackage{rotating}
\usepackage[vlined,ruled]{algorithm2e} 
 \SetAlFnt{\small}
\usepackage[numbers,square,comma,sort&compress]{natbib}
\usepackage[a4paper,left=2cm, right=2cm, top=0.5cm, bottom=1cm, includeheadfoot]{geometry}

\renewcommand{\emph}[1]{{\normalsize\rmfamily\itshape #1}}
\newcommand{\colsquare}[1]{\raisebox{-0.6mm}[0mm][0.6mm]{\color{#1}\scalebox{1.6}{$\blacksquare$}}}
\newcommand{\mylittlesquare}[1]{{\color{#1}$\blacksquare$}}
\newcommand{\opensquare}[1]{\raisebox{-0.6mm}[0mm][0.6mm]{\color{#1}\scalebox{1.6}{$\square$}}}
\newcommand{\coltriangle}[1]{\raisebox{-0.7mm}[0mm][0.6mm]{\color{#1}\scalebox{1.3}{$\blacktriangle$}}}
\newcommand{\coltriangledown}[1]{\raisebox{0.1mm}[0mm][0.6mm]{\color{#1}\scalebox{1.3}{$\blacktriangledown$}}}

\title{A system-wide network reconstruction of gene regulation and metabolism in \textit{Escherichia coli}}
\author[1]{Anne Grimbs}
\author[2]{David F. Klosik}
\author[2]{Stefan Bornholdt}
\author[1]{Marc-Thorsten Hütt\thanks{m.huett@jacobs-university.de}}
\affil[1]{Computational Systems Biology, Department of Life Sciences \& Chemistry, Jacobs University, Bremen, 28759, Germany}
\affil[2]{Institute for Theoretical Physics, University of Bremen, Bremen, 28359, Germany}
\date{}

\begin{document}
 \maketitle
 
 \begin{abstract}
   Genome-scale metabolic models have become a fundamental tool for
   examining metabolic principles. However, metabolism is not solely
   characterized by the underlying biochemical reactions and
   catalyzing enzymes, but also affected by regulatory events. Since
   the pioneering work of Covert and co-workers as well as Shlomi and
   co-workers it is debated, how regulation and metabolism
   synergistically characterize a coherent cellular state. The first
   approaches started from metabolic models which were extended by
   the regulation of the encoding genes of the catalyzing enzymes. By
   now, bioinformatics databases in principle allow addressing the
   challenge of integrating regulation and metabolism on a
   system-wide level. Collecting information from several databases
   we provide a network representation of the integrated gene
   regulatory and metabolic system for \textit{Escherichia coli},
   including major cellular processes, from metabolic processes via
   protein modification to a variety of regulatory events. Besides
   transcriptional regulation, we also take into account regulation
   of translation, enzyme activities and reactions. Our network model
   provides novel topological characterizations of system components
   based on their positions in the network. We show that network
   characteristics suggest a representation of the integrated system
   as three network domains (regulatory, metabolic and interface
   networks) instead of two. This new three-domain representation
   reveals the structural centrality of components with known high
   functional relevance. This integrated network can serve as a
   platform for understanding coherent cellular states as active
   subnetworks and to elucidate crossover effects between metabolism
   and gene regulation.
 \end{abstract}

\section*{Introduction}
So far, metabolic processes and gene regulatory events are typically considered individually in system-level investigations. However, ample evidence exists that the majority of cellular processes involves both, metabolism and gene regulation, and thus requires their joint examination \cite{Kochanowski2015}. One of the best-investigated individual examples in \textit{Escherichia coli} (\textit{E.~coli}) is the phosphoenolpyruvate--carbohydrate phosphotransferase system (PTS) which is responsible for import and phosphorylation of sugars \cite{Escalante2012}. Additionally, the PTS is involved in the regulation of the import process depending on the available carbohydrate mixtures in the growth medium. By carbon catabolite repression and inducer exclusion, primarily the uptake of a preferred carbon source to be metabolized, such as glucose, is selected from other carbohydrates present in the growth medium. In order to understand the underlying principles, not only the effects of both 'layers', metabolism and regulation, need to be taken into account, but also their interface \cite{Goncalves2013}.\par
\begin{figure}[htb]
  \centering
  \resizebox{\textwidth}{!}{
\begin{tikzpicture}
 \coordinate (x) at (1,0); \coordinate (y) at (0,1);
 \node[gene,fill=myyellow] (G0){};
 \node[gene,fill=myyellow] (G1) at ($(G0)+1.*(x)$) {};
 \node[gene,fill=myyellow] (G2) at ($(G1)+2.*(x)$) {};
 \node[gene,fill=myyellow] (G3) at ($(G2)+2.*(x)$) {};
 \node[protein,fill=myred] (P0) at ($(G0)-1.*(y)$) {};
 \node[protein,fill=myred] (P1) at ($(G1)-1.*(y)$) {};
 \node[protein,fill=mygray] (PC01) at ($0.5*(G0)+0.5*(G1)-2.*(y)$) {};
 \node[protein,fill=myred] (P2) at ($(G2)-1.25*(y)$) {};
 \node[protein,fill=myred] (P3) at ($(G3)-1.25*(y)$) {};
 \draw[encodes] (G0.south) to (P0);
 \draw[encodes] (G1.south) to (P1);
 \draw[encodes] (P0.south) to (PC01);
 \draw[encodes] (P1.south) to (PC01);
 \draw[encodes] (G2.south) to (P2);
 \draw[encodes] (G3.south) to (P3);
 
 \draw[activates] ($(P2.west)+0.125*(y)$) to ($(P2.west)-0.75*(x)+0.125*(y)$) to ($(P2.west)-0.75*(x)+2.*(y)$) -| (G0);
 \draw[activates] (P2.west) to ($(P2.west)-1.*(x)$) |- (G1);
 \draw[activates] (PC01.west) to ($(PC01.west)-0.75*(x)$) to ($(PC01.west)-0.75*(x)+2.875*(y)$) -| (G2);
 \draw[inhibites] ($(P2.east)+0.125*(y)$) to ($(P2.east)+0.5*(x)+0.125*(y)$) |- (G2);
 \draw[activates] (P2.east) to ($(P3.west)-0.5*(x)$) |- (G3);
 \draw[inhibites] (P3.east) to ($(P3.east)+0.5*(x)$) |- (G3);
 
 \node[gene,fill=myyellow] (G4) at ($(PC01)-1.75*(x)-1.*(y)$) {};
 \node[gene,fill=myyellow] (G5) at ($(PC01)-0.75*(x)-1.*(y)$) {};
 \node[gene,fill=myyellow] (G6) at ($(PC01)+0.75*(x)-1.*(y)$) {};
 \node[gene,fill=myyellow] (G7) at ($(PC01)+1.75*(x)-1.*(y)$) {};
 \node[gene,fill=myyellow] (G8) at ($(PC01)+2.5*(x)-1.*(y)$) {};
 \node[gene,fill=myyellow] (G9) at ($(PC01)+3.75*(x)-1.*(y)$) {};
 \node[gene,fill=myyellow] (G10) at ($(PC01)+6.25*(x)-1.*(y)$) {};
 \draw[activates] (PC01.south) to node (GeneRegulation) {} (G4);
 \draw[inhibites] (PC01.south) to (G7);
 \draw[activates] (P2.south) to (G5);
 \draw[activates] (P2.south) to (G7);
 \draw[inhibites] (P2.south) to (G8);
 \draw[inhibites] (P2.south) to (G10);
 \draw[activates] (P3.south) to (G8);
 \draw[activates] (P3.south) to (G10);
 
 \node[protein,fill=myred] (P45) at ($0.5*(G4)+0.5*(G5)-1.625*(y)$) {};
 \node[protein,fill=myred] (P6) at ($(G6)-1.125*(y)$) {};
 \node[protein,fill=myred] (P7) at ($(G7)-1.125*(y)$) {};
 \node[protein,fill=myred] (P8) at ($(G8)-1.625*(y)$) {};
 \node[protein,fill=myred] (P9) at ($(G9)-1.125*(y)$) {};
 \node[protein,fill=myred] (P10) at ($(G10)-2.125*(y)$) {};
 \draw[encodes] (G4.south) to (P45);
 \draw[encodes] (G5.south) to (P45);
 \draw[encodes] (G6.south) to (P6);
 \draw[encodes] (G7.south) to (P7);
 \draw[encodes] (G8.south) to (P8);
 \draw[encodes] (G9.south) to (P9);
 \draw[encodes] (G10.south) to node (Encode) {} (P10);
 
 \node[protein,fill=mygray] (PC67) at ($0.5*(P6)+0.5*(P7)-1.*(y)$) {};
 \draw[encodes] (P6.south) to (PC67);
 \draw[encodes] (P7.south) to (PC67);
 \node[reaction,fill=mygreen] (R8) at ($(P9)-1.*(y)$) {};
 \node[protein,fill=myindigo] (PC10) at ($(R8)+1.25*(x)$) {};
 \draw[encodes] (P9.south) to (R8);
 \draw[encodes] ($(P8.east)-0.05*(x)+0.075*(y)$) to (R8);
 \draw[encodes] (R8) to ($(PC10.west)+0.05*(x)+0.065*(y)$);
 \draw[inhibites] ($(PC10.east)-0.05*(x)+0.065*(y)$) to ($(PC10.east)+0.5*(x)+0.065*(y)$) |- (G9);
 
 \node[reaction,fill=mygreen] (R45) at ($(P45)-1.6*(y)$) {};
 \node[reaction,fill=mygreen] (R67) at ($(PC67)-1.6*(y)$) {};
 \node[reaction,fill=mygreen] (R9f) at ($(R8)-1.6*(y)$) {};
 \node[reaction,fill=mygreen] (R9r) at ($(R8)-2.6*(y)$) {};
 \node[reaction,fill=mygreen] (R10) at ($(P10)-2.1*(y)$) {};
 \draw[encodes] (P45.south) to node (Catalysis) {} (R45);
 \draw[encodes] (PC67.south) to (R67);
 \draw[encodes] (P10.south) to (R10);
 
 \node[compound,fill=myblue] (C1) at ($(R45)-1.25*(x)$) {};
 \node[compound,fill=myblue] (C2) at ($(R45)+1.25*(x)$) {};
 \node[compound,fill=myblue] (C3) at ($(R67)-1.25*(x)-0.5*(y)$) {};
 \node[compound,fill=myblue] (C4) at ($(R67)+1.25*(x)+0.5*(y)$) {};
 \node[compound,fill=myblue] (C5) at ($(R67)+1.25*(x)-0.5*(y)$) {};
 \node[compound,fill=myblue] (C6) at ($(R9f)+1.25*(x)-0.5*(y)$) {};
 \node[compound,fill=myblue] (C7) at ($(R10)+1.25*(x)+0.5*(y)$) {};
 \node[compound,fill=myblue] (C8) at ($(R10)+1.25*(x)-0.5*(y)$) {};
 \draw[encodes] ($(C1.west)-0.5*(x)$) to (C1.west);
 \draw[encodes] (C1.east) to (R45);
 \draw[encodes] (R45) to (C2.west);
 \draw[encodes] (C2.east) to (R67);
 \draw[encodes] ($(C3.west)-3.0*(x)$) to (C3.west);
 \draw[encodes] (C3.east) to (R67);
 \draw[encodes] (R67) to (C4.west);
 \draw[encodes] (R67) to (C5.west);
 \draw[encodes] (C4.east) to (R8);
 \draw[encodes] (C5.east) to (R9f);
 \draw[encodes] (R9f) to (C6.west);
 \draw[encodes] (C6.west) to (R9r);
 \draw[encodes] (R9r) to (C5.east);
 \draw[encodes] (C6.east) to (R10);
 \draw[encodes] (R10) to (C7.west);
 \draw[encodes] (R10) to (C8.west);
 \draw[encodes] (C7.east) to ($(C7.east)+0.5*(x)$);
 \draw[encodes] (C8.east) to ($(C8.east)+0.5*(x)$);
 \draw[activates] (C2) |- ($(P45.east)-0.05*(x)+0.065*(y)$);
 
 \node[below left=0em and 8.75em of G0] (G) {Gene};
 \draw[dotted, mygray] (G) -- (G0);
 \node[above left=-0.5em and 6em of GeneRegulation] (TR) {\begin{minipage}{7em}\centering\textsl{Transcriptional regulation}\end{minipage}};
 \draw[dotted, mygray] (TR) -- (GeneRegulation);
 \node[above left=-0.5em and 6em of Catalysis] (EC) {\begin{minipage}{3.8em}\centering\textsl{Enzyme catalysis}\end{minipage}};
 \draw[dotted, mygray] (EC) -- (Catalysis);
 \node[below left=1em and 5.75em of R45] (R) {Reaction};
 \draw[dotted, mygray] (R) -- (R45.south west);
 \node[below right=0em and 9.4em of P3] (TF) {\begin{minipage}{5.9em}\centering Transcription factor\end{minipage}};
 \draw[dotted, mygray] (TF) -- (P3);
 \node[above right=-0.5em and 5.75em of Encode] (GE) {\begin{minipage}{4.5em}\centering\textsl{Gene expression}\end{minipage}};
 \draw[dotted, mygray] (GE) -- (Encode);
 \node[below right=0em and 6em of P10] (E) {Enzyme};
 \draw[dotted, mygray] (E) -- (P10);
 \node[below right=0em and 2em of C7] (C) {Compound};
 \draw[dotted, mygray] (C) -- (C7);
 \begin{pgfonlayer}{background}
  \node[box,fit=($(G1)+0.75*(y)$)(G4)($(G10)+0.75*(x)$),label={[anchor=north east]26:{\bfseries\large Gene regulation}}] (RBox) {};
  \node[box,fill=black!3,fit=(P45)(PC67)(P9)(P10)] (PBox) {};
  \node[box,fit=(C1)(C8),label={[anchor=south west]-167:{\bfseries\large Metabolism}}] (MBox) {};
 \end{pgfonlayer}
\end{tikzpicture}}
\caption{Schematic representation of the involved processes and biological elements in the integrative metabolic-regulatory \textit{E.~coli} network. Gene regulatory processes primarily comprise genes (\protect\tikz\protect\node[gene,fill=myyellow,scale=0.4]{};) and several proteins (monomers \protect\tikz\protect\node[protein,fill=myred,scale=0.4]{}; as well as complexes \protect\tikz\protect\node[protein,fill=mygray,scale=0.4]{};), mainly transcription factors. In contrast, metabolic processes are predominantly defined by small molecules (\protect\tikz\protect\node[compound,fill=myblue,scale=0.4]{};) and the catalyzing biochemical reactions (\protect\tikz\protect\node[reaction,fill=mygreen,scale=0.4]{};). The interactions between regulatory and metabolic processes can be mainly characterized by proteins (also modified proteins \protect\tikz\protect\node[protein,fill=myindigo,scale=0.4]{};) serving as enzymes and regulators, respectively. While regulatory links are represented as dashed lines, the encoding and reaction-associated links are shown as solid lines.}\label{fig:IntegrativeApproach}
\end{figure}
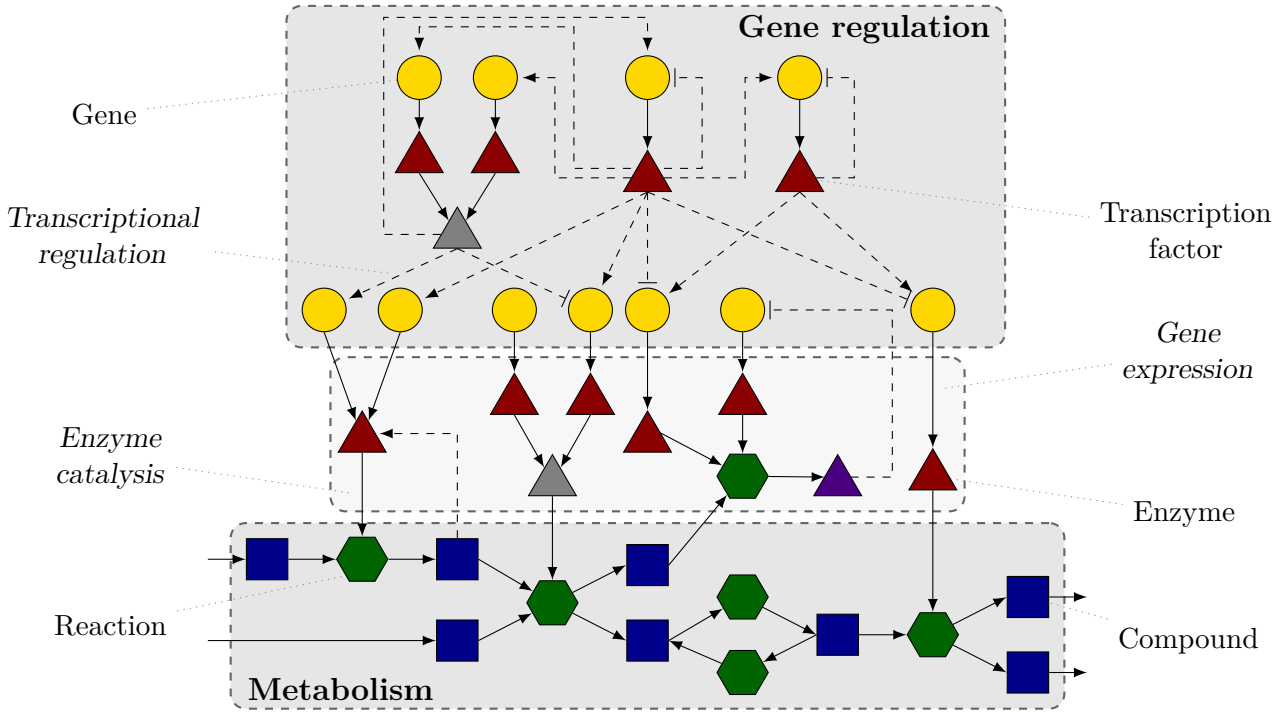

On a more qualitative level, the importance of the interface of metabolism and gene regulation can be illustrated by having a closer look at their most prominent representatives, namely, enzymes and metabolic transcriptional regulators. Both examples are proteins and can be thought of as a component type organizing the interplay of genes and metabolic reactions (Figure~\ref{fig:IntegrativeApproach}). For enzymes the connection is straightforward: The majority of metabolic reactions can only take place if the corresponding genes of the catalyzing enzymes are expressed. These genes, in turn, are often involved in regulatory processes, especially if they are associated with central biochemical reactions. In contrast, metabolic transcriptional regulators can be illustrated by looking at transcription factors, the probably best-investigated transcriptional regulators. Some of them require the binding of a metabolite to be active and are therefore called metabolic transcriptional regulators. In the context of the integrative view discussed here, it is noteworthy that only the interaction with a metabolic component enables their functionality as gene expression regulators.

Conventional reconstructions of \textit{E.~coli}'s metabolism as well as of its gene regulation thoroughly describe the process itself but usually lack information on interacting elements of the other biological system. While there are numerous genome-scale metabolic reconstructions available \cite{Edwards2000,Reed2003,Feist2007,Orth2010,Orth2011,monk2017iml1515}, only a few large-scale transcriptional regulatory networks exist that are mainly based on the information from RegulonDB \cite{Gama-Castro2015}. First attempts to integrate both cellular processes started from metabolic reconstructions which were expanded by regulatory genes and stimuli of the associated encoding metabolic genes \cite{Covert2004,Shlomi2007}. Both studies started from the metabolic model of \citet{Reed2003} and include 104 regulatory genes and 583 regulatory rules regulating approximately 50~\% of the metabolic genes. In this manner, the close proximity of regulatory events was captured but more far-reaching and global effects, \textit{e.g.}, self-contained regulatory dynamics among genes, could not be considered. Further approaches examine the regulatory processes of the metabolic network based on the aforementioned pioneering attempts \cite{Samal2008,Gianchandani2009}. For this purpose, the information about regulatory events was assembled in terms of Boolean rules as a variant of Boolean network models. \par

More recently, \citet{Chandrasekaran2010} introduced a method called probabilistic regulation of metabolism, a new variant of regulatory flux balance analysis, i.e., the class of approaches behind some of the pioneering integrative models discussed above \cite{Covert2004,Shlomi2007}. 
The state of the many variants of integrating regulatory information into flux-balance analysis models has  been reviewed in \cite{Imam2015} and \cite{vivek2016advances}. 
The necessity of achieving such data integration, even on the network level, has recently been discussed in \cite{hao2018genome}. 
To a certain degree all these studies consider the regulation of metabolism but only cover the proximity rather than a genome scale.\par
Understanding the interplay of metabolism and gene regulation will help to gain insight in cellular, system-wide responses such as to changing environmental conditions. Here, we present the database-assisted reconstruction of an integrative \textit{E.~coli} network capturing metabolic as well as regulatory processes. 
The attribution of network components (in terms of individiual vertices) to the metabolic and regulatory domains, as well as the protein interface enables the further characterization of the network in terms of its modular organization, its path statistics and the vertex centrality.\par
In particular, we formulate a new measure by evaluating \textsl{domain-traversing paths}, in order to quantitatively assess the role of components in the interface domain and thus identify cross-systemic key elements contributing to both regulatory and metabolic processes. In all cases, these topological assessments highlight system components and functional subsystems, which are well known for their biological relevance, thus emphasizing the predictive power of network topology. Employing observations on the topological (structural, network-architectural) level, in order to identify components in the system of particular functional relevance has a long tradition in network biology (and in network science in general).\par
The main results of our investigation are: We present an integrated network representation of gene regulation and metabolism of \textit{E.~coli} and illustrate how it is a promising starting point for the structural investigation of system-wide phenomena. In particular, the network perspective suggests the explicit consideration of a protein interface between the genetic and metabolic realms of the cell. Employing network metrics we argue (1) that a three-domain partitioning is architecturally and functionally plausible, and (2) show that prominent components of the network according to the structural investigation tend to be of evident biological importance. Especially, the evaluation of possible paths through the interface domain of the network reconstruction yields well-known functional subsystems. The overlap of structural and biological relevance, here, suggests that a careful analysis of such a structural model can guide biological investigations by focusing on a limited number of structurally outstanding components.
This network model can also serve as a starting point for a range of topological analyses with methods developed in statistical physics (see, \textit{e.g}., \cite{Radde2016} for a recent review).\par
Summarizing, in contrast to the separate analyses of (\textit{e.g.}, the metabolic or gene regulatory) subsystems, we expect that the integrative network model shown here will draw the attention to system-wide feedback loops not contained in the individual subsystems and to different roles of individual components, which become only visible from the perspective of interdependent networks. 

\section*{Results}
\subsection*{Database-assisted network reconstruction}
By now, the dramatic growth of bioinformatics databases \cite{Galperin2017}, both in content and in diversity, allows addressing the challenge of integrating regulation and metabolism on a system-wide level. We devised a semi-automated framework to integrate information from EcoCyc database \cite{Keseler2012} and RegulonDB \cite{Gama-Castro2015} into a network for \textit{E.~coli} including major cellular processes, from metabolic processes via protein modifications to a variety of regulatory events (see Methods). Networks are an efficient data structure for integrating this wealth of information \cite{Ideker2012,Pratt2015,KuYu2016}. In this way, the vast amount of information contained in the bioinformatics databases provide an 'architectural embedding' for metabolic-regulatory networks and guides subsequent steps of model refinement and validation. We augmented and validated the resulting network based on existing reconstructions of metabolic \cite{Feist2007,Orth2011,OBrien2013,Smallbone2013,Liu2014} as well as of gene regulatory processes \cite{Gama-Castro2015}.\par
\begin{figure}[htb]
 \begin{minipage}{.46\textwidth}
  \includegraphics[width=.95\textwidth]{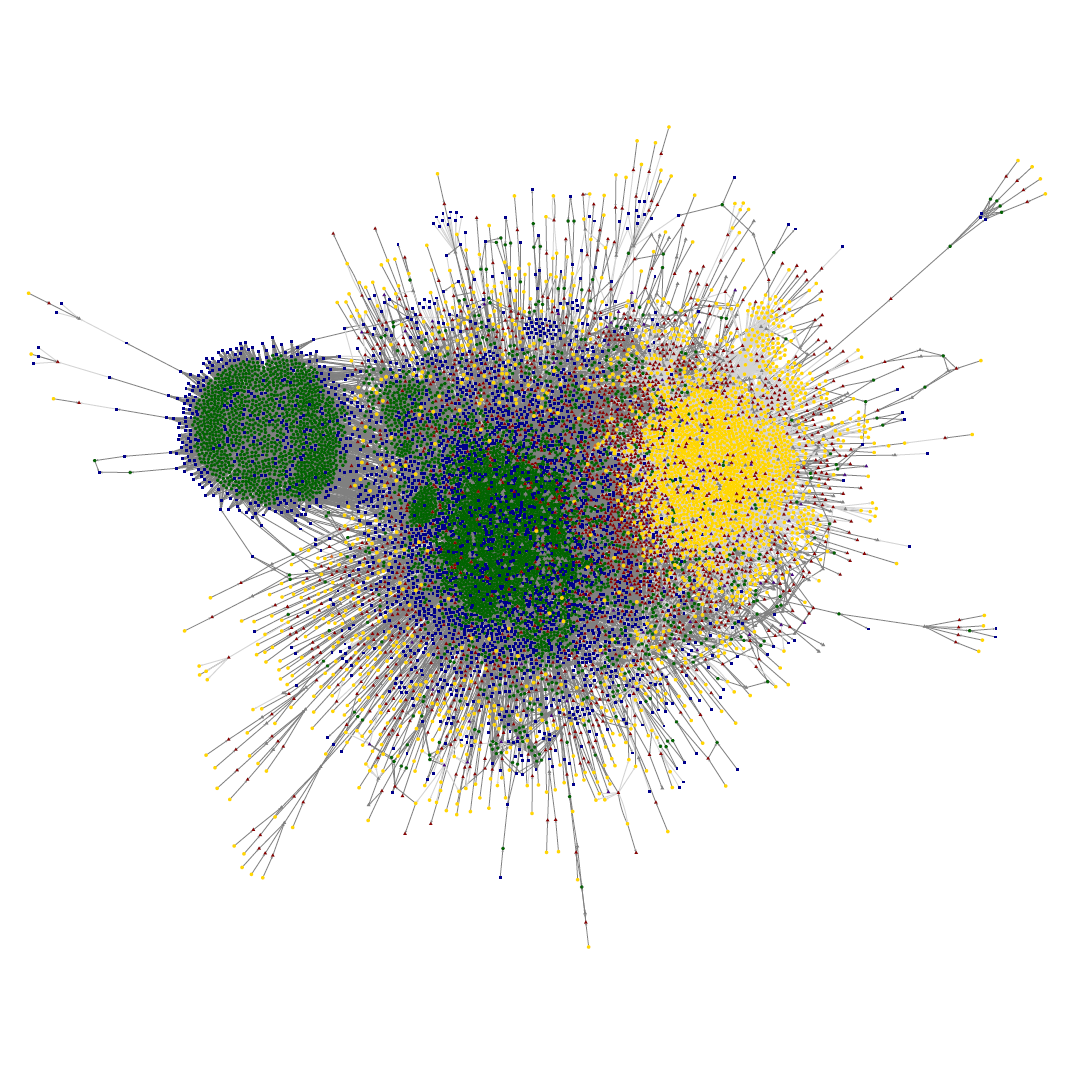}
 \end{minipage}
 \begin{minipage}{.535\textwidth}
  \tabcolsep1.5mm
  \begin{tabular}{C{1.25em}l*{2}{r}}
  \toprule
  \multicolumn{3}{l}{Vertex composition} & \textit{i}MC1010* \\
  \midrule
  \tikz\node[reaction,fill=mygreen,scale=0.55]{}; & Reaction & 4693 & 569/\hphantom{0}767 \\
  \tikz\node[compound,fill=myblue,scale=0.55]{}; & Compound & 2681 & 557/\hphantom{0}615 \\
  \tikz\node[gene,fill=myyellow,scale=0.55]{}; & Gene & 2545 & 971/1010 \\
  \tikz\node[protein,fill=myred,scale=0.55]{}; & Protein monomer & 1917 & \multirow{4}{*}{771/\hphantom{0}817} \\
  \tikz\node[protein,fill=mygray,scale=0.55]{}; & Protein-protein complex & 929 \\
  \tikz\node[protein,fill=myindigo,scale=0.55]{}; & Protein-compound complex & 100 \\
  \tikz\node[protein,fill=myorange,scale=0.55]{}; & Protein-RNA \mbox{complex} & 3 \\
  & & 12868 & 2868/3209 \\
  \bottomrule
  \multicolumn{4}{L{22em}}{\footnotesize *~accounted only for enzymatic reactions and unique metabolites (1076 and 762 in total)}
  \end{tabular}
 \end{minipage}
 \caption{Spring-block graph representation (using a scalable force directed placement algorithm) and vertex composition of the integrative \textit{E.~coli} network. The coverage of the pioneer model from \citet{Covert2004} is provided in column \textit{i}MC1010.}\label{fig:GraphVertexComposition}
\end{figure}
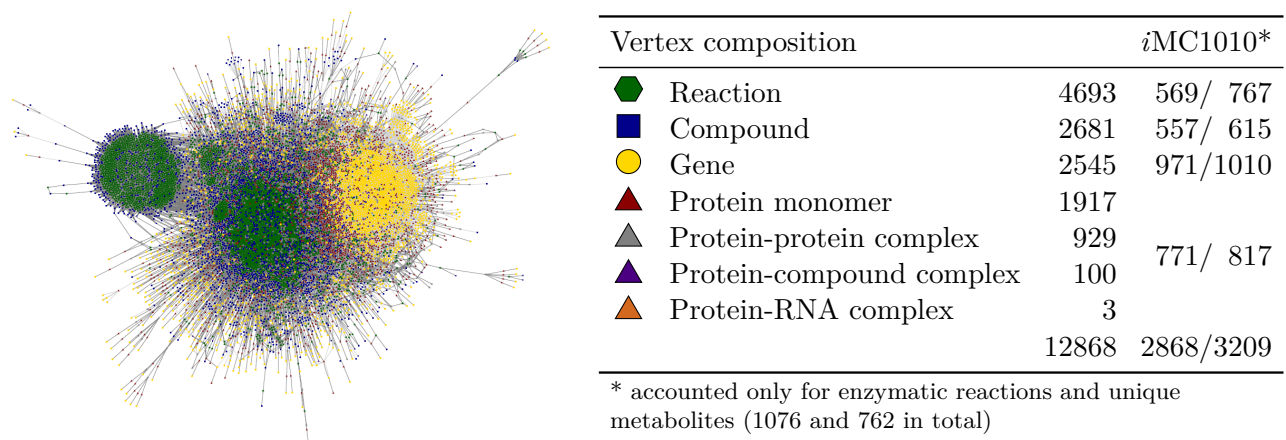

The integrative \textit{E.~coli} network constructed here comprises the three major biological components, genes, proteins, and metabolites, as well as the metabolizing reactions summing up to more than 12,000 components. Represented as a graph the network has seven types of vertices (Figure~\ref{fig:GraphVertexComposition}, Table~\ref{tab:Supp-VertexCompositionModules}) and seven different types of edges including two types of encoding associations, four reaction-associated relations, and regulatory links (Table~\ref{tab:Supp-EdgeCompositionModules}). The graph representation facilitates the mapping of reactions and their catalyzing enzymes, as both are depicted as vertices. In contrast, metabolic systems are often represented as hypergraphs to illustrate the Boolean 'AND' association of reaction educts and the fixed stoichiometric ratio of the involved metabolites. Those aspects are assigned explicitely as edge properties in the graph representation. Besides the associations of reaction educts, the encoding relations of protein complexes are of Boolean 'AND' type, termed \textsl{conjunct} links. On the contrary, associations representing isoforms of protein subunits, isoenzymes as well as reaction products are implemented by Boolean 'OR' links, called \textsl{disjunct}. The third linkage type, \textsl{regulation}, covers approximately 7,300~regulatory associations, \textit{i.e.}, transcriptional, translational as well as metabolic ones (Table~\ref{tab:Supp-RegulationTypes}).

\subsection*{The metabolic and regulatory processes}
The comparison with existing models reveals that the presented integrative network is a comprehensive representation of the metabolic and regulatory processes in \textit{E.~coli}. The very first approach of embedding metabolic processes in the regulatory context of \citet{Covert2004}, the \textit{i}MC1010 model, started from a metabolic model which was extended by the regulation of the encoding genes of the catalyzing enzymes. For the purpose of determining the overlap of the integrative metabolic-regulatory network and the \textit{i}MC1010 model, transport reactions as well as the artificial biomass reaction have been disregarded and, moreover, only unique metabolites (neglecting compartmentation) have been taken into account. Else, the different levels of details of the transport systems such as PTS as well as of the compound compartmentation would render a correct mapping impossible. Overall, the \textit{i}MC1010 model is covered by our model to more than 89~\% (Figure~\ref{fig:GraphVertexComposition}, see Table~\ref{tab:Supp-ComparisonVertexComposition}, column~3).\par
To assess the coverage of \textit{E.~coli}'s metabolic processes, the embedded metabolic processes of the integrative \textit{E.~coli} network have been associated to the ones of an established \textit{E.~coli} metabolic reconstruction, namely the \textit{i}AF1260 model from \citet{Feist2007}. About 67~\% of the involved biochemical reactions, compounds and genes could be mapped directly (see Table~\ref{tab:Supp-ComparisonVertexComposition}, column~4). Particularly, these two thirds capture almost all biologically relevant components in terms of \textit{in silico} viability. Using flux balance analysis for simulating the biomass production capacity of the \textit{i}AF1260 model and taking the overlap with mapped components of the integrative \textit{E.~coli} network revealed that for the default medium setup approximately 75~\% of the essential reactions (to yield 1~\% biomass) are covered by the integrative \textit{E.~coli} network.\par
Analogous to the metabolic processes, the coverage of \textit{E.~coli}'s gene regulation has been determined using the transcriptional regulatory network from RegulonDB \cite{Gama-Castro2015}. This model has been assembled in a similar fashion but is accounting only for transcription factors and their regulated genes. With a coverage of more than 98~\%, the transcription-related regulatory processes are considered as completely recorded in the integrative \textit{E.~coli} network (see Table~\ref{tab:Supp-ComparisonVertexComposition}, column~5). Apart from that, for this assessment of overlap a comparison of regulatory processes associated with RNA translation as well as metabolic regulatory events is not possible since the RegulonDB transcriptional regulatory network does not consider protein and metabolic interaction processes.

\subsection*{The interface of metabolic and regulatory processes}
The most conspicuous links between metabolic and gene regulatory processes are metabolic transcription factors, \textit{i.e.}, gene expression regulators binding metabolites, and metabolic genes, \textit{i.e.}, genes with significant and coordinated response on the metabolic level such as encoding enzymes. Intuitively, the interface is considered so far as the direct interactions of metabolic elements and gene regulatory elements, and the integrative \textit{E.~coli} network can be partitioned into metabolic and regulatory domain (MD~--~RD).\par
However, by examining those interactions in more detail the topological role of proteins becomes apparent. Regarding the metabolic transcription factors, the respective metabolite binds to a protein and this metabolite-protein complex then subsequently regulates the gene expression. In the case of metabolic genes, ultimately the respective gene encodes a protein which either by itself or as a complex serves as an enzyme. In line with this, the interface of metabolic and gene regulatory processes should be considered as the series of interactions of metabolites and genes, respectively, with proteins and subsequent protein modifications. Thus, the interface does not only comprise interactions (edges) but also components (vertices), and the integrative \textit{E.~coli} network will in the following be divided into a metabolic domain, a protein interface and a regulatory domain (MD~--~PI~--~RD).\par
In the next section, the plausibility of the three-domain partition (and the set of biologically motivated rules devised to create it) will be assessed in comparison to the likewise proposed two-domain (MD~--~RD) representation.

\subsubsection*{The interface structure -- a matter of network partition}
In order to assess the large-scale structure of the reconstructed network we apply a set of rules that assign each vertex of the network to one of two and three domains, respectively, by considering the biological types of the vertices themselves as well as those of their neighbors
(as outlined in the Methods section). Since these rules have been designed to group together vertices connected to the same biological
processes we expect them to result in biologically plausible network partitions.\par
\begin{figure}[p]
 \centering
 \begin{tikzpicture}
  \node[label={[anchor=south]-89:{Functional three-domain partition}}] (MD-PI-RD){\includegraphics[width=.45\textwidth]{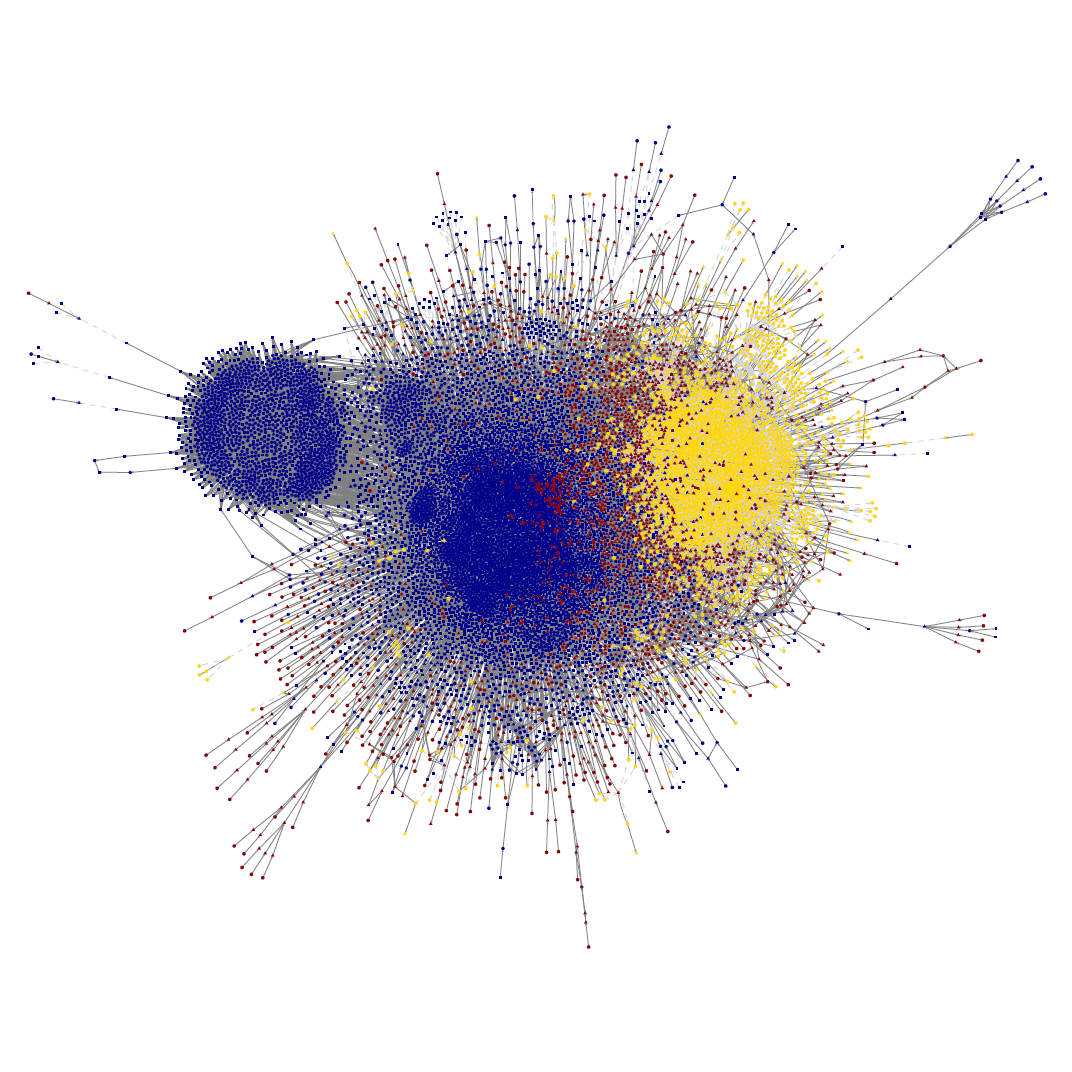}};
  \node[above right=-3.5em and -6.2em of MD-PI-RD] {\includegraphics[scale=0.21]{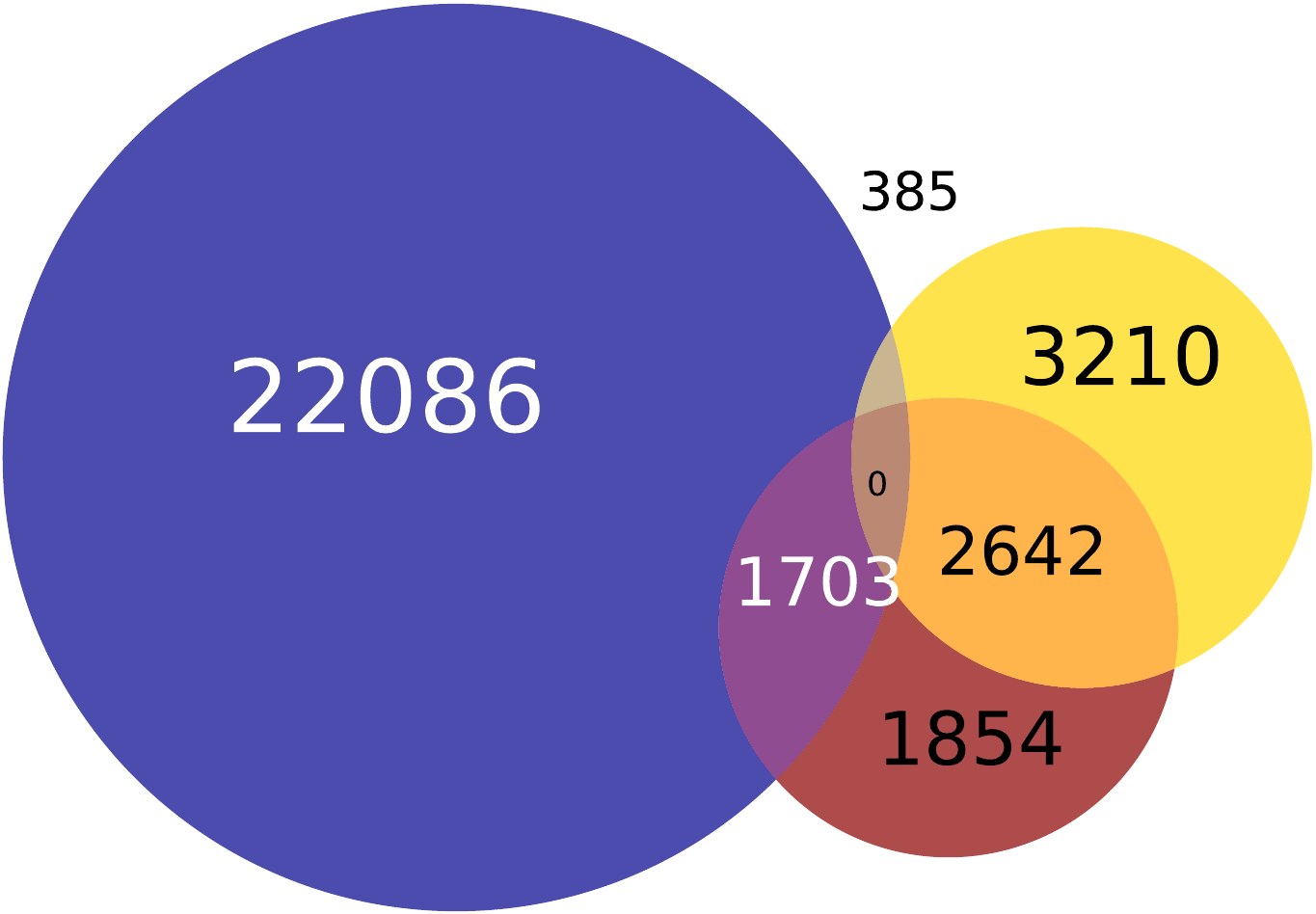}};
  \node[above left=0.5em and -2em of MD-PI-RD] {\textbf{A}};
  \node[label={[anchor=south]-89:{Functional two-domain partition}},right=1.6em of MD-PI-RD] (MD-RD) {\includegraphics[width=.45\textwidth]{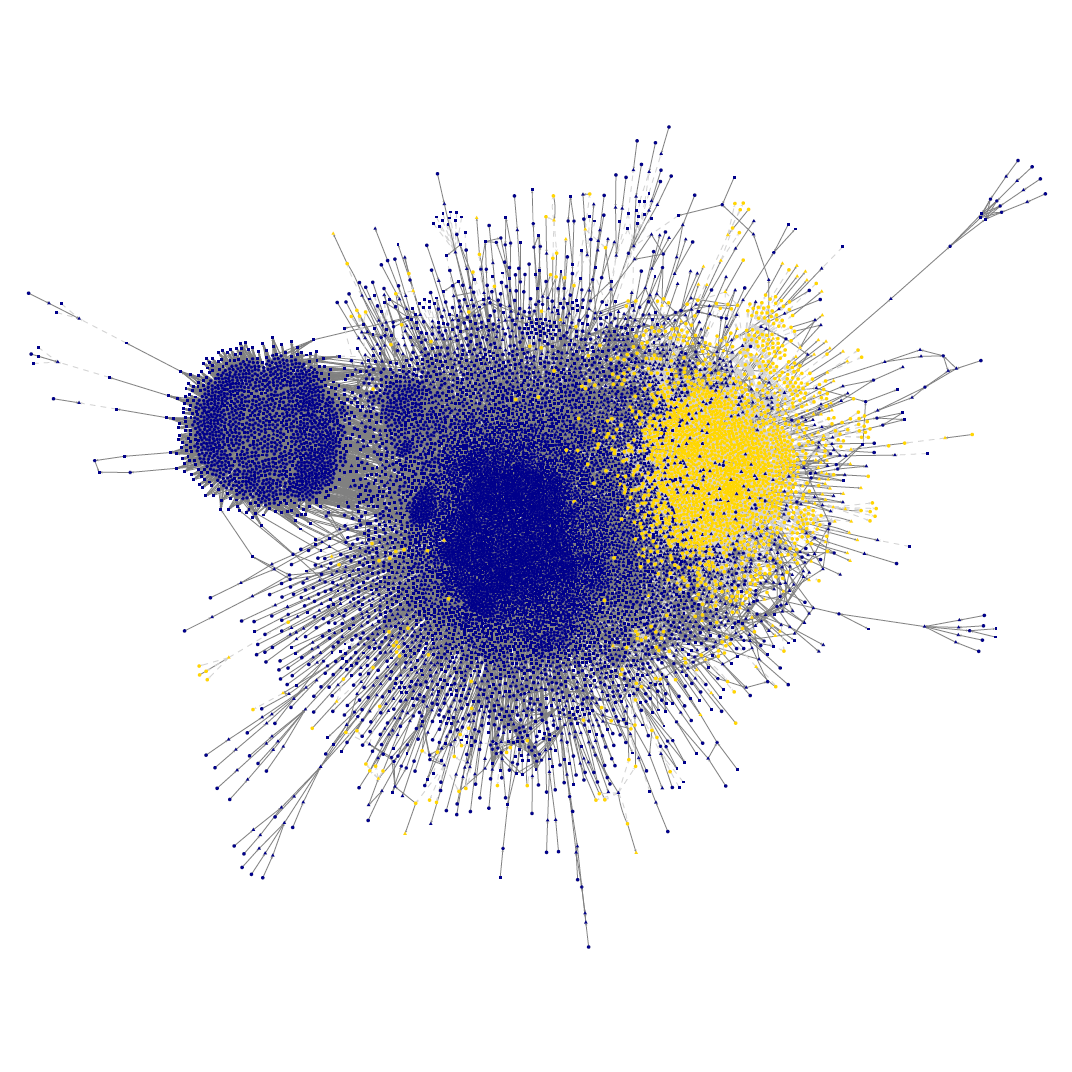}};
  \node[above right=-3.5em and -6.2em of MD-RD] {\includegraphics[scale=0.18]{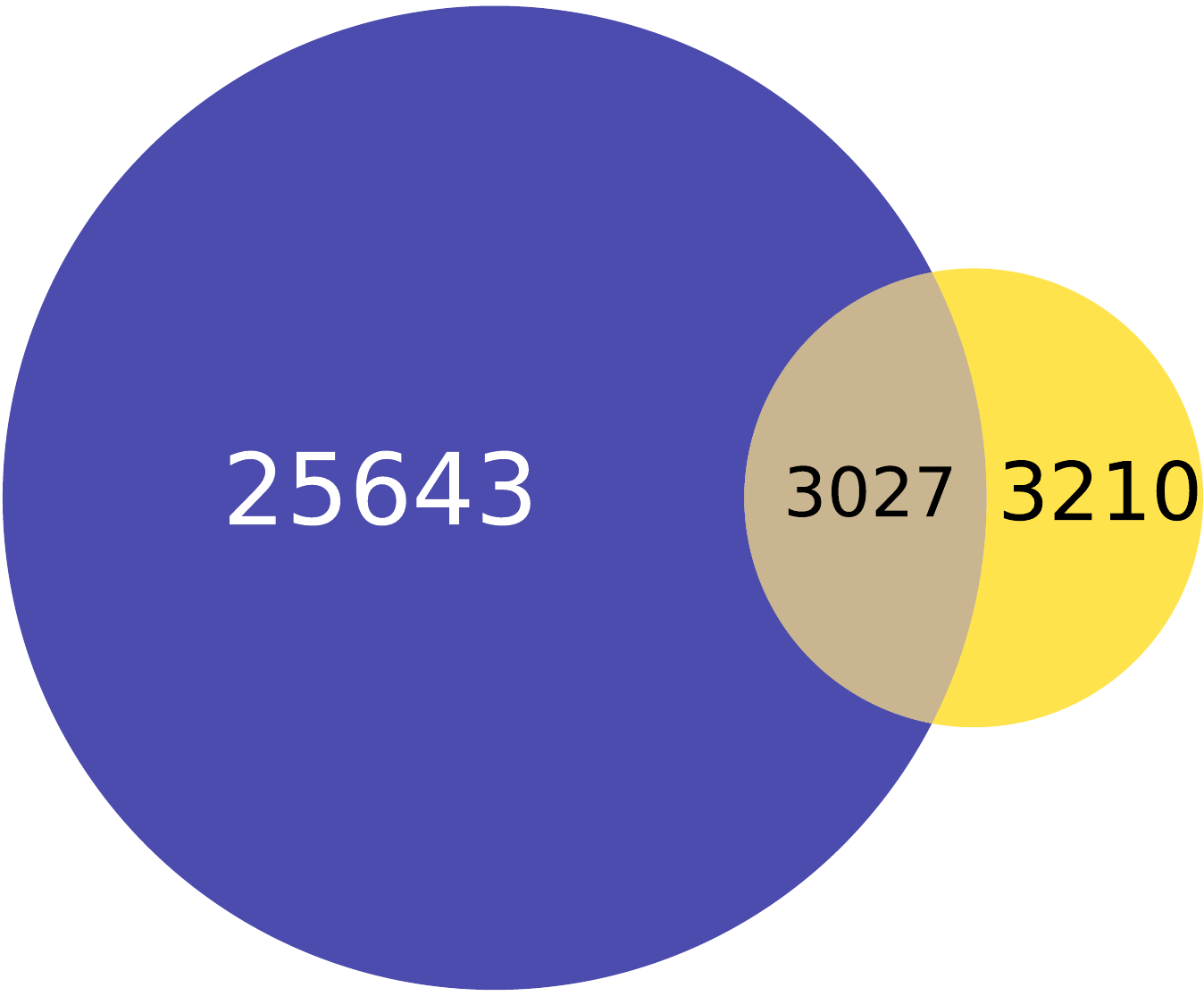}};
  \node[above left=0.5em and -2em of MD-RD] {\textbf{B}};
  \node[label={[anchor=south]-89:{Vertex type-driven three-domain partition}},below=3.5em of MD-PI-RD] (V3) {\includegraphics[width=.45\textwidth]{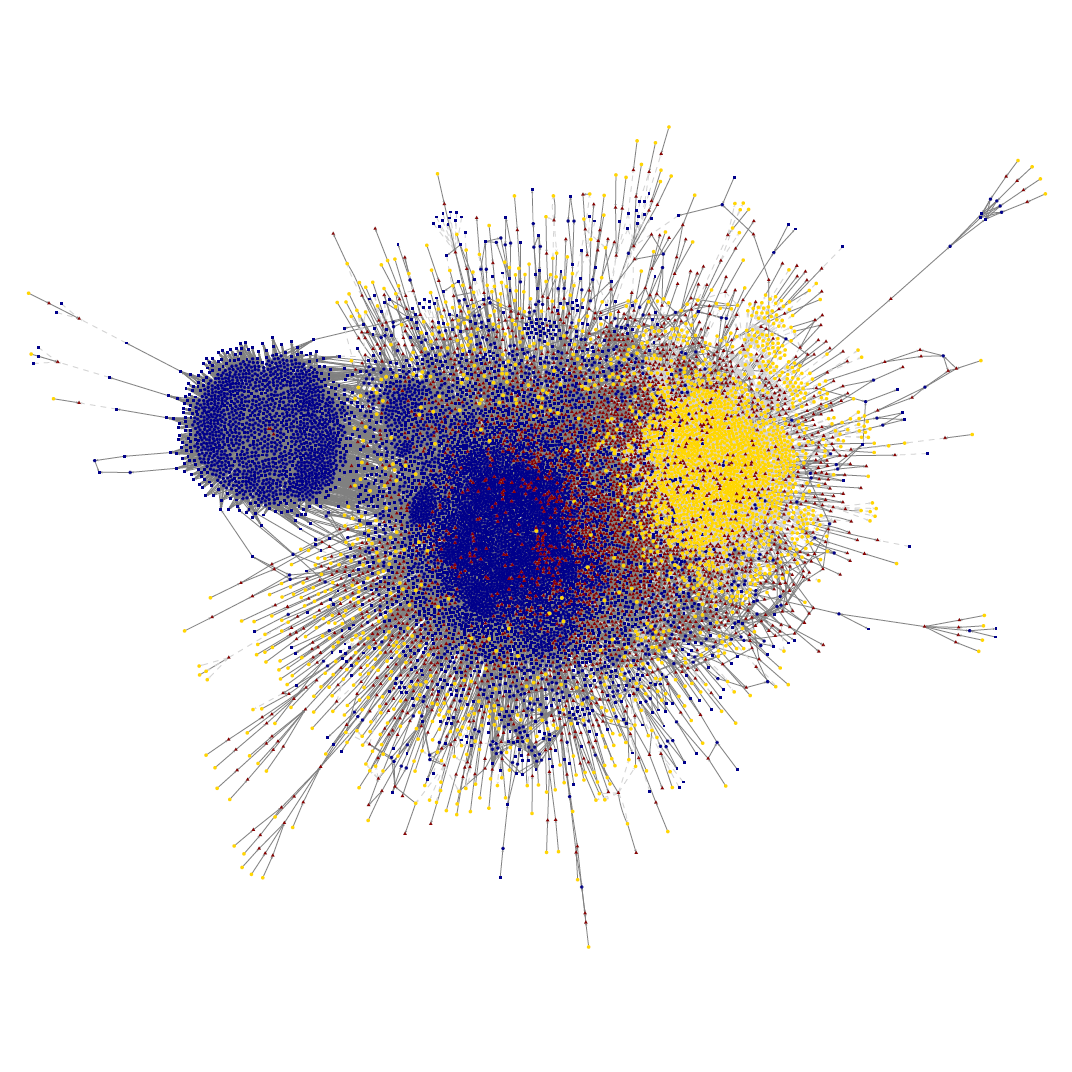}};
  \node[above right=-3.5em and -6.2em of V3] {\includegraphics[scale=0.185]{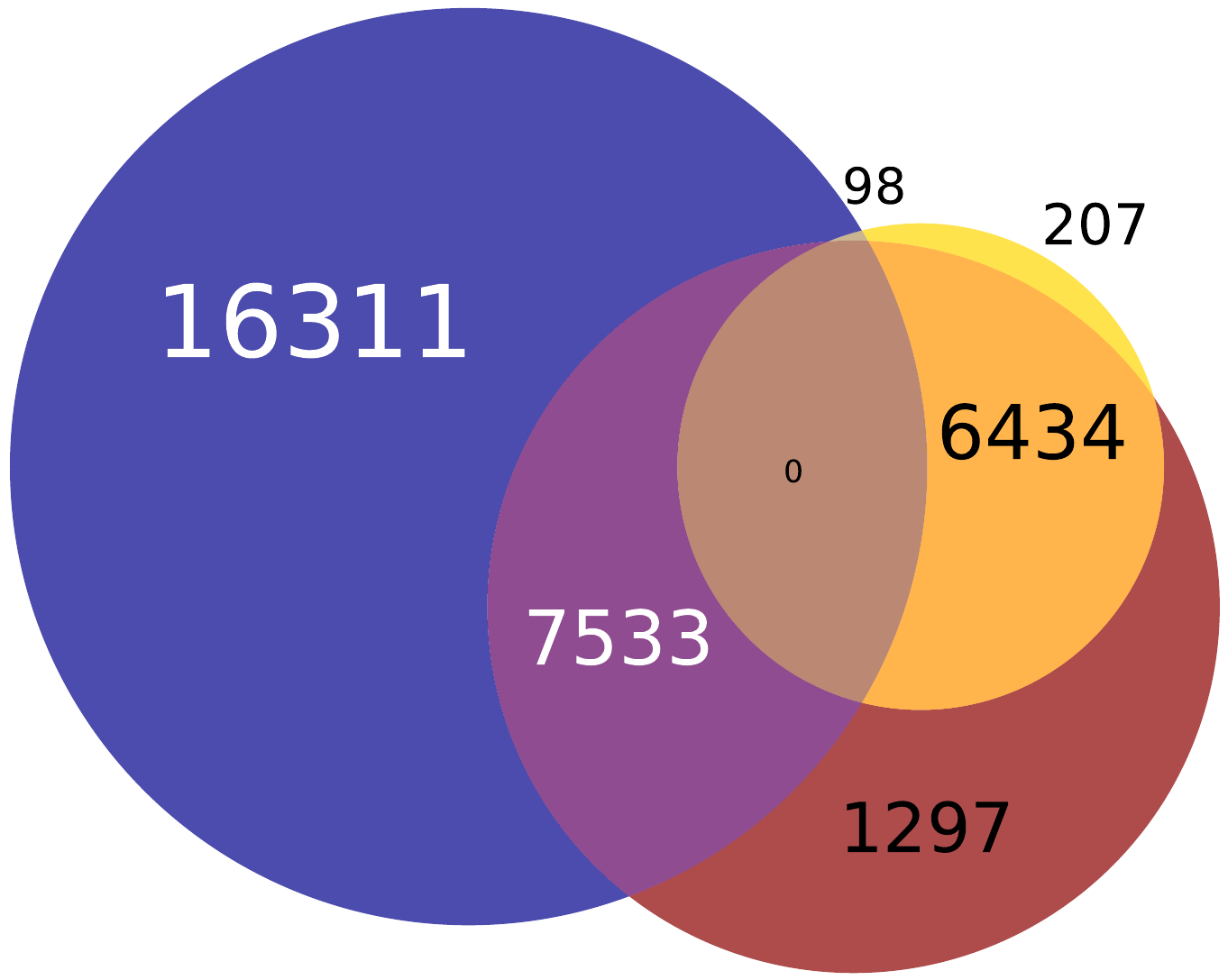}};
  \node[above left=0.5em and -2em of V3] {\textbf{C}};
  \node[label={[anchor=south]-89:{Vertex type-driven two-domain partition}},below=3.5em of MD-RD] (V2) {\includegraphics[width=.45\textwidth]{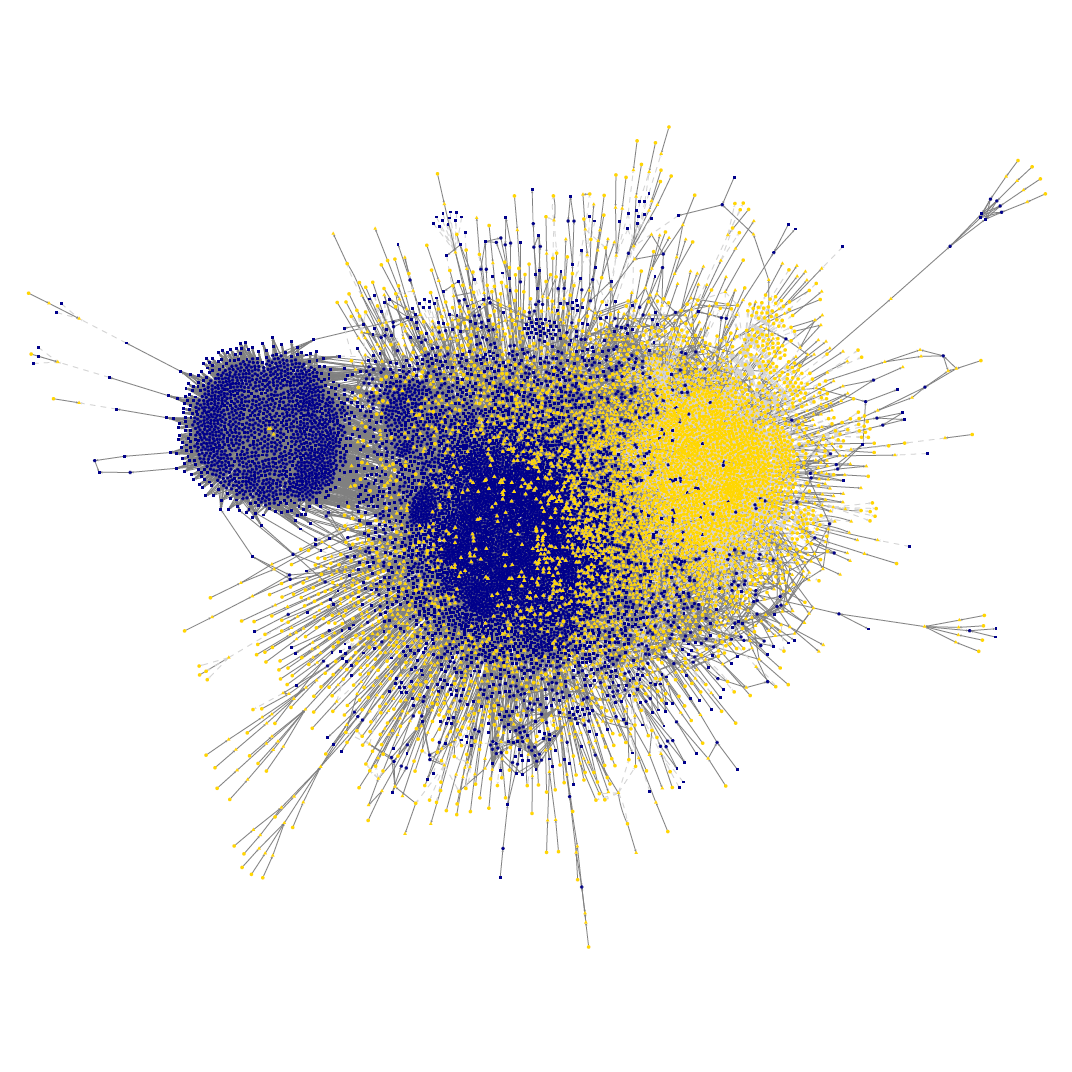}};
  \node[above right=-3.5em and -6.2em of V2] {\includegraphics[scale=0.195]{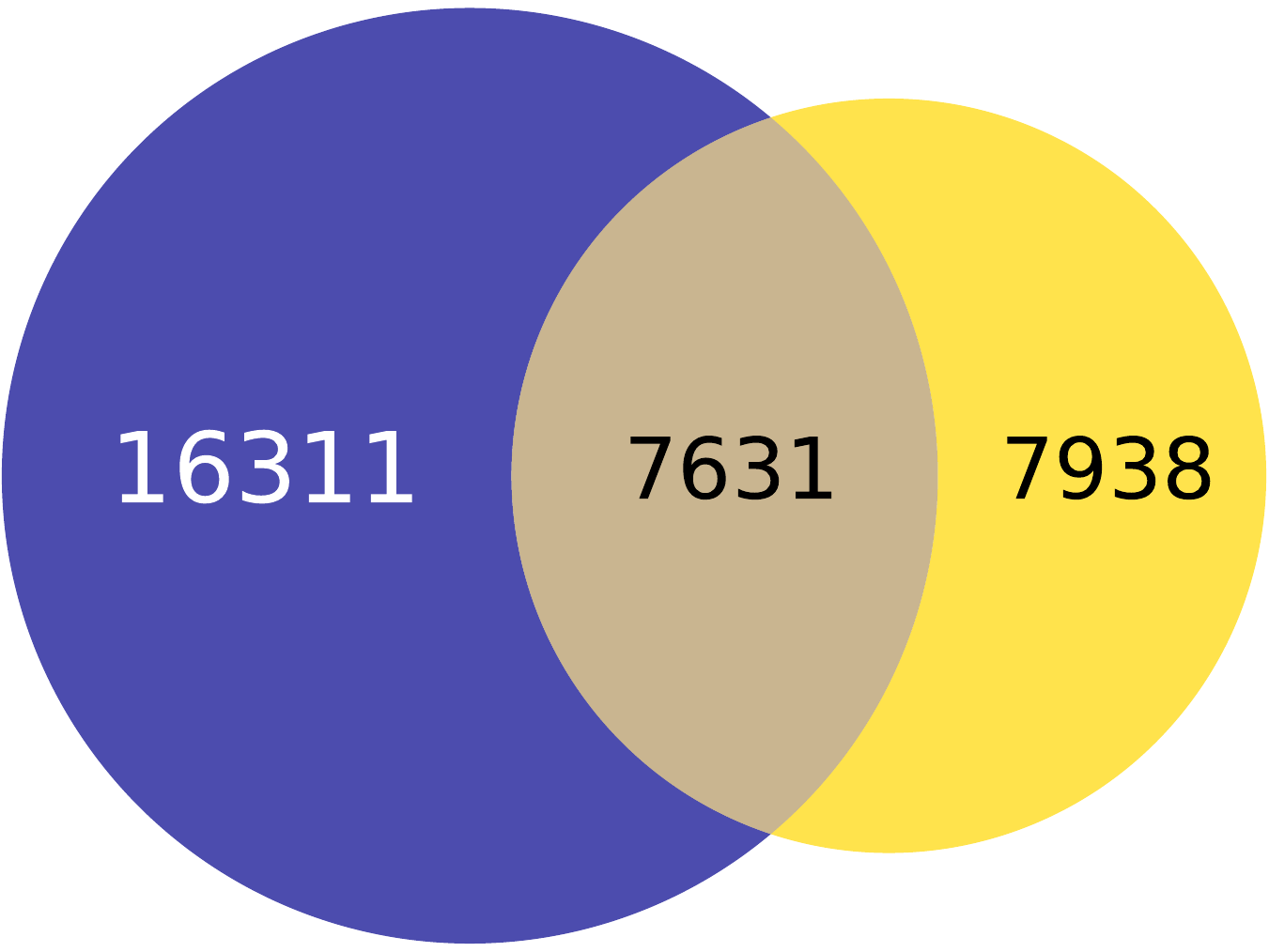}};
  \node[above left=0.5em and -2em of V2] {\textbf{D}};
 \end{tikzpicture}
 \caption{Graph snapshots of the four partitions: the functional three-domain partition into metabolic
 and regulatory domains and protein interface (MD~--~PI~--~RD) (\textbf{A}), the functional two-domain partition into metabolic and regulatory domains (MD~--~RD) (\textbf{B}), vertex-driven three-domain partition into compounds/reactions, proteins and genes (\textbf{C}), vertex-driven two-domain partition into compounds/reactions, and proteins/genes (\textbf{D}). Vertices are colored according to their domain-affiliation: yellow -- (pseudo) regulatory and gene-focused domain, respectively, and blue -- (pseudo) metabolic  and compound-focused domain, respectively. The interface domain in the three-domain partitions are drawn in red. The diagrams in the top right corners of each panel show the edge composition of the system in terms of intra-domain and 
  inter-domain edges.}\label{fig:Partitions}
\end{figure}
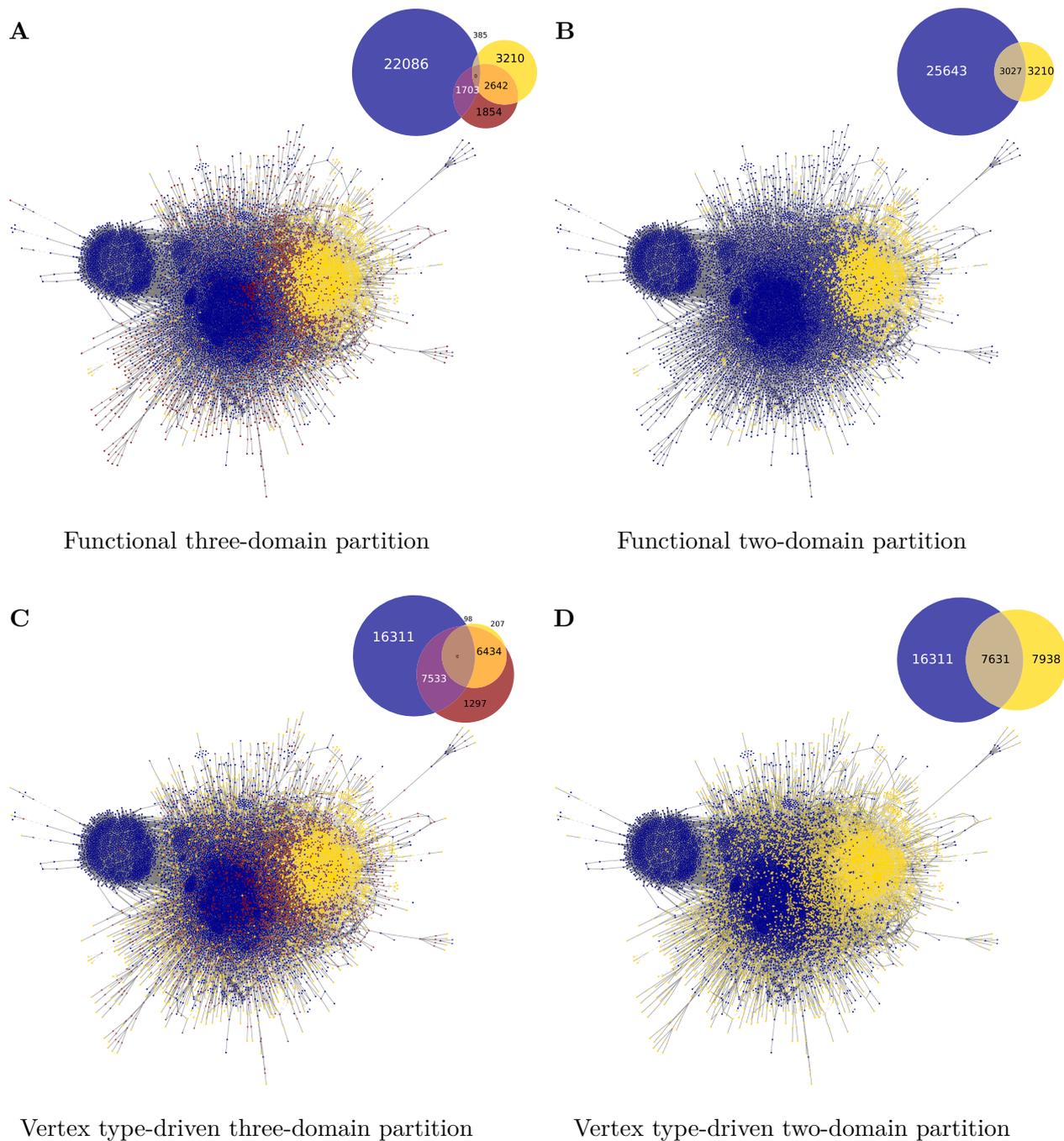

To complement the two \textsl{functional} partitions, MD~--~RD and MD~--~PI~--~RD, two partitions that solely take into account the vertex types have been analyzed, also representing a metabolic-regulatory division into two and three domains, respectively. For the \textsl{vertex-driven} two-domain partition, the sets of gene and protein vertices denote the regulatory processes while in the three-domain partition regulation is given by the set of genes and the interface domain only consists of the protein vertices. In both cases, metabolism is represented by the sets of reactions and compounds. In the three-domain case, the vertex-driven three-domain partition, the vertex set of proteins form an interface similar to the MD~--~PI~--~RD partition (Figure~\ref{fig:Partitions}). The functional and vertex-driven three-domain partitions are of roughly similar size in terms of vertex count, while the respective two-domain partitions have a metabolic-regulatory vertex ratio of 5:1 and 4:3, respectively (see Table~\ref{tab:PartitionProperties}).

First, the two three-domain partitions will be compared, \textit{i.e.}, the functional partition, MD~--~PI~--~RD, and the vertex-driven partition. In the following, we will argue that the additional third domain acts as an interface between the regulatory and metabolic domains in the functional partition, while we will see that the vertex-driven partition fails to give a coherent picture of the domain-level organization of the biological system. 

 Especially, it will become clear, also in later sections, that the interface domain in the functional partition contains processes that are known to play prominent roles in system-scale communication within the cell, and may therefore be considered an important component of the large-scale organizational structure of the combined regulation and metabolism of \textit{E.~coli}.\par
A simple quantity to illustrate the domain-level picture is the fraction of inter-module edges (linking to a vertex of a different domain) over all edges connected to vertices of a specific domain (\textit{i.e.}, external and internal edges). Of course, there is no objectively 'correct' partition the result of our procedure could be measured against, but there are a number of fundamental properties that a biologically plausible partition in the given context should possess. On the one hand, a proper interface provides the main means of communication between the regulatory and the metabolic processes, \textit{i.e.}, the majority of paths between the outer two domains should run through the interface. Indeed, the interface of the functional partition shows a considerably larger inter-module edge fraction than the remaining domains (0.7 compared to 0.5 and 0.1, Table~\ref{tab:PartitionProperties}), stressing its special character as a bridging module. A high inter-module edge fraction of the interface is also found in the vertex-driven partition, however, its regulatory domain shows an even higher inter-module edge fraction which indicates an entanglement between the two groups rather than one domain acting as a bridging module to another domain. This exactly give rise for the second criteria, that the domains should capture actual processes (here, structures on the level of several vertices). Unambiguously, regulatory or metabolic processes should be contained within the respective domain so that system-wide interaction takes place between processes. In the following chapter, Interface characterization, it will be shown that this actually is also the case for the interface in the functional partition. In contrast, in the vertex-driven partition already the regulatory domain show deficiencies with respect to that criterion. Since this regulatory domain solely contains gene-gene interactions the intermediate transcription factor steps are not within the domain which become visible in the almost exclusively inter-module edges, linking it to the interface domain.

\begin{table}[htb]
 \centering
 \caption{Topological properties of the functional and vertex type-driven network partitions. The functional partitions are denoted by the respective modules, metabolic domains (MD), regulatory domain (RD) and protein interface (PI). The vertex type-driven partitions are represented by the comprising vertex types, reaction (\protect\tikz\protect\node[reaction,fill=mygreen,scale=0.4]{};), compound (\protect\tikz\protect\node[compound,fill=myblue,scale=0.4]{};), gene (\protect\tikz\protect\node[gene,fill=myyellow,scale=0.4]{};), and protein (\protect\tikz\protect\node[protein,fill=myred,scale=0.4]{};). For each property, the module-specific coefficients and contributions (I, II, III) are presented, respectively. For the modularity, $M$, the overall network coefficient (Total) is shown as well as the best coefficient is underlined,  the module-specific values correspond to the terms in the sum of equation (\ref{eq:Modularity}).}\label{tab:PartitionProperties} 
  \begin{tabular}{lc*{4}{r}}
  \toprule
  & & \multicolumn{2}{l}{Functional partitions} & \multicolumn{2}{l}{Vertex-driven partitions} \\
  \cmidrule(r){3-4}\cmidrule(l){5-6}
  & & MD~--~PI~--~RD & MD~--~RD & \tikz\node[reaction,fill=mygreen,scale=0.55]{}; \tikz\node[compound,fill=myblue,scale=0.55]{};~--~\tikz\node[protein,fill=myred,scale=0.55]{};~--~\tikz\node[gene,fill=myyellow,scale=0.55]{}; & \tikz\node[reaction,fill=mygreen,scale=0.55]{}; \tikz\node[compound,fill=myblue,scale=0.55]{};~--~\tikz\node[protein,fill=myred,scale=0.55]{}; \tikz\node[gene,fill=myyellow,scale=0.55]{}; \\
  & & \makebox[1.65em][c]{I}~--~\makebox[1.1em][c]{II}~--~\makebox[1.5em][c]{III} & \makebox[1.65em][c]{I}~--~\makebox[1.5em][c]{III} & \makebox[2.2em][c]{I}~--~\makebox[.9em][c]{II}~--~\makebox[.9em][c]{III} & \makebox[2.2em][c]{I}~--~\makebox[2.1em][c]{III} \\
  \midrule
  Vertices & I & 8369 & 10655 & 7374 & 7374 \\
  & II & 2286 &  & 2949 & \\
  & III & 2213 & 2213 & 2545 & 5494\\
  \rowcolor{black!10}& I & 0.086 & 0.106 & 0.319 & 0.319 \\
  \rowcolor{black!10}\multirow{-2}{6.5em}{Inter-module edge fraction}& II & 0.701 &  & 0.915 & \\
  \rowcolor{black!10}& III & 0.485 & 0.485 & 0.969 & 0.49 \\
  Modularity, $M$ & Total & \underline{0.287} & 0.157 & 0.081 & 0.226 \\
  & I & 0.166 & 0.079 & 0.113 & 0.113 \\
  & II & 0.042 &  & -0.027 & \\
  & III & 0.078 & 0.079 & -0.005 & 0.113 \\
  \bottomrule
 \end{tabular}
\end{table}

Next, we compare the three-domain partitions with the two-domain partitions. While the introduction of a third domain allows to study the system in terms of an explicit interface, the partitions into two domains is much closer to common biological intuition. The question which needs to be answered is whether metabolism and gene regulation are solely interfaced by the linking processes such as gene expression, and activation or inhibition of transcription factors and genes, so that the system can appropriately be described with two domains. Or whether there is an actual interface that preferably comprises entire processes additionally including protein modifications and suchlike. Here, this question will be assessed from a topological perspective.\par
A relevant topological quantity is the network modularity \cite{Guimera2005} of a given network partition. For a biologically meaningful classification, one would expect on the network level that the regulatory and the metabolic domains show high intra-module connectivity (a large number of links are within a domain) and sparse inter-module linkages (a small number of links are between domains). Accordingly, the network modularity should be high for a successful partition. The results for the modularity are listed in Table~\ref{tab:PartitionProperties}. The functional partitions clearly outperform the vertex type-driven partitions. Also, when going from MD~--~RD to MD~--~PI~--~RD there is a notable increase in the modularity of the network ($M_2 = 0.157$, $M_3 = 0.287$). Note that here we consider specific candidates for biologically plausible partitions, while a purely topological analysis of the module structure of this large network yields a much larger set of significant modules. Here, a detailed biological interpretation is still missing and will be discussed elsewhere.\par
Altogether, the functional partition into metabolic domain, protein interface and regulatory domain reflects a biologically reliable classification in two delimited domains linked by a bridging module. Reinforced by the topological properties, the interface structure including full protein modification processes will be used subsequently.

\subsubsection*{Interface characterization}
The interface of metabolic and gene regulatory processes of the integrative \textit{E.~coli} network comprises, as expected, predominantly proteins, \textit{i.e.}, monomers and complexes (Table~\ref{tab:Supp-VertexCompositionModules}), and mainly protein modification processes such as protein translation, protein complex formation and biochemical protein conversion (Table~\ref{tab:Supp-EdgeCompositionModules}). On closer examination, the covered processes can be divided in internal and peripheral ones. According to the bridging role of the interface, the majority of these are peripheral processes (Figure~\ref{fig:Partitions}, Table~\ref{tab:Supp-EdgeCompositionModules}). The peripheral processes, in turn, can be subdivided according to their directionality meaning from regulatory to metabolic domain (subsequently termed 'downwards') and from metabolic to regulatory domain ('upwards'), respectively. To enumerate the portion of peripheral processes forming complete paths across the interface, direct downwards and upwards links and the new topological concept of \textsl{domain-traversing paths} (or short: \textsl{traversing paths}) have to be considered. A traversing path connects regulatory and metabolic domain via the protein interface, whereby only starting and end vertex are not affiliated to the bridging domain  and the path direction is considered carefully (see Methods).\par
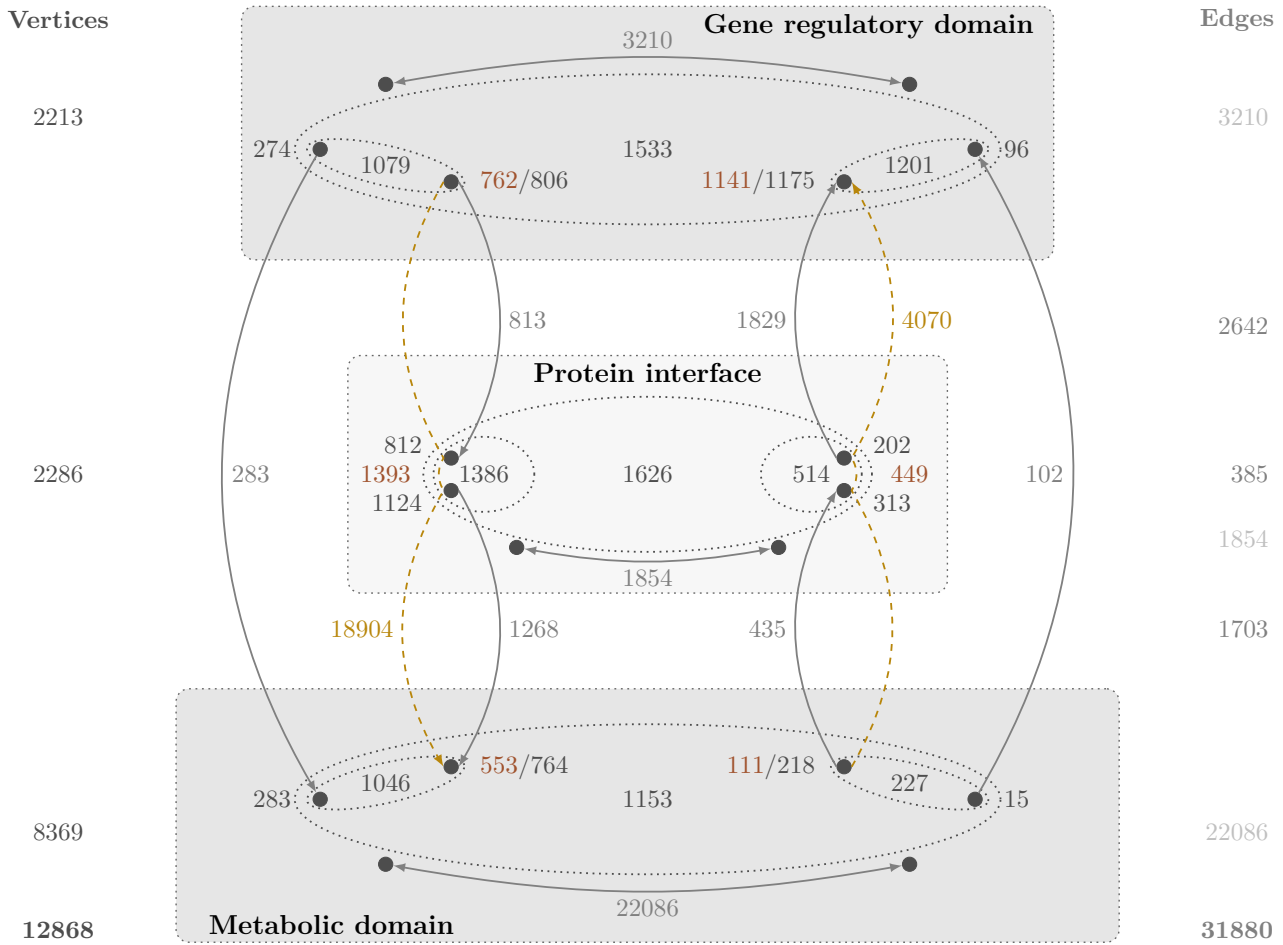
\begin{figure}[htb]
 \centering
 \resizebox{\textwidth}{!}{\begin{tikzpicture}
 \coordinate (x) at (1,0); \coordinate (y) at (0,1);
 \coordinate (r1) at (0,0); \coordinate (r2) at (12,3.5);
 \coordinate (p1) at (2,-2.25); \coordinate (p2) at (10,-4.75);
 \coordinate (m1) at (-1,-7); \coordinate (m2) at (13,-10.5);
 \node[box,dotted,semithick,fit=(r1)(r2),label={[anchor=north east]18:{\bfseries\large Gene regulatory domain}}] {};
 \node[box,dotted,semithick,fill=black!3,inner sep=1.5em,fit=(p1)(p2),label={[anchor=north]90:{\bfseries\large Protein interface}}] {};
 \node[box,dotted,semithick,fit=(m1)(m2),label={[anchor=south west]-164:{\bfseries\large Metabolic domain}}] {};
 
 \node[black!70,font=\bfseries] (V) at ($(r2)-15*(x)$) {Vertices};
 \node[black!70] (VR) at ($(V)-1.5*(y)$) {2213};
 \node[black!70] (VI) at ($(V)-7*(y)$) {2286};
 \node[black!70] (VM) at ($(V)-12.5*(y)$) {8369};
 \node[black!70,font=\bfseries] (aV) at ($(V)-14*(y)$) {12868};
 
 \node[mygray,font=\bfseries] (E) at ($(r2)+3*(x)$) {Edges};
 \node[mylightgray] (ER) at ($(E)-1.5*(y)$) {\hphantom{0}3210};
 \node[mygray] (ERI) at ($(E)-4.725*(y)$) {\hphantom{0}2642};
 \node[mygray] (ERM) at ($(E)-7*(y)$) {\hphantom{00}385};
 \node[mylightgray] (EI) at ($(E)-8*(y)$) {\hphantom{0}1854};
 \node[mygray] (EMI) at ($(E)-9.375*(y)$) {\hphantom{0}1703};
 \node[mylightgray] (EM) at ($(E)-12.5*(y)$) {22086};
 \node[mygray,font=\bfseries] (aE) at ($(E)-14*(y)$) {31880};

 \node[node,label={[xshift=-0.5em,black!70]180:{274}}] (R1) at ($(r1)+(x)+1.5*(y)$) {};
 \node[node,label={[xshift=-0.5em,black!70]180:{283}}] (M1) at ($(m1)+2*(x)-1.5*(y)$) {}; 
 \node[node,label={[xshift=0.5em,black!70]0:{\textcolor{mycol1}{762}/806}}] (R2) at ($(r1)+3*(x)+(y)$) {};
 \node[node,label={[xshift=0.5em,black!70]0:{\textcolor{mycol1}{553}/764}}] (M2) at ($(m1)+4*(x)-1.0*(y)$) {};
 \node[node,label={[xshift=-0.5em,black!70,yshift=0.5em]180:{812}}] (I1) at ($(p1)+(x)-1.0*(y)$) {};
 \node[node,label={[xshift=-0.5em,black!70,yshift=-0.5em]180:{1124}}] (I2) at ($(p1)+(x)-1.5*(y)$) {};
 \node[mycol1] at ($0.5*(I1)+0.5*(I2)-1.*(x)$) {1393};
 
 \node[node,label={[xshift=-0.5em,black!70]180:{\textcolor{mycol1}{111}/218}}] (M3) at ($(m1)+10*(x)-1.0*(y)$) {};
 \node[node,label={[xshift=-0.5em,black!70]180:{\textcolor{mycol1}{1141}/1175}}] (R3) at ($(r1)+9*(x)+(y)$) {};
 \node[node,label={[xshift=0.5em,black!70]0:{15}}] (M4) at ($(m1)+12*(x)-1.5*(y)$) {};
 \node[node,label={[xshift=0.5em,black!70]0:{96}}] (R4) at ($(r1)+11*(x)+1.5*(y)$) {};
 \node[node,label={[xshift=0.5em,black!70,yshift=-0.5em]0:{313}}] (I3) at ($(p1)+7*(x)-1.5*(y)$) {};
 \node[node,label={[xshift=0.5em,black!70,yshift=0.5em]0:{202}}] (I4) at ($(p1)+7*(x)-1.0*(y)$) {};
 \node[mycol1] at ($0.5*(I3)+0.5*(I4)+1.*(x)$) {449};
 
 \node[node] (R5) at ($(r1)+2*(x)+2.5*(y)$) {};
 \node[node] (R6) at ($(r1)+10*(x)+2.5*(y)$) {};
 \node[node] (I5) at ($(p1)+2*(x)-2.375*(y)$) {};
 \node[node] (I6) at ($(p1)+6*(x)-2.375*(y)$) {};
 \node[node] (M5) at ($(m1)+3*(x)-2.5*(y)$) {};
 \node[node] (M6) at ($(m1)+11*(x)-2.5*(y)$) {};
 
 \draw[line,rotate=-13] ($0.5*(R1)+0.5*(R2)$) ellipse [x radius=3.2em,y radius=0.8em] node {1079};
 \draw[line,rotate=13] ($0.5*(R3)+0.5*(R4)$) ellipse [x radius=3.2em,y radius=0.8em] node {1201};
 \draw[line] ($0.5*(R1)+0.5*(R4)$) ellipse [x radius=14.em,y radius=3.em] node {1533};
 \draw[line] ($0.5*(I1)+0.5*(I2)+0.5*(x)$) ellipse [x radius=2.0em,y radius=1.5em] node {1386};
 \draw[line] ($0.5*(I3)+0.5*(I4)-0.5*(x)$) ellipse [x radius=2.0em,y radius=1.5em] node {514};
 \draw[line] ($0.5*(I1)+0.5*(I3)$) ellipse [x radius=8.9em,y radius=3.1em] node {1626};
 \draw[line,rotate=13] ($0.5*(M1)+0.5*(M2)$) ellipse [x radius=3.2em,y radius=0.8em] node {1046};
 \draw[line,rotate=-13] ($0.5*(M4)+0.5*(M3)$) ellipse [x radius=3.2em,y radius=0.8em] node {227};
 \draw[line] ($0.5*(M1)+0.5*(M4)$) ellipse [x radius=14.em,y radius=3.em] node {1153};
 
 \draw[edge] (R1) to[out=-120,in=120]node[right] (DDE) {283} (M1);
 \draw[edge,dashed,mycol] (R2.west) to[out=-120,in=120] (I1.west) to[out=-120,in=120] (I2.west) to[out=-120,in=120]node[mycol,left] (IDEA) {18904} (M2.west);
 \draw[edge] (R2.east) to[out=-60,in=60]node[right] (rpDE) {813} (I1.east);
 \draw[edge] (I2.east) to[out=-60,in=60]node[right] (pmDE) {1268} (M2.east);
 \draw[edge] (M4) to[out=60,in=-60]node[left] (DUE) {102} (R4);
 \draw[edge,dashed,mycol] (M3.east) to[out=60,in=-60] (I3.east)  to[out=60,in=-60] (I4.east) to[out=60,in=-60]node[mycol,right] (IUEA) {4070} (R3.east);
 \draw[edge] (M3.west) to[out=120,in=-120]node[left] (mpUE) {435} (I3.west);
 \draw[edge] (I4.west) to[out=120,in=-120]node[left] (prUE) {1829} (R3.west);
 \draw[edge,latex-latex] (R5) to[out=10,in=170]node[above] (rIE) {3210} (R6);
 \draw[edge,latex-latex] (I5) to[out=-10,in=-170]node[below] (pIE) {1854} (I6);
 \draw[edge,latex-latex] (M5) to[out=-10,in=-170]node[below] (mIE) {22086} (M6);
 \end{tikzpicture}
 }\caption{Schematic overview of the components and connections of the integrative \textit{E.~coli} network, especially those involved in the protein interface. The information about edges are presented in gray and about traversing paths are shown in dark goldenrod while the number of vertices are shown in dark blue and the traversing paths-related ones are given in dark brown, in addition. The solid lines denote direct link connections while the dashed lines the traversing paths connections.}\label{fig:InterfaceTruePaths}
\end{figure}

Examination of the downwards-upwards subdivision, especially the traversing paths, reveals a considerable (though biologically expected) asymmetry of the interface (Figure~\ref{fig:InterfaceTruePaths}): The downwards interface is much more pronounced comprising predominantly the transcription of enzymes, \textit{i.e.}, metabolic genes, and the formation of enzymatic protein complexes. On the contrary, the upwards interface is comparably sparse with roughly half the direct (102/283) and quarter the traversing paths (4,070/18,904) connections of the downwards interface. These few upwards processes mainly include the formation of metabolic regulators, especially transcription factors, and the corresponding regulatory events.

\begin{figure}[htb]
 \includegraphics[width=\textwidth]{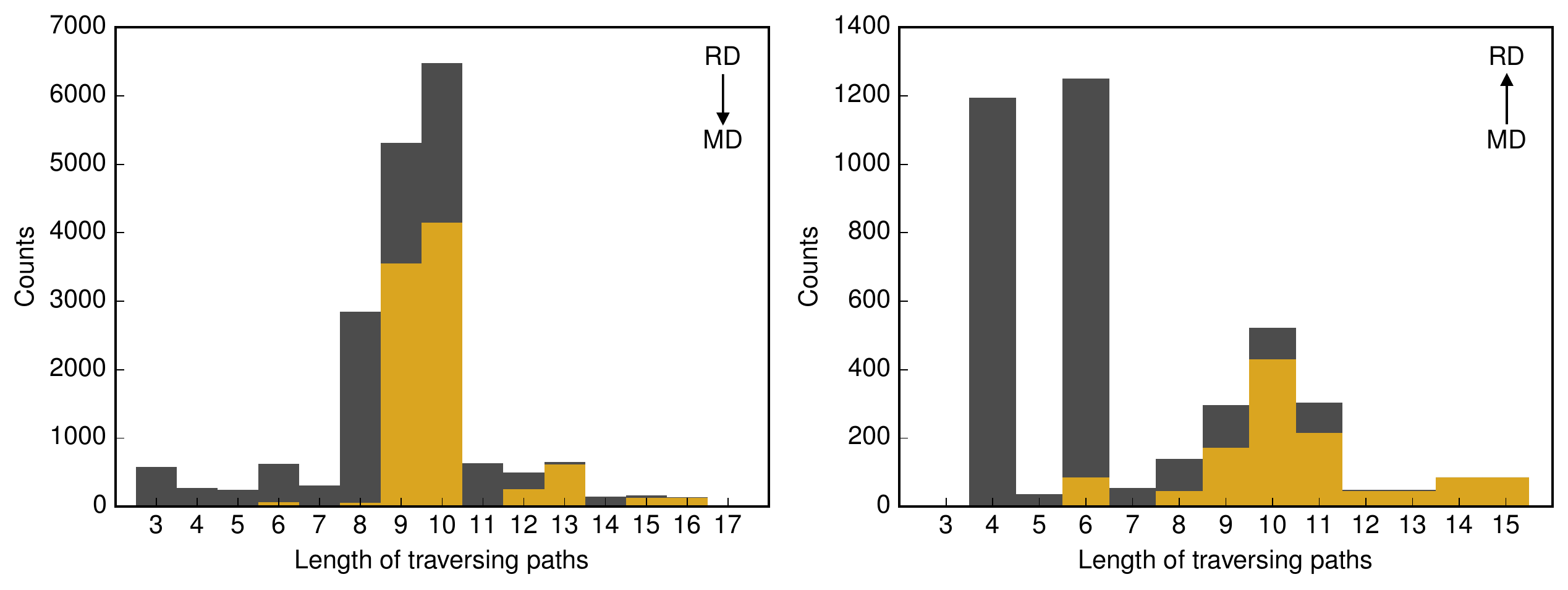} 
 \caption{Distribution of the path lengths for the downwards (RD~\raise0.25em\hbox{\protect\tikz\protect\draw[->](0,0)to(0.4,0);}~MD)    and upwards traversing paths (MD~\raise0.25em\hbox{\protect\tikz\protect\draw[->](0,0)to(0.4,0);}~RD), respectively (dark blue). The golden bars represent the fraction of downwards and upwards traversing paths comprising the PTS and RNR, and the NtrBC system associated    vertices.}\label{fig:TruePathsHistogram}
\end{figure}

In addition to confirming the interface asymmetry, the traversing paths reveal the bottleneck characteristic of the interface. First indications for this special property are (1) the low number of involved vertices and (2) the distribution of traversing path lengths. For both, downwards and upwards traversing paths, the number of distinct interface vertices in the traversing paths is low compared to the total number, \textit{i.e.}, 1,393 and 449 interface vertices of 2,286 in total, respectively (Figure~\ref{fig:InterfaceTruePaths}). On the other hand, for both, downwards and upwards traversing paths, emerges a remarkable clustering of paths of length 8--10 and four, six, and 9--11, respectively (Figure~\ref{fig:TruePathsHistogram}). 
This is in contrast to a smooth distribution one would expect in random graphs.
By enumerating the involved vertices it is striking that more than 44~\% of traversing paths contain one of five three-vertex-combinations, respectively. The respective combinations of downwards and upwards traversing paths pertain to three functional systems, the phosphoenolpyruvate-dependent sugar phosphotransferase system, PTS, the ribonucleotide reducing system, RNR system, as well as the nitrogen regulation two-component signal transduction system, NtrBC system (Table~\ref{tab:Supp-TruePathVertices}).

\begin{figure}[htb]
 \centering
 \begin{tikzpicture}
  \tikzset{vertexA/.style={draw,ellipse,thick,color=black!40,text=black,minimum height=2.2em,minimum width=4.6em},
           vertexB/.style={draw,ellipse,thick,color=black!40,text=black,minimum height=2.2em,minimum width=6.5em},
           vertexC/.style={draw,ellipse,thick,color=black!40,text=black,minimum height=2.2em,minimum width=5.2em},
           vertexD/.style={draw,rectangle,thick,color=black!40,text=black,minimum height=1.5em,minimum width=3.0em},
           edge/.style={-latex,thick,black!40}}
  \node[vertexA] (Glc) {\small Glc$_{\mathrm{p}}$};
  \coordinate (V0) at (1,-2.5);
  \node[vertexA] (G6P) at ($(Glc)+(2,-1.5)$) {\small G6P};
  \coordinate (V1) at ($(G6P)+(1.5,0.75)$);
  \node[vertexA,color=black!90,fill=black!3] (PTSH-P) at ($(Glc)+(5,0)$) {\small HPr-P};
  \node[anchor=west,vertexA,color=black!90,fill=black!3] (PTSH) at ($(G6P.east)+(1.2,0)$) {\small HPr};
  \node[anchor=west,vertexA] (PTSI-P) at ($(PTSH.east)+(1.2,0)$) {\small EI-P};
  \node[anchor=west,vertexA] (PTSI) at ($(PTSH-P.east)+(1.2,0)$) {\small EI};
  \coordinate (V2) at ($0.5*(PTSH)+0.5*(PTSI-P)+(0,0.75)$);
  \node[anchor=west,vertexA] (Pyr) at ($(PTSI-P.east)+(1.2,0)$) {\small Pyr};
  \node[anchor=west,vertexA] (PEP) at ($(PTSI.east)+(1.2,0)$) {\small PEP};
  \coordinate (V3) at ($0.5*(PTSI-P)+0.5*(Pyr)+(0,0.75)$);
  \draw[edge] (Glc.east) to[out=0,in=90] (V1) to[out=-90,in=15] (G6P.east);
  \draw[edge] (PTSH-P.west) to[out=195,in=90] (V1) to[out=-90,in=165] (PTSH.west);
  \draw[edge,black!90] (PTSH.east) to[out=15,in=-90] (V2) to[out=90,in=-15] (PTSH-P.east);
  \draw[edge,black!90] (PTSI-P.west) to[out=165,in=-90] (V2) to[out=90,in=195] (PTSI.west);
  \draw[edge] (PTSI.east) to[out=-15,in=90] (V3) to[out=-90,in=15] (PTSI-P.east);
  \draw[edge] (PEP.west) to[out=195,in=90] (V3) to[out=-90,in=165] (Pyr.west);
  \draw[thick,mygray,decoration={snake,amplitude=.15mm,segment length=2.2mm},decorate] (V0) -- (1,0.5);
  \node[left=1em of V0,yshift=2] {\footnotesize Periplasm};
  \node[right=1em of V0,yshift=2] {\footnotesize Cytosol};
  \node at ($(Glc)+(-1.5,0.5)$) {\textbf{A}};
 
  \node[vertexB,below right=4em and -0.1em of V0] (NADP) {\small NADP};
  \node[vertexB] (NADPH) at ($(NADP)+(0,-1.5)$) {\small NADPH};
  \node[anchor=west,vertexB,color=black!90,fill=black!3] (rTRX) at ($(NADP.east)+(1.2,0)$) {\small red. TRX};
  \node[anchor=west,vertexB,color=black!90,fill=black!3] (oTRX) at ($(NADPH.east)+(1.2,0)$) {\small ox. TRX};
  \coordinate (V1) at ($0.5*(NADPH)+0.5*(oTRX)+(0,0.75)$);
  \node[anchor=west,vertexB] (NDP) at ($(rTRX.east)+(1.2,0)$) {\small NDP};
  \node[anchor=west,vertexB] (dNDP) at ($(oTRX.east)+(1.2,0)$) {\small dNDP};
  \coordinate (V2) at ($0.5*(oTRX)+0.5*(dNDP)+(0,0.75)$);
  \node[anchor=north,vertexD,color=black!90,fill=black!3] (RNR) at ($(V2)+(0,-1.25)$) {\small RNR};
  \draw[edge] (NADP.east) to[out=-15,in=90] (V1) to[out=-90,in=15] (NADPH.east);
  \draw[edge] (rTRX.west) to[out=195,in=90] (V1) to[out=-90,in=165] (oTRX.west);
  \draw[edge] (oTRX.east) to[out=15,in=-90] (V2) to[out=90,in=-15] (rTRX.east);
  \draw[edge] (dNDP.west) to[out=165,in=-90] (V2) to[out=90,in=195] (NDP.west);
  \draw[edge,shorten >= 0.75em,black!90] (RNR) to (V2);
  \path let \p1 = (Glc), \p2 = (NADP) in node at ($(\x1,\y2)+(-1.5,0.5)$) {\textbf{B}};

  \node[vertexC,below right=16em and -2.2em of V0] (ATP) {\small ATP};
  \node[vertexC] (ADP) at ($(ATP)+(0,-1.5)$) {\small ADP};
  \node[vertexC] (NRII) at ($(ATP.east)+(2.1,0)$) {\small NtrB};
  \node[vertexC,color=black!90,fill=black!3] (NRIIP) at ($(ADP.east)+(2.1,0)$) {\small NtrB-P};
  \coordinate (V1) at ($0.5*(ADP)+0.5*(NRIIP)+(0,0.75)$);
  \node[vertexC,color=black!90,fill=black!3] (NRIP) at ($(NRII.east)+(2.1,0)$) {\small NtrC-P};
  \node[vertexC] (NRI) at ($(NRIIP.east)+(2.1,0)$) {\small NtrC};
  \coordinate (V2) at ($0.5*(NRIIP)+0.5*(NRI)+(0,0.75)$);
  \node[vertexC] (Water) at ($(NRIP.east)+(2.1,0)$) {\small H$_2$O};
  \node[vertexC] (Pi) at ($(NRI.east)+(2.1,0)$) {\small Pi};
  \coordinate (V3) at ($0.5*(NRI)+0.5*(Pi)+(0,0.75)$);
  \draw[edge] (ATP.east) to[out=-15,in=90] (V1) to[out=-90,in=15] (ADP.east);
  \draw[edge] (NRII.west) to[out=195,in=90] (V1) to[out=-90,in=165] (NRIIP.west);
  \draw[edge,black!90] (NRIIP.east) to[out=15,in=-90] (V2) to[out=90,in=-15] (NRII.east);
  \draw[edge,black!90] (NRI.west) to[out=165,in=-90] (V2) to[out=90,in=195] (NRIP.west);
  \draw[edge] (NRIP.east) to[out=-15,in=90] (V3) to[out=-90,in=15] (NRI.east);
  \draw[edge] (Water.west) to[out=195,in=90] (V3) to[out=-90,in=165] (Pi.west);
  \draw[edge,dashed,mygray] (NRIP) tonode[right] (T) {\footnotesize transcriptional activation} ($(NRIP)+(0,1.25)$);
  \path let \p1 = (Glc), \p2 = (T) in node at ($(\x1,\y2)+(-1.5,0.5)$) {\textbf{C}};
 \end{tikzpicture}
 \caption{Classical representation of the three major interface systems of the integrative \textit{E.~coli} network, the    phosphoenolpyruvate-dependent sugar phosphotransferase system (PTS, \textbf{A}), the ribonucleotide reducing system (RNR system,    \textbf{B}) and the nitrogen regulation two-component signal transduction system (NtrBC system, \textbf{C}). The edges represent biochemical reactions and the vertices denote the involved compounds and proteins. The reactions and proteins highlighted in dark blue are the most abundant vertices determining nearly half of the traversing paths (Table~\ref{tab:Supp-TruePathVertices}).}\label{fig:PTS-RNR-NtrBC}
\end{figure}
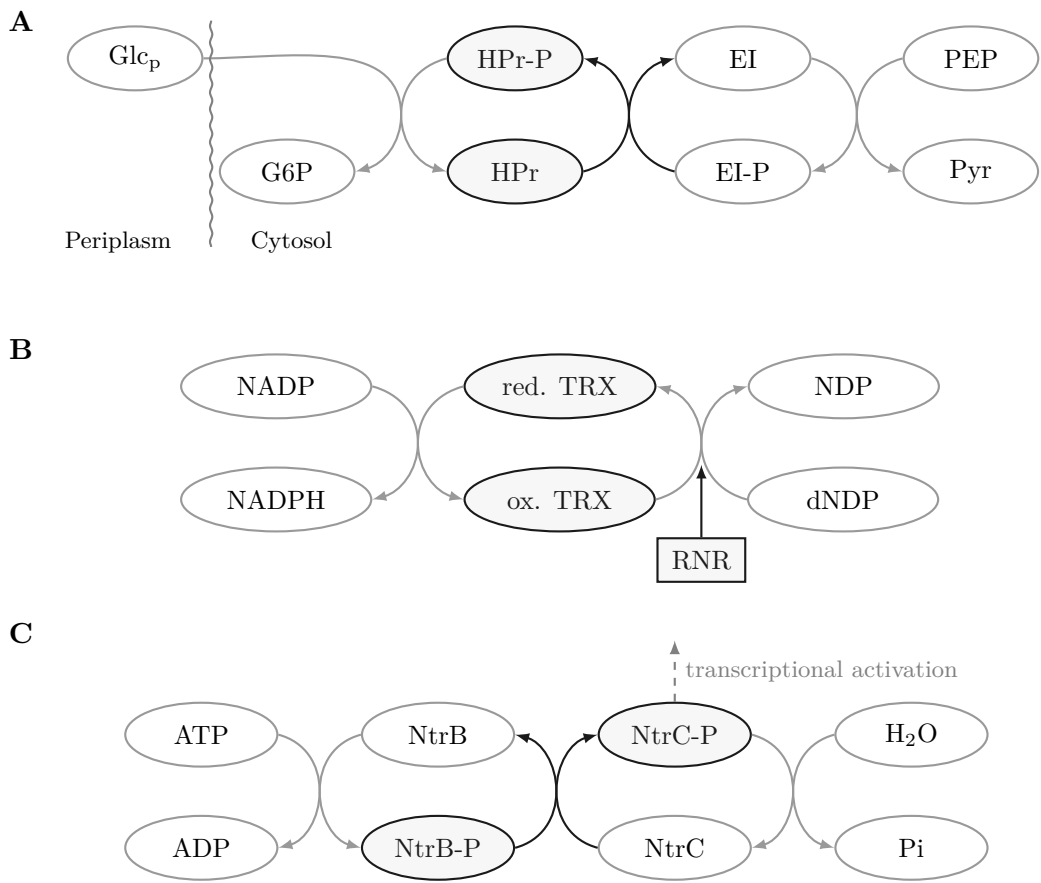

All three biological subsystems, the PTS \cite{Deutscher2008,Escalante2012}, the RNR \cite{Thelander1979,Fontecave1992,Jordan1998,Torrents2014} as well as the NtrBC system \cite{Jiang1999,Reitzer2003,Brown2014} are well-studied with respect to their functionality and their cellular context. A schematic representation of the three subsystems is provided in Figure~\ref{fig:PTS-RNR-NtrBC}. The PTS is an enzymatically active protein complex involved in the transport and phosphorylation of several sugars, so-called PTS-sugars \cite{Escalante2012}. In the integrative \textit{E.~coli} network more than 18 different sugars serve as potential substrates which are imported from peroxisome to cytosol at the same time (Table~\ref{tab:Supp-PTS-sugars}). The substrate variety together with the manifold usage of the associatively produced pyruvate point out the key role of the PTS in \textit{E.~coli}'s metabolism and, moreover, suggest that the PTS acts as a bottleneck in the interface.\par
The RNR system, the second system dominating the downwards traversing paths, provides the major DNA building blocks \cite{Jordan1998}. Each of the different core enzyme classes, ribonucleotide reductase class I--III, are capable of catalyzing the reduction of all four nucleotides. Its transcriptional and metabolic regulation ensures the balanced supply and, thus, avoid the increase of mutation rates and the loss of DNA replication fidelity \cite{Mathews2006}. The central cellular role which is reflected in its regulatory embedding, together with its alternate substrates point to its special position in the interface.\par
The NtrBC system is a two-component signal transduction system initiating the nitrogen starvation response regulation. More precisely, depending on the nitrogen availability NtrB can autophosphorylate and the transfer of the NtrB phosphate group activates the global transduction regulator, NtrC. In \textit{E.~coli}, more than 40 genes known to be activated are involved in the nitrogen-response reaction such as active transport and mobilization of nitrogen in terms of N-containing compounds (for integrative \textit{E.~coli} network see Table~\ref{tab:Supp-NtrBC-regulatedentities}). The extensive regulatory function and the linkage to metabolism due to the allocation of ATP for NtrB autophosphorylation indicate that also the NtrBC system acts as a bottleneck in the interface, in the opposite direction to the PTS and RNR system.

The three central traversing paths systems and their biological relevance suggest that a topologically prominent position can be indicative of a biologically important functional entity. To corroborate the general validity of this indication, in the following section different topological properties have been analyzed and the prominent elements have been further characterized from a functional perspective.

\subsection*{Cross-systemic key elements of \textit{E.~coli}}
The integration of metabolic and regulatory events allows us to determine the key elements of \textit{E.~coli}, especially those beyond the individual processes. In particular, the functional three-domain partition facilitates to recover network components (in terms of individual vertices) of evident biological relevance, \textit{e.g.}, by means of simple centrality measures. In the following, two different aspects of centrality have been examined \cite{Opsahl2010}: degree centrality depicting the direct linkage of a vertex, and betweenness centrality which can be thought of as the participation of a vertex in the network flow \cite{Newman2010}.\par
Starting with the prominent local vertex structure, the so-called hubs (here, vertices with a total degree larger than 50), it is noticable that they are primarily compounds and proteins, in particular protein complexes and appear in all three domains (see Table~\ref{tab:Supp-DegreeCentrality}, columns~3--5). In the metabolic domain, hubs include trivial compounds such as H\textsuperscript{+} and H\textsubscript{2}O and, so-called, currency metabolites, \textit{e.g.}, ATP, NAD(P)H and coenzyme A, while hubs of regulatory processes are obviously global regulators which characteristically exhibit a remarkably strong asymmetry of in-degree and out-degree. Particularly, well-known transcriptions factors top this list such as FNR (fumarate and nitrate reduction) \cite{Kiley1998}, Fis (factor for inversion stimulation) and H-NS (histone-like nucleoid structuring protein) \cite{Travers2005}.
As stated above, hubs predominantly occur in metabolic and gene regulatory domain while only a few are affiliated to the protein interface. However, it was not to be expected to identify cross-systemic elements solely based on their degree.\par
To assess/detect cross-systemic key elements an extended approach of degree centrality has been used that additionally accounts for the domain boundaries. The \textsl{intra-domain degree fraction} $\xi$, also termed embeddedness \cite{Fortunato2016}, denotes the ratio of the internal degree of a vertex, within a domain, and the total degree in the network.
This measure very clearly distinguishes between, on the one hand, metabolic and regulatory hubs which show intra-domain degree fractions $\xi>0.87$ (except one single compound with $\xi=0.185$) and hubs in the interface which in contrast have $\xi\le0.06$ (see Table \ref{tab:Supp-DegreeCentrality}, last column). Thus, while metabolic and regulatory hubs are embedded in their respective domains, hubs in the protein interface are mainly connected to vertices in the neighboring domains. In total, seven hubs show a significant low intra-domain degree fraction pointing to their prevalent interactions with the other two domains (Figure~\ref{fig:Supp-IntraDegree} and Table~\ref{tab:Supp-IntraDegree}, column~5). Six of them are affiliated to the protein interface exhibiting numerous interactions with the regulatory domain. Their linkages to the metabolic domain become visible when considering their composition, in case of the protein complexes, and their modes of action, respectively. The former involve the four protein-compound complexes Crp-cAMP (cyclic-AMP receptor protein binding cyclic-AMP) \cite{Kolb1993,Deutscher2008,Fic2009}, DksA-ppGpp (dnaK suppressor binding guanosine 3'-diphosphate 5'-diphosphate) \cite{Magnusson2005,Potrykus2008,Srivatsan2008}, NsrR-NO (nitrite-sensitive repressor binding nitric oxide) \cite{Spiro2007,Partridge2009,Tucker2010} and Lrp-Leu (leucine-responsive regulatory protein binding leucine) \cite{Ernsting1992,Calvo1994,Brinkman2003} whose naming schemes already indicate the metabolic link. The latter, namely, protein complex Cra (catabolite repressor activator) and protein monomer Lrp (leucine-responsive regulatory protein) form in the presence of appropriate metabolites, \textit{i.e.}, fructose 1,6-bisphosphate/fructose 1-phosphate and leucine, complexes affecting their regulatory effect. The remaining hub is the metabolic-domain vertex representing guanosine 5'-diphosphate 3'-diphosphate (ppGpp). Besides its special domain-affiliation among the low intra-domain degree hubs, ppGpp acts as an important regulator of both, metabolism and transcriptional processes. More precisely, it regulates several enzyme activities as well as numerous transcription initiations by allosterically binding to RNA polymerase.\par
So far, we demonstrated that the protein interface of the \textit{E.~coli} network reconstruction acts as a bridging module between regulatory and metabolic domain enabling their interaction and communication. Therefore, we expect the betweenness centrality to directly highlight vertices from the interface. Indeed, ten out of the top-25-ranked (still including currency metabolites) vertices are from the interface (see Table~\ref{tab:Supp-BetweennessCentrality}, column~5), while overall the interface only accounts for about 18~\% of the vertices of the network. Especially, the already mentioned protein-compound complexes Crp-cAMP and DksA-ppGpp are among these compounds. In general, currency metabolites and trivial compounds (see above) as well as global regulators are among the central components with respect to betweenness. Overall, the compliance of the most central components regarding degree and betweenness accounts approximately 50~\%. Apart from that, biochemical reactions building up and/or breaking down these metabolites and proteins as well as the other involved reactants pertain to the most betweenness-central components. Component association to functional systems allows to assess the systemic feature and by considering the corresponding network affiliation to depict the candidates for cross-systemic key elements. In this manner the network analysis allows us to detect the central role of Crp-cAMP, Lrp-Leu and ppGpp on purely topological grounds, as each component is the focus of such a functional system with high betweenness. Additionally among the top-ranked vertices with respect to betweenness centrality are five further cross-systemic components which are assigned to the protein interface, namely, phosphorylated PhoB (PhoB-P), Fur-Fe\textsuperscript{2+}, and three outer membrane proteins (Omp), OmpC, OmpE and OmpF (Table~\ref{tab:Supp-BetweennessCentrality}). The former two components are transcription factors and therefore acting in the gene regulatory domain, while at the same time they are protein complexes binding a metabolic small molecule depicting the connection to the metabolic processes. The latter three, the outer membran porins, form hydrophilic channels, enabling non-specific diffusion of small molecules across the outer membrane \cite{Schulz1993,Jap1996,Schirmer1998}. In this role these proteins represent the most obvious connections of gene regulatory and metabolic domain -- their encoding genes are highly regulated while the porins enable numerous metabolic transport reactions.\par
By focusing on the connecting domain of gene regulation and metabolism, the two centrality measures reinforce the key role of further cross-systemic elements. Considering the protein interface-induced subgraph both centralities point out the vertices that top the list of the above-discussed downwards traversing paths (Table~\ref{tab:Supp-KeyPIElements}). In more detail, both major systems contributing to the downwards traversing paths are represented each by three vertices, namely, PTS and RNR system (Figure~\ref{fig:PTS-RNR-NtrBC}, panels A and B). Having a look at the intra-domain degree fraction, which put the focus on protein interface vertices as described above, additionally highlights a representative of the upwards traversing path system NtrBC (Figure~\ref{fig:PTS-RNR-NtrBC}, panel C), as the second non-hub (Table~\ref{tab:Supp-IntraDegree}). This corroborates the predictions from the traversing paths and, thus, shows that our new topological measure reveals cross-systemic elements which otherwise only stand out under detailed scrutiny of a large amount of biological information.

\section*{Discussion}
Here, we present an integrative network covering metabolic processes as well as regulatory events of \textit{E.~coli} but, especially, the interaction between both systems. With more than 10,000 vertices, it comprises around two third of the metabolic processes currently integrated in metabolic reconstructions \cite{Feist2007} and concerning regulatory events, the presented network incorporates more than 95~\% of the established transcription-related processes \cite{Gama-Castro2015}. Both, metabolic and gene regulatory processes are integrated on a genome scale rather than one of the two providing the network basis which is then expanded by closely related processes in the other subsystem, as it has been done, for example, in conventional metabolic reconstructions which solely involve the encoding genes indirectly. Hitherto, integration of transcriptomics data could only be achieved using the so-called gene-protein-reaction (GPR) associations. On the one hand, this procedure limits the applicable data set to metabolic genes and, on the other hand, it acts on the assumption that all expressed enzymes are present in their active form. Starting from the integrative \textit{E.~coli} network, integrating transcriptomics data is much more straightforward and, more importantly, the complete data set can be applied. In this way, multi-domain variants of the frequently employed network-based interpretation of 'omics' data \cite{Marr2008,Sonnenschein2011,Sonnenschein2012,Knecht2016,Beber2016} can be formulated and indirect and regulatory impacts on metabolism can be examined.\par
The novelty of the reconstruction, the connection of metabolism and gene regulation, allows us not only to investigate the separate systems but also to assess their interactions. The most relevant connecting links are proteins, on the one hand, those acting as enzymes and, on the other hand, metabolic transcription factors. The functional classification, together with the topological analysis, suggests a network division into three domains: metabolic domain, protein interface and regulatory domain. This partition was corroborated by different connectivity measures and reflects a biologically reliable categorization in two delimited modules linked by a bridging module.\par
The principal structural feature of the network model, the three-domain organization, is reminicient of the 'bow-tie' architectures frequently discussed in the theory of complex systems, where an input and an output layer are connected via a (typically much smaller) intermediate network \cite{Kitano2004,Kitano2007,Friedlander2014}. Such a bow-tie structure (or, rather, the presence of several nested bow-tie architectures) has for example been discussed for metabolic networks \cite{Csete2004}, where the diversity of inputs (nutrients) and outputs (biomass components) is much larger than the intermediate processing layer. It has been hypothesized that such a bow-tie organization is a prerequisite for the robust operation of a complex system \cite{Kitano2004,Kitano2007}. Here we observe a bow-tie organization in a system consisting of a rich 'material flow' system (metabolism) and a similarly rich 'control' system (gene regulation) connected via a protein interface.\par
As our topological assessment shows, the bridging character of the protein interface entails a bottleneck functionality. The analysis of the new topological measure, termed \textsl{traversing paths}, highlighted three major biological systems represented by 12 vertices forming more than 40~\% of these paths (comprising in total 1465 distinct vertices). These traversing path systems, namely phosphotransferase system (PTS), ribonucleotide reducing (RNR) and nitrogen regulation two-component signal transduction (NtrBC) system, are well-investigated ones with key biological relevance for \textit{E.~coli}'s metabolism as well as its gene regulation suggesting that a topologically prominent position points to an important biologically functional entity.\par
Further detection of cross-systemic key elements in the network was accomplished using additional topological measures. In particular, two centrality measures were studied to account for different aspects of importance in terms of direct linkage and participation in network flow. Apart from conspicuous components, such as trivial compounds, currency metabolites and global regulators, a group of seven hubs were revealed by degree centrality whose characteristic is a significant low intra-domain degree fraction what numerically reflects the bridging feature of the protein interface. As expected, these components are located in the interface except for one, the vertex representing guanosine 5'-diphosphate 3'-diphosphate (ppGpp) which is affiliated to the metabolic domain. On the other hand, the inspection of betweenness centrality highlights rather biological systems than single components and as such point to key components detected before in their functional context. Besides trivial compounds and currency metabolites, this includes Crp-cAMP (cyclic-AMP receptor protein binding cyclic-AMP), Lrp-Leu (leucine-responsive regulatory protein binding leucine) and ppGpp which stand out due to their intra-domain degree fraction as well as seven further components already revealed as hubs.\par
Intriguingly, the interface-specific key elements of the network could be corroborated by exactly these two centrality measures. The assessment of the interface-induced subgraph using both centralities emphasizes altogether eight vertices of the downwards traversing paths discussed above contributing to the two major systems PTS and RNR. Taking into account the intra-domain degree fraction point out a representative of the upwards traversing path system NtrBC. In conclusion, the importance of vertices revealed by the here presented traversing paths could be reinforced by well-established topological measures showing the predictive power of the new measure.\par
Eventually, the key elements of the integrative \textit{E.~coli} network according to both centralities illustrate the importance of the different domains and their combined consideration (Table~\ref{tab:KeyElements}). Unsurprisingly, the majority of key elements are affiliated to the metabolic domain and represent trivial compounds and currency metabolites, \textit{e.g.}, H\textsuperscript{+}, H\textsubscript{2}O, ATP and NAD(P)\textsuperscript{+}. Moreover, predominantly cross-systemic components top this combined list of central elements. First of all, the vertices emphasized also by their low intra-domain degree fraction attract attention, namely, Crp-cAMP, Lrp-Leu and ppGpp. These vertices demonstrate the value of the integrative approach: Only when embedded in domain context their vertex importance emerged. In case of the former two components, additionally, the composition unveils the cross-systemic role, \textit{i.e.}, a transcriptional factor protein binding a metabolic small molecule affecting its regulatory activity. Likewise the two regulatory key elements, Fur-Fe\textsuperscript{2+} and PhoB-P, exhibit this conspicuous linkage to the metabolic domain illustrating their cross-systemic property. In other words, they belong to the so-called metabolic transcription factors and, thus, are related to the upwards interface. The opposite is the case for the three metabolic Omp (outer membran porin) transporters that are among the key elements. While their metabolic linkage is more than obvious, the relation to the regulatory domain appears when the encoding genes are examined. These are highly regulated amongst others by the global regulators Crp-cAMP, Fur-Fe\textsuperscript{2+}, Lrp-Leu and PhoB-P. In this manner, the Omp's are classical representatives of proteins related to the downwards interface, even though they are not affiliated with it. The remaining key elements are three 
metabolic small molecules which are counter-intuitively also related to the interface and the cross-systemic elements detected by the traversing paths. While in case of pyruvate the connection to PTS is
apparent at first glance (Figure~\ref{fig:PTS-RNR-NtrBC}, panel A), the link of 
glutamate and ammonium and the NtrBC system is less perceptible. The actual connecting element is glutamine which is the ligase product of glutamate and ammonium. It activates the (de)uridylylation of the regulatory protein PII which, in turn, inhibits NtrB autophosphorylation \cite{Jiang1999,Heeswijk2013}. Altogether, the links to the three major traversing path systems are certainly not the only important processes these elements are involved in but they reinforce their biologically central roles. Remarkably, these connecting elements show up when considering the entire network while to acknowledge their importance the interface-specific analysis is needed.

\begin{table}[htb]
 \caption{Key elements of the integrative \textit{E.~coli} network
   with respect to degree (DC) and betweenness centrality (BC) rank
   as well as their functional characteristic and cross-systemic
   property, respectively. Squares denote trivial compounds
   (\protect\opensquare{black}) and currency metabolites
   (\protect\colsquare{black!30}) while the colored arrows depict the
   cross-systemic contribution --
   \protect\coltriangledown{black!70}~downwards interface-related,
   \protect\coltriangle{black!70}~upwards interface-related. The
   orange arrows emphasize the cross-systemic components with
   significant low intra-domain degree fraction and the golden ones
   point out elements indirectly linking to one of the major traversing
   paths systems.}\label{tab:KeyElements}
 \centering
 \begin{tabular}{lrrc}
  \toprule
  Vertex name & DC & BC & Property \\
  \midrule
  Proton & 1 & 1 & \opensquare{black} \\ 
  \rowcolor{black!5}H$_2$O & 2 & 2 & \opensquare{black} \\ 
  ATP & 5 & 3 & \colsquare{black!30} \\ 
  \rowcolor{black!5}Phosphate (P) & 4 & 4 & \colsquare{black!30} \\ 
  Proton (periplasmic) & 6 & 5 & \opensquare{black} \\ 
  \rowcolor{black!5}Crp-cAMP, transcriptional dual regulator & 3 & 9 & \coltriangle{chocolate} \\ 
  ADP & 10 & 6 & \colsquare{black!30} \\ 
  \rowcolor{black!5}outer membran porin F & 7 & 24 & \coltriangledown{black!70} \\ 
  outer membran porin C & 7 & 29 & \coltriangledown{black!70} \\ 
  \rowcolor{black!5}H$_2$O (periplasmic) & 12 & 31 & \opensquare{black} \\ 
  outer membran porin E & 9 & 34 & \coltriangledown{black!70} \\ 
  \rowcolor{black!5}Fur-Fe$^{2+}$, transcriptional dual regulator & 25 & 21 & \coltriangle{black!70} \\ 
  Pyrophosphate & 18 & 30 & \colsquare{black!30} \\ 
  \rowcolor{black!5}NAD$^+$ & 13 & 35 & \colsquare{black!30} \\ 
  Phosphate (periplasmic) & 19 & 36 & \colsquare{black!30} \\ 
  \rowcolor{black!5}PhoB-P, transcriptional dual regulator & 45 & 12 & \coltriangle{black!70} \\ 
  Lrp-Leucine, transcriptional dual regulator & 43 & 17 & \coltriangle{chocolate} \\ 
  \rowcolor{black!5}Guanosine 5'-diphosphate 3'-diphosphate & 35 & 26 & \coltriangle{chocolate} \\ 
  NADP$^+$ & 21 & 41 & \colsquare{black!30} \\ 
  \rowcolor{black!5}Glutamate & 24 & 40 & \coltriangle{goldenrod} \\ 
  Pyruvate & 30 & 39 & \coltriangledown{goldenrod} \\ 
  \rowcolor{black!5}Coenzyme A & 23 & 48 & \colsquare{black!30} \\ 
  CO$_2$ & 32 & 41 & \colsquare{black!30} \\ 
  \rowcolor{black!5}NH$_4^+$ & 31 & 46 & \coltriangle{goldenrod} \\ 
  \bottomrule
 \end{tabular}
\end{table}

Beyond the detection of key elements, the integrative approach will allow to examine the interplay and distribution of short-term and long-term regulation in \textit{E.~coli}'s metabolism. While metabolic regulation of, for instance, enzyme activities occurs on a short time-scale, regulation of gene expression is a long-term control process. Both types of regulation have been incorporated in the network even though only on a qualitative level, \textit{i.e.}, as activator or inhibitor. Like this, the different effective ranges in metabolism can be assessed and, thus, its covering by one or both regulation types where central metabolism is said to be highly controlled.

\subsection*{From the perspective of recent advances in network theory}
With their balance of structural detail and functional simplicity, network models are capable of revealing organizational principles, which are hard to recognize on a smaller systemic scale (\textit{e.g.}, by analyzing individual pathways) or in functionally richer system representations (\textit{e.g.}, in dynamical models). One purpose of the network provided here is to enable work at the interface of statistical physics and systems biology, where the rich toolbox of complex network analysis is employed to identify functionally relevant non-random features of such biological networks.\par
The recent work of \citet{Jensen2016}, for example, showed that network structure can reveal, whether an enzyme is susceptible rather to genetic knockdown or pharmacologic inhibition. While in the present study, the network measures do not distinguish between different kinds of vertices or links, the rich biological meta data concerning the different biological roles of the components could be translated into distinct vertex and edge classes. In our own investigation \cite{Klosik2017} we used this fact to study, in a further example of such an interdisciplinary effort, the balance of robustness and sensitivity in the interdependent network of gene regulation and metabolism, based on the reconstructed network provided here.\par
In general, we expect that our network reconstruction can serve as a relevant data resource for the application of methods from the analysis of multiplex \cite{Radicchi2017} and other multilayer networks \cite{Kivelae2014,Radde2016}. Recently, there has been a growing interest in the properties of these systems, especially in the presence of explicit interdependencies between vertices \cite{Gao2012,Radicchi2017}. In contrast to monoplex networks interdependent networks can show a qualitatively different robustness against failures, \textit{i.e.}, cascading failures leading to a
sudden system breakdown at a critical initial attack size \cite{Buldyrev2010,Son2012}. The case of different vertex types (as opposed to different edge types) has been considered, for example, in the context of secure communication in a network where eavesdroppers control sets of vertices \cite{Krause2016}.\par
On a general level, analyzing statistics of paths with respect to the network's large-scale structure, like the domain-traversing paths used here, might prove useful for the evaluation of other networks
that show (possibly more than one) interface-like features. 

\subsection*{Concluding remarks}
In summary, the analysis of network topology allows to determine key system components in the integrative \textit{E.~coli} network. In line with expectations, trivial compounds as well as currency metabolites showed up regardless of the measure that has been applied. In addition, further obvious components including several global regulators were identified. More striking is the detection of components and systems which solely emerge when analyzing specifically the interface. These hidden elements are associated to two of the biologically well-investigated functional subsystems, PTS and NtrBC. Both well-established and newly designed measures of the interface point out the same subsystems, and even the analysis of the entire network discloses components indirectly related to these hidden subsystems.\par
Apart from trivial and currency metabolites, every detected key element of the entire network contributes to some extent to the downwards and/or upwards interface. This unlooked-for cross-systemic property is reflected either in the complex composition, the intra-domain degree fraction, the proximity to key systems, and/or the interplay with regulatory and metabolic processes. The biological relevance of these components supports their detection and reinforces the predictive power of the novel traversing path measure. In general, we believe that the presented integrative \textit{E.~coli} network allows further investigations of the interplay of metabolism and gene regulation which will provide insights into cellular, system-wide responses.

\section*{Methods}
\small
The interconnected \textit{E.~coli} network is based on the EcoCyc database \cite{Keseler2012}, release 20.0, which includes verified information of metabolic and regulatory processes (corresponds to RegulonDB 8.6 \cite{Gama-Castro2015}) for \textit{E.~coli} K-12 substr. MG1655. The network is represented as a graph comprising five different types of vertices, encoding genes, protein monomers and complexes (including enzymes), small compounds, and (bio)chemical reactions (Table~\ref{tab:Supp-VertexCompositionModules}), as well as three types of edges, encoding and catalyzing associations, reaction connections to educts and products, and regulatory links to sources and targets (Table~\ref{tab:Supp-EdgeCompositionModules}).

\subsection*{Extraction of database information}
First, relevant information of the database has been extracted and arranged (Algorithm~\ref{alg:Database}). For each regulatory process, the respective source and target were specified and converted to match one of the vertex types ('regulation.dat', file name of the EcoCyc-archive). To this end, the transcript units were separated into promoter, genes and terminator (if applicable), and the regulatory processes were multiplied per comprising gene. Moreover, each regulating RNA has been translated into its encoding gene to meet the vertex types. In case of the metabolic processes, the reaction educts and products as well as the catalyzing enzymes have been assembled and converted to match one of the vertex groups, the respective educt and product stoichiometry have been assigned and the reaction compartmentation and reversiblity have been assessed ('reactions.dat'). Thereby, as cell compartments the periplasmic space, the inner membrane, and the cytosol have been taken into account and reversible reactions have been split up.

\begin{algorithm}[htb]
 \newcommand{\Code}[1]{\textcolor{mygray}{\ttfamily#1}}
 \newcommand{\ReturnVal}[1]{\upshape\ttfamily{\textcolor{mygray}{#1}}~=~}
 \newcommand{\Text}[1]{\rmfamily\slshape{\textcolor{black}{#1}}}
 \SetArgSty{Code}
 \SetProgSty{Code}
 \SetFuncArgSty{Code}
 \SetCommentSty{textnormal}
 \SetKwProg{Fnc}{}{}{}
 \SetKwFunction{ARP}{AssembleRegulatoryProcesses}
 \SetKwFunction{AMP}{AssembleCompartmentedMetabolicProcesses}
 \SetKwFunction{VPV}{ValidateProcessesVertices}
 \SetKwFunction{KwRange}{\textnormal{range}}
 \SetKw{KwIn}{\textcolor{black}{in}}
 \SetKw{KwDef}{Definition:}
 \DontPrintSemicolon
 \Fnc{\ReturnVal{regprocs}\ARP{}}{
  Extract information on regulatory processes ('regulation.dat')\;
  \Code{REG\_type,REG\_ID,REG\_source,REG\_target,REG\_mode = ParseRegulation()}\;
  \For{rT \KwIn REG\_type}{
   \For{rID \KwIn REG\_ID[rT]}{
    Convert regulation source and target to match one of the vertex groups\;
    \If{REG\_source[iT][iID] == \Text{RNA}}{
     Translate RNA into corresponding genes
    }
    \If{REG\_target[iT][iID] ==  \Text{transcript unit}}{
     Split up transcription units into promoters, genes and terminators\;
     Translate promoters and terminators into corresponding genes
    }
    \Code{s,t = Match2Vertex(REG\_source[rT][rID],REG\_target[rT][rID])}\;
    \Code{regprocs.append(rT,rID,s,t,REG\_mode[rT][rID])}\;
   }
  }
 }
 \Fnc{\ReturnVal{metprocs}\AMP{}}{
  Extract information on metabolic processes ('reaction.dat')\;
  \Code{Rxn\_ID,Rxn\_enz,Rxn\_dir,Rxn\_l,Rxn\_r,Rxn\_loc = ParseReactions()}\;
  \For{mID \KwIn Rxn\_ID}{
   Match catalyzing enzyme, educts and products to one of the vertex groups\;
   \Code{enz,left,right = Match2Vertex(Rxn\_enz[mID],Rxn\_l[mID],Rxn\_r[mID]])}\;
   Annotate reaction compartmentation\;
   \Code{left\_comp,right\_comp = AnnotateCompartment(Rxn\_loc[mID],left,right)}\;
   Split up reversible reactions and reverse 'right-to-left' reactions\;
   \uIf{Rxn\_dir == \textnormal{'reversible'}}{
    \Code{metprocs.append(mID+'\_f',enz,left\_comp,right\_comp)\;
    metprocs.append(mID+'\_r',enz,right\_comp,left\_comp)}
   }
   \uElseIf{Rxn\_dir == \textnormal{'right-to-left'}}{
    \Code{metprocs.append(mID,enz,right\_comp,left\_comp)}
   }
   \Else{
    \Code{metprocs.append(mID,enz,left\_comp,right\_comp)}
   }
  }
 }
 \Fnc{\ReturnVal{valReg,valMet,valVertices}\VPV{regprocs,metprocs}}{
  Compile vertex candidates from regulation sources and targets as well as metabolic reactions, the corresponding educts and products, and encoding enzymes\;
  \Code{vertCands = AssembleCandidates(regprocs,metprocs)}\;
  Assign vertex candidates to the seven types: \textsl{reaction}, \textsl{compound}, \textsl{gene}, \textsl{protein monomer}, \textsl{protein-protein-complex}, \textsl{protein-compound-complex}, \textsl{protein-rna-complex}\;
  \Code{rxns,cmps,gns,prts,ppc,pcc,prc = ComposeVertexLists(vertCands)}\;
  Decode generic terms ('classes.dat') and double annotations\;
  \Code{curReg,curMet,curRxns,curCmps = Curation(regprocs,metprocs,rxns,cmps)}\;
  Prune vertex lists regarding unmapped candidates\;
  \Code{valVertices = PruneVertices(curRxns,curCmps,gns,prts,ppc,pcc,prc)}\;
  Prune regulatory and metabolic processes with respect to validated vertex lists\;
  \Code{valReg,valMet = PruneRegMetProcesses(curReg,curMet,valVertices)}\;
  Export lists of validated vertices and links, \textit{i.e.}, processes
 }
 \caption{Extraction of database information (EcoCyc, release 20.0) on regulatory and metabolic processes.}\label{alg:Database}
\end{algorithm}

Second, vertex candidates have been validated ('reactions.dat', 'compounds.dat', 'proteins.dat', 'genes.dat', 'rnas.dat') and divided into \textsl{reaction}, \textsl{compound}, \textsl{protein monomer},
\textsl{protein-protein complex}, \textsl{protein-compound complex}, \textsl{protein-RNA complex}, and \textsl{gene}. In doing so, generic terms such as \texttt{DIPEPTIDES} have been substituted ('classes.dat') and double annotations, \textit{e.g.}, \texttt{CPD-15709} and \texttt{FRUCTOSE-6P} have been decoded. Thereupon, the compositions and the encoding genes of the assembled proteins have been gathered and matched to the vertex groups and the respective logical operation and stoichiometry have been annotated ('protcplxs.col'). Based on the validated vertex lists, the regulatory and metabolic processes have been updated whereby each process was removed with at least one unidentified vertex resulting in the final edge lists.

\subsection*{Network implementation}
With the validated vertex and edge lists the graph has been assembled and its largest weakly connected component has been extracted. The three domain partition MD~--~PI~--~RD (Tables~\ref{tab:DetailedVertexComposition} and~\ref{tab:Supp-VertexCompositionModules}) as well as the two-domain partition are implemented as vertex properties \textit{affiliation} and \textsl{metabolic}. Algorithms~\ref{alg:NetworkAffilation1} and \ref{alg:NetworkAffilation2} show how the domain affiliation of a vertex is determined by its type and its neighbors' types and affiliations.

\renewcommand{\thealgocf}{\arabic{algocf}A}
\begin{algorithm}[htb]
 \newcommand{\Code}[1]{\textcolor{mygray}{\ttfamily#1}}
 \newcommand{\ReturnVal}[1]{{\upshape\ttfamily{\textcolor{mygray}{#1}}~=~}}
 \newcommand{\Text}[1]{{\textcolor{black}{\rmfamily\slshape #1}}}
 \SetArgSty{Code}
 \SetProgSty{Code}
 \SetFuncArgSty{Code}
 \SetCommentSty{textnormal}
 \SetKwProg{Fnc}{}{}{}
 \SetKwProgOpen{FncOpen}{}{}{}
 \SetKwFunction{ANV}{AssignNonAmbiguousVertices}
 \SetKwFunction{KwRange}{\textnormal{range}}
 \SetKw{KwIn}{\textcolor{black}{in}}
 \SetKw{KwNm}{\textcolor{black}{\slshape\#\!}}
 \SetKw{KwDef}{Definition:}
 \DontPrintSemicolon
 \FncOpen{\ReturnVal{aff}\ANV{}}{
  Metabolic regulatory processes will be interpreted as \Text{metabolic} links\;
  \Code{MetRegs = [\Text{Regulation-of-Enzyme-Activity},\Text{Regulation-of-Reactions}]}\;
  \For{v \KwIn Vtypes(network.vertices) == \Text{reaction}}{
   \lIf{Vtypes(Educts(v) $\wedge$ Products(v)) == \Text{compound}}{\Code{aff(v)~=~\Text{metabolic}}}
   \lElseIf{Vtypes(Educts(v) $\wedge$ Products(v)) == (\Text{compound} $\vee$ \Text{protein})}{\Code{aff(v)~=~\Text{interface}}}
   \lElse{\Code{aff(v)~=~\Text{tba}}}
  }
  \For{v \KwIn Vtypes(network.vertices) == \Text{compound}}{
   Compounds which participate in at least one reaction\;
   \uIf{\KwNm InvolvedRxns(v) > 0}{
    \lIf{\KwNm aff(InvolvedRxns(v)) == 1}{\Code{aff(v) = aff(InvolvedRxns(v))}}
    \lElseIf{\Text{metabolic} \KwIn aff(InvolvedRxns(v))}{\Code{aff(v)~=~\Text{metabolic}}}
    \lElseIf{\Text{interface} \KwIn aff(InvolvedRxns(v))}{\Code{aff(v)~=~\Text{interface}}}
    \lElse{\Code{aff(v)~=~\Text{ambiguous}}}
   }
   Compounds which adjacent vertices are either compounds or proteins\;
   \uElseIf{Vtypes(Neighbors(v)) == (\Text{compound} $\vee$ \Text{protein})}{
    \uIf{\Text{regulation} \KwIn Etypes(OutEdges(v))}{
     \lIf{REGtypes(OutEdges(v)) \KwIn MetRegs}{\Code{aff(v)~=~\Text{metabolic}}}
     \lElse{\Code{aff(v)~=~\Text{ambiguous}}}
    }
    \lElse{\Code{aff(v)~=~\Text{ambiguous}}}
   }
   \lElse{\Code{aff(v)~=~\Text{tba}}}
  }
  \For{v \KwIn Vtypes(network.vertices) == \Text{protein}}{
   Proteins with enzymatic function\;
   \uIf{\Text{enzyme - reaction} \KwIn Etypes(OutEdges(v))}{
    \uIf{Vtypes(OutNeighbors(v)) == \Text{reaction}}{
     \lIf{\KwNm aff(OutNeighbors(v)) == 1}{\Code{aff(v) = aff(OutNeighbors(v))}}
     \lElseIf{\Text{metabolic} \KwIn aff(OutNeighbors(v)))}{\Code{aff(v)~=~\Text{metabolic}}}
     \lElseIf{\Text{interface} \KwIn aff(OutNeighbors(v)))}{\Code{aff(v)~=~\Text{interface}}}
     \lElse{\Code{aff(v)~=~\Text{tba}}}
    }
    \uElseIf{\Text{regulation} \KwIn Etypes(OutEdges(v))}{
     \lIf{REGtypes(OutEdges(v)) \KwIn MetRegs}{\Code{aff(v)~=~\Text{metabolic}}}
     \lElse{\Code{aff(v)~=~\Text{regulatory}}}
    }
    \lElseIf{\KwNm aff(InvolvedRxns(OutNeighbors(v))) == 1}{\Code{aff(v) = aff(InvolvedRxns(OutNeighbors(v)))}}
    \lElseIf{\Text{metabolic} \KwIn aff(InvolvedRxns(OutNeighbors(v)))}{\Code{aff(v)~=~\Text{metabolic}}}
    \lElseIf{\Text{interface} \KwIn aff(InvolvedRxns(OutNeighbors(v)))}{\Code{aff(v)~=~\Text{interface}}}
    \lElse{\Code{aff(v)~=~\Text{tba}}}
   }
   Proteins involved in metabolic reactions\;
   \lElseIf{\Text{educt - reaction} \KwIn Etypes(OutEdges(v))}{\Code{aff(v)~=~\Text{interface}}}
   Proteins involved in regulatory processes\;
   \uElseIf{\Text{regulation} \KwIn Etypes(OutEdges(v))}{
    \lIf{REGtypes(OutEdges(v)) \KwIn MetRegs}{\Code{aff(v)~=~\Text{metabolic}}}
    \lElse{\Code{aff(v)~=~\Text{regulatory}}}
   }
   Proteins involved in protein complex formation\;
   \uElseIf{\Text{protein - complex} \KwIn Etypes(OutEdges(v))}{
    \lIf{\Text{reaction - product} \KwIn Etypes(InEdges(v))}{\Code{aff(v)~=~\Text{interface}}}
    \uElseIf{\Text{regulation} \KwIn Etypes(InEdges(v))}{
     \lIf{REGtypes(InEdges(v)) \KwIn MetRegs}{\Code{aff(v)~=~\Text{metabolic}}}
     \lElse{\Code{aff(v)~=~\Text{regulatory}}}
    }
    \lElse{\Code{aff(v)~=~\Text{interface}}}
   }
   \lElse{\Code{aff(v)~=~\Text{ambiguous}}}
  }
 }
 \caption{Network affilition compilation based on vertex type, and the vertex neighbors types and affiliations. Affiliation assignment for non-ambiguous reactions, compounds and proteins.\hfill\textcolor{white}{.}\hspace*{1em}\hfill\mbox{\textbf{continued~in~Algorithm~\ref{alg:NetworkAffilation2}}}}\label{alg:NetworkAffilation1}
\end{algorithm}

\renewcommand{\thealgocf}{\arabic{algocf}B}
\addtocounter{algocf}{-1}
\begin{algorithm}[htb]
 \newcommand{\Code}[1]{\textcolor{mygray}{\ttfamily#1}}
 \newcommand{\ReturnVal}[1]{{\upshape\ttfamily{\textcolor{mygray}{#1}}~=~}}
 \newcommand{\Text}[1]{{\textcolor{black}{\rmfamily\slshape #1}}}
 \SetArgSty{Code}
 \SetProgSty{Code}
 \SetFuncArgSty{Code}
 \SetCommentSty{textnormal}
 \SetKwProg{Fnc}{}{}{}
 \SetKwProgOpen{FncOpen}{}{}{}
 \SetKwFunction{ANV}{AssignNonAmbiguousVertices}
 \SetKwFunction{AAV}{AssignAmbiguousVertices}
 \SetKwFunction{KwRange}{\textnormal{range}}
 \SetKw{KwIn}{\textcolor{black}{in}}
 \SetKw{KwNm}{\textcolor{black}{\slshape\#\!}}
 \SetKw{KwDef}{Definition:}
 \DontPrintSemicolon
 \Fnc{\ReturnVal{aff}\ANV{}{\hfill\Text{\bfseries function resumption}}}{
  \For{v \KwIn Vtypes(network.vertices) == \Text{gene}}{
   \lIf{\Text{regulation} \KwIn Etypes(OutEdges(v))}{\Code{aff(v)~=~\Text{regulatory}}}
   \lElseIf{\Text{regulation} \KwIn Etypes(InEdges(v))}{\Code{aff(v)~=~\Text{regulatory}}}
   \lElseIf{\Text{regulatory} \KwIn aff(OutNeighbors(v))}{\Code{aff(v)~=~\Text{regulatory}}}
   \lElseIf{\Text{metabolic} \KwIn aff(OutNeighbors(v))}{\Code{aff(v)~=~\Text{metabolic}}}
   \lElse{\Code{aff(v)~=~\Text{ambiguous}}}
  }
 }
 Assign affiliation for vertices formerly denoted as \textsl{ambiguous}\;
 \Fnc{\ReturnVal{aff}\AAV{aff}}{
  \Code{additionalRun = True}\;
  \While{additionalRun}{
   \Code{additionalRun = False}\;
   \For{v \KwIn aff(network.vertices) == \Text{ambiguous}}{
    \uIf{\KwNm aff(AllNeighbors(v)) == 1}{
     \Code{aff(v)~=~aff(AllNeighbors(v))}\;
     \Code{additionalRun = True}
    }
    \lElseIf{(\Text{regulatory} $\wedge$ \Text{metabolic}) \KwIn aff(AllNeighbors(v))}{\Code{aff(v)~=~\Text{ambiguous}}}
    \uElseIf{\Text{regulatory} \KwIn aff(AllNeighbors(v))}{
     \Code{aff(v)~=~\Text{regulatory}}\;
     \Code{additionalRun = True}
    }
    \uElseIf{\Text{metabolic} \KwIn aff(AllNeighbors(v))}{
     \Code{aff(v)~=~\Text{metabolic}}\;
     \Code{additionalRun = True}
    }
    \lElse{\Code{aff(v)~=~\Text{ambiguous}}}
   }
  }
  \For{v \KwIn aff(network.vertices) == \Text{ambiguous}}{\Code{aff(v)~=~\Text{interface}}}
 }
 \caption{Network affilition compilation based on vertex type, and the vertex neighbors types and affiliations. Affiliation assignment for non-ambiguous genes and vertices assigned as \textsl{ambiguous}.}\label{alg:NetworkAffilation2}
\end{algorithm}
\renewcommand{\thealgocf}{\arabic{algocf}}

Moreover, the mapping to the \textit{E.~coli} model of \citet{Covert2004} has been annotated which integrates the metabolic network \textit{i}JR904 published by \citet{Reed2003} and the transcription regulatory events related to the encoding genes of the catalzying enzymes. To this end, genes, proteins, metabolites as well as biochemical reactions of the metabolic model have been mapped to the EcoCyc database (release 20.0), in a first step automatically based on their identifier and the resulting dictionaries have been manually curated. As the EcoCyc database does not account for compartmentation of compounds and reaction as well as for exchange reactions, unique metabolites and internal reactions have been considered resulting in a coverage of more than 93~\%. By additionally disregarding internal transport reactions a coverage of 96.5~\% can be achieved (Table~\ref{tab:DetailedVertexComposition}).

\begin{table}[htb]
 \centering
 \caption{Comparison of vertex composition and the coverage to the
   model from \citet{Covert2004} of the integrative \textit{E.~coli}
   network (Largest WCC), the underlying full graph and the EcoCyc
   database (release 20.0).}\label{tab:DetailedVertexComposition}
 \begin{tabular}{l*{6}{r}}
  \toprule
  & \multicolumn{2}{l}{Largest WCC} & \multicolumn{2}{l}{Full graph} & \multicolumn{2}{l}{Database**} \\
  \cmidrule(lr){2-3}\cmidrule(lr){4-5}\cmidrule(lr){6-7}
  Vertices & EcoCyc & \textit{i}MC1010* & EcoCyc & \textit{i}MC1010* & EcoCyc & \textit{i}MC1010* \\
  \midrule
  Reaction & 4693 & 569/\hphantom{0}767 & 7251 & 601/\hphantom{0}767 & \textcolor{black!50}{2617} & 717/\hphantom{0}767 \\
  Compound & 2681 & 557/\hphantom{0}615 & 2785 & 558/\hphantom{0}615 & 2678 & 614/\hphantom{0}615 \\
  Gene & 2545 & 971/1010 & 2801 & 981/1010 & 4506 & 1010/1010 \\
  Protein monomer & 1917 & \multirow{4}{*}{771/\hphantom{0}817} & 2012 & \multirow{4}{*}{775/\hphantom{0}817} & \multirow{4}{*}{5708} & \multirow{4}{*}{815/\hphantom{0}817} \\
  Protein-protein complex & 929 & & 986 \\
  Protein-compound complex & 100 & & 103 \\
  Protein-RNA complex & 3 & & 4 \\
  & 12868 & 2868/3209 & 15942 & 2915/3209 & 15509 & 3151/3209 \\
  \bottomrule
  \multicolumn{7}{L{42em}}{\footnotesize *~accounted only for enzymatic reactions and unique metabolites (1076 and 762 in total); **~reactions in EcoCyc database are case insensitive for compartmentation and reversibility}
 \end{tabular}
\end{table}

Integrating the manually curated Covert dictionaries, each vertex has attributed (1) a unique identifier, according to the EcoCyc identifier but also indicating the compartment, (2) a unique type reference, (3) a unique assignment of the model components from \citet{Covert2004}, if applicable, and (4) the affiliations of the two- and three-domain partition. Furthermore, vertices of types \textsl{gene} and \textsl{reaction} have (5a) a name assigned, the blattner ID and the EC number, if applicable. The remaining vertices have additionally (5b) a compartment assigned, where cytosol (\texttt{c}), extracellular space (\texttt{e}), periplasmic space (\texttt{p}), inner membrane (\texttt{i}), outer membrane (\texttt{o}) and membrane in general (\texttt{m}) were taken into account. Similarly, each edge of the network has the attribute (1) type, specifying the connected vertices, and the corresponding (2) stoichiometry, where zero is assigned if not applicable or ambiguous. For edges depicting regulatory processes the stoichiometry actually denotes the mode of regulation, namely activation ($+$,1) inhibition ($-$,-1) or combined (0). These edges additionally have assigned (3) an identifier, according to the EcoCyc identifier and (4) a name, specifying the regulation type. All other edge types can be classified as either representing conjunct or disjunct links in the sense that all or solely one incoming link is required for functionality (Table~\ref{tab:Supp-EdgeCompositionModules}).\par
The fully annotated integrative reconstruction of \textit{E.~coli}'s metabolic and regulatory processes is provided as a graph representation in Supplementary File 1.

\subsection*{Graph properties concerning intra- and inter-module connectivity}
The following measures have been used in the assessment of the graph
partitioning scheme.

\paragraph{Inter-module edge fraction $c$:} %
Given the set of vertices with the domain label $D$, edges connecting
these vertices to a vertex of a different label are considered
\textit{external}, while edges between vertices of the same label are
\textit{internal}. We call
$$c_D=\frac{(\#\text{external edges of D)}}{\#(\text{external + internal edges of D})}$$
the inter-module edge fraction of domain $D$.

\paragraph{Network modularity $M$} denotes the degree to which a given partition divides the network in highly connected groups, modules, which are comparably sparsely connected among each other. Therefore, the intra-module links are counted against the total degree of the module vertices (Equation~\ref{eq:Modularity}),
\begin{align}
 M &= \sum_{j=1}^{N_M} \left( \frac{L(v_{M_j},w_{M_j})}{L_G} - \left( \frac{\deg(v_{M_j})}{2L_G} \right)^2 \right) \label{eq:Modularity}
 \intertext{with $N_M$ -- \# of modules, $L_G$ -- \# of links of graph $G$, $L(v_{M},w_{N}) = \sum\limits_{v \in M}\sum\limits_{w \in N} \mathrm{link}\left( v,w \right)$, $\deg(v_M) = \sum\limits_{v \in M} \deg(v)$.} \notag
\end{align}\vspace*{-5em}

\subsection*{Domain-traversing paths}
A \textsl{traversing path} connects the regulatory and the metabolic domains via the protein interface, specifically, a traversing path of length $k$ is of the form
\begin{equation} \label{eq:tp}
  \left[ (u,v_1), (v_1,v_2), \ldots (v_{k-1},w) \right]
\end{equation}
where the vertices $u$ and $w$ are from the regulatory and the metabolic domain (and vice versa) and the vertices $v_i$ are distinct and part of the protein interface. Starting from the set of edges
directly at the intersection of two domains iteratively the vertex successors of the interface domain as well as the final, first successor in the third domain have been determined (Algorithm~\ref{alg:TraversingPaths}).

\begin{algorithm}[htb]
 \newcommand{\Code}[1]{\textcolor{mygray}{\ttfamily#1}}
 \newcommand{\ReturnVal}[1]{{\upshape\ttfamily{\textcolor{mygray}{#1}}~=~}}
 \newcommand{\Text}[1]{{\textcolor{black}{\rmfamily\slshape #1}}}
 \SetArgSty{Code}
 \SetProgSty{Code}
 \SetFuncArgSty{Code}
 \SetCommentSty{textnormal}
 \SetKwProg{Fnc}{}{}{}
 \SetKwProgOpen{FncOpen}{}{}{}
 \SetKwFunction{TP}{TraversingPaths}
 \SetKwFunction{IS}{InterfaceSuccessors}
 \SetKwFunction{KwRange}{\textnormal{range}}
 \SetKw{KwIn}{\textcolor{black}{in}}
 \SetKw{KwAnd}{\textcolor{black}{and}}
 \SetKw{KwNot}{\textcolor{black}{not}}
 \SetKw{KwDef}{Definition:}
 \SetKwInOut{Input}{input}\SetKwInOut{Output}{output}
 \DontPrintSemicolon
 \Input{graph \Code{G = \{V,E\}}, \\ map \Code{aff : V $\rightarrow$ \{\Text{regulatory},\Text{interface},\Text{metabolic}\} }}\;
 %
 \FncOpen{\ReturnVal{downTP,upTP}\TP{ G, aff }}{

   \Code{downTP,upTP~=~list()}\;
   
  \For{(source,target) \KwIn \Code{E}}{
   \If{aff(target)~==~\Text{interface}}{
    \uIf{aff(source)~==~\Text{regulatory}}{
     \Code{\textcolor{black}{InterfaceSuccessors}( G, [source,target], \Text{metabolic}, downTP )}
    }
    \ElseIf{aff(source)~==~\Text{metabolic}}{
     \Code{\textcolor{black}{InterfaceSuccessors}( G, [source,target], \Text{regulatory}, upTP )}
    }
   }
  }
 } 
 \emph{recursively determine all successors}\;
 \Fnc{\IS{ G, vList, aimAff, sucList}}{
  \For{suc \KwIn outneighbours( vList[-1] )}{
   \Code{vL = list(vList)}\;
    \If{suc \KwNot \KwIn vL}{
    \Code{vL.append( suc )}\;
    \lIf{aff(suc)~==~\Text{interface}}{\Code{\textcolor{black}{InterfaceSuccessors}( G, vL, aimAff, sucList )}}
    \ElseIf{aff(suc)~==~aimAff}{\Code{sucList.append( vL )}}
   }
  }
 }
 \caption{Recursive algorithm for the determination of the, so-termed, domain-traversing paths from regulatory to metabolic domain and \textit{vice versa} truly passing the interface domain.}\label{alg:TraversingPaths} 
\end{algorithm}

\subsection*{Vertex centrality}
The key elements of the integrative \textit{E.~coli} network have been determined based on two graph properties.
\paragraph{Degree Centrality $DC$} is a local centrality measure and denotes the total number of in- and out-going edges of a vertex, (Equation~\ref{eq:Degree}),
\begin{align}
 DC(v) &= k_v = k_v^\text{in} + k_v^\text{out}\text{.} \label{eq:Degree}
\end{align}
Here, the vertices with a total degree greater than 50 are termed hubs.\par
By additionally accounting for the domain boundaries, the intra-domain degree fraction $\xi$ (also termed embeddedness \cite{Fortunato2016}) have been defined as ratio of internal degree, within domain $D$, and total degree of a vertex, (Equation~\ref{eq:IntraDegree}),
\begin{align}
 \xi_{D}(v) &= \frac{k_v^{\mathrm{int}}}{k_v} = \frac{1}{k_v} \sum_{w \in D} \left( A_{vw} + A_{wv} \right) \label{eq:IntraDegree}
\end{align}
where $A$ denotes the adjacency matrix of the graph.
\paragraph{Betweenness Centrality $BC$} describes the impact on the flux through the network, under the assumption that the transfer follows the shortest paths. In particular, it quantifies the fraction of shortest paths between all pairs of vertices which involve the designated vertex (Equation~\ref{eq:Betweenness}),
\begin{align}
 BC(v) &= \sum_{s \neq v \neq t \in V}\frac{\sigma_{st}(v)}{\sigma_{st}} \label{eq:Betweenness}
\end{align}
where $\sigma_{st}$ is the number of all shortest-paths between the vertices $s$ and $t$ while $\sigma_{st}(v)$ yields the number of these paths that run through $v$ \cite{Newman2010}. 

\clearpage
\bibliographystyle{Myunsrtnat}

\clearpage
\setcounter{figure}{0}
\renewcommand{\thefigure}{S\arabic{figure}}
\setcounter{table}{0}
\renewcommand{\thetable}{S\arabic{table}}
\section*{Supplementary information}
\begin{table}[htb]
 \centering
 \caption{Vertex composition of the integrative \textit{E.~coli} network in total (Total) and for the partition regulatory domain -- protein interface -- metabolic domain (RD,PI,MD).}\label{tab:Supp-VertexCompositionModules}
 \begin{tabular}{C{1.25em}l*{4}{r}}
  \toprule
  \multicolumn{2}{l}{Vertices} & Total & RD & PI & MD \\
  \midrule
  \tikz\node[reaction,fill=mygreen,scale=0.55]{}; & \textsl{reaction} & 4693 & 0 & 477 & 4216 \\
  \tikz\node[compound,fill=myblue,scale=0.55]{}; & \textsl{compound} & 2681 & 0 & 26 & 2655 \\
  \tikz\node[gene,fill=myyellow,scale=0.55]{}; & \textsl{gene} & 2545 & 1949 & 311 & 285 \\
  \tikz\node[protein,fill=myred,scale=0.55]{}; & \textsl{protein monomer} & 1917 & 198 & 1129 & 590 \\
  \tikz\node[protein,fill=mygray,scale=0.55]{}; & \textsl{protein-protein complex} & 929 & 65 & 243 & 621 \\
  \tikz\node[protein,fill=myindigo,scale=0.55]{}; & \textsl{protein-compound complex} & 100 & 0 & 100 & 0 \\
  \tikz\node[protein,fill=myorange,scale=0.55]{}; & \textsl{protein-rna complex} & 3 & 1 & 0 & 2 \\
  & & 12868 & 2213 & 2286 & 8369 \\
  \bottomrule
 \end{tabular}
\end{table}

\begin{table}[htb]
 \centering\tabcolsep2mm
 \caption{Edge composition of the integrative \textit{E.~coli} network in total (Total), for the partition regulatory domain -- protein interface -- metabolic domain (RD,PI,MD) and for the peripheral edges between the three domains.}\label{tab:Supp-EdgeCompositionModules}
 \begin{tabular}{lllc*{7}{r}}
  \toprule
  \multicolumn{3}{l}{Edges} & Link & Total & \multicolumn{1}{c}{MD} & \multicolumn{1}{c}{$\nicefrac{\mathrm{MD}}{\mathrm{PI}}$} & \multicolumn{1}{c}{PI} & \multicolumn{1}{c}{$\nicefrac{\mathrm{PI}}{\mathrm{RD}}$} & \multicolumn{1}{c}{RD} & \multicolumn{1}{c}{$\nicefrac{\mathrm{RD}}{\mathrm{MD}}$} \\
  \midrule
  \tikz\draw[-latex,shorten >=0.35em](0,0)node[gene,fill=myyellow,scale=0.55]{}to(0.8,0)node[protein,fill=myred,scale=0.55]{}; & \multicolumn{2}{l}{\textsl{gene - protein}} & D & 1916 & 312 & 0 & 325 & 803 & 198 & 278 \\
  \rowcolor{black!10}\tikz\draw[-latex,shorten <=0.35em,shorten >=0.35em](0,0)node[protein,fill=myred,scale=0.55]{}to(0.8,0)node[protein,fill=mygray,scale=0.55]{}; & \multicolumn{2}{l}{\textsl{protein - complex}} & C & 1182 & 6 & 809 & 291 & 68 & 5 & 3 \\
  \rowcolor{black!10}\tikz\draw[-latex,shorten <=0.35em,shorten >=0.35em](0,0)node[protein,fill=myred,scale=0.55]{}to(0.8,0)node[protein,fill=myindigo,scale=0.55]{}; & & & & 4 & 0 & 0 & 4 & 0 & 0 & 0 \\
  \rowcolor{black!10}\tikz\draw[-latex,shorten <=0.35em,shorten >=0.35em](0,0)node[protein,fill=mygray,scale=0.55]{}to(0.8,0)node[protein,fill=mygray,scale=0.55]{}; & & & & 80 & 8 & 42 & 22 & 7 & 1 & 0 \\
  \rowcolor{black!10}\tikz\draw[-latex,shorten <=0.35em,shorten >=0.35em](0,0)node[protein,fill=myindigo,scale=0.55]{}to(0.8,0)node[protein,fill=mygray,scale=0.55]{}; & & & & 6 & 0 & 0 & 0 & 6 & 0 & 0 \\
  \rowcolor{black!10}\tikz\draw[-latex,shorten <=0.35em,shorten >=0.5em](0,0)node[protein,fill=myred,scale=0.55]{}to(0.8,0)node[reaction,fill=mygreen,scale=0.55]{}; & \multicolumn{2}{l}{\textsl{enzyme - reaction}} & D & 1272 & 1225 & 5 & 42 & 0 & 0 & 0 \\
  \rowcolor{black!10}\tikz\draw[-latex,shorten <=0.35em,shorten >=0.5em](0,0)node[protein,fill=mygray,scale=0.55]{}to(0.8,0)node[reaction,fill=mygreen,scale=0.55]{}; & & & & 2775 & 2657 & 6 & 109 & 1 & 0 & 2 \\
  \tikz\draw[-latex,shorten >=0.5em](0,0)node[compound,fill=myblue,scale=0.55]{}to(0.8,0)node[reaction,fill=mygreen,scale=0.55]{}; & \multicolumn{2}{l}{\textsl{educt - reaction}} & C & 7707 & 7374 & 298 & 35 & 0 & 0 & 0 \\
  \tikz\draw[-latex,shorten <=0.35em,shorten >=0.5em](0,0)node[protein,fill=myred,scale=0.55]{}to(0.8,0)node[reaction,fill=mygreen,scale=0.55]{}; & & & & 246 & 0 & 1 & 245 & 0 & 0 & 0 \\
  \tikz\draw[-latex,shorten <=0.35em,shorten >=0.5em](0,0)node[protein,fill=mygray,scale=0.55]{}to(0.8,0)node[reaction,fill=mygreen,scale=0.55]{}; & & & & 181 & 0 & 3 & 178 & 0 & 0 & 0 \\
  \tikz\draw[-latex,shorten <=0.35em,shorten >=0.5em](0,0)node[protein,fill=myindigo,scale=0.55]{}to(0.8,0)node[reaction,fill=mygreen,scale=0.55]{}; & & & & 100 & 0 & 0 & 100 & 0 & 0 & 0 \\
  \rowcolor{black!10}\tikz\draw[-latex,shorten >=0.5em](0,0)node[reaction,fill=mygreen,scale=0.55]{}to(0.8,0)node[compound,fill=myblue,scale=0.55]{}; & \multicolumn{2}{l}{\textsl{reaction - product}} & D & 8303 & 7892 & 398 & 13 & 0 & 0 & 0 \\
  \rowcolor{black!10}\tikz\draw[-latex,shorten >=0.35em](0,0)node[reaction,fill=mygreen,scale=0.55]{}to(0.8,0)node[protein,fill=myred,scale=0.55]{}; & & & & 171 & 0 & 1 & 170 & 0 & 0 & 0 \\
  \rowcolor{black!10}\tikz\draw[-latex,shorten >=0.35em](0,0)node[reaction,fill=mygreen,scale=0.55]{}to(0.8,0)node[protein,fill=mygray,scale=0.55]{}; & & & & 252 & 0 & 7 & 210 & 35 & 0 & 0 \\
  \rowcolor{black!10}\tikz\draw[-latex,shorten >=0.35em](0,0)node[reaction,fill=mygreen,scale=0.55]{}to(0.8,0)node[protein,fill=myindigo,scale=0.55]{}; & & & & 102 & 0 & 0 & 102 & 0 & 0 & 0 \\
  \tikz\draw[-latex,shorten >=0.5em](0,0)node[compound,fill=myblue,scale=0.55]{}to(0.8,0)node[compound,fill=myblue,scale=0.55]{}; & \multicolumn{2}{l}{\textsl{transport}} & C & 291 & 281 & 8 & 2 & 0 & 0 & 0 \\
  \rowcolor{black!10}\tikz\draw[-latex,dashed,shorten >=0.5em](0,0)node[gene,fill=myyellow,scale=0.55,solid]{}to(0.8,0)node[gene,fill=myyellow,scale=0.55,solid]{}; & \textsl{regulation} & \footnotesize{3,8} & R & 207 & 0 & 0 & 0 & 0 & 207 & 0 \\
  \rowcolor{black!10}\tikz\draw[-latex,dashed,shorten <=0.35em,shorten >=0.5em](0,0)node[protein,fill=myred,scale=0.55,solid]{}to(0.8,0)node[gene,fill=myyellow,scale=0.55,solid]{}; & & \footnotesize{1,2,9} & & 1274 & 0 & 0 & 0 & 376 & 898 & 0 \\
  \rowcolor{black!10}\tikz\draw[-latex,dashed,shorten <=0.35em,shorten >=0.35em](0,0)node[protein,fill=myred,scale=0.55,solid]{}to(0.8,0)node[protein,fill=myred,scale=0.55,solid]{}; & & \footnotesize{12,14} & & 9 & 4 & 1 & 2 & 1 & 1 & 0 \\
  \rowcolor{black!10}\tikz\draw[-latex,dashed,shorten <=0.35em,shorten >=0.35em](0,0)node[protein,fill=myred,scale=0.55,solid]{}to(0.8,0)node[protein,fill=mygray,scale=0.55,solid]{}; & & \footnotesize{12} & & 5 & 2 & 2 & 1 & 0 & 0 & 0 \\
  \rowcolor{black!10}\tikz\draw[-latex,dashed,shorten <=0.35em,shorten >=0.5em](0,0)node[protein,fill=myred,scale=0.55,solid]{}to(0.8,0)node[reaction,fill=mygreen,scale=0.55,solid]{}; & & \footnotesize{13} & & 4 & 0 & 2 & 2 & 0 & 0 & 0 \\
  \rowcolor{black!10}\tikz\draw[-latex,dashed,shorten <=0.35em,shorten >=0.5em](0,0)node[protein,fill=mygray,scale=0.55,solid]{}to(0.8,0)node[gene,fill=myyellow,scale=0.55,solid]{}; & & \footnotesize{1,4,9} & & 2082 & 0 & 0 & 0 & 185 & 1897 & 0 \\
  \rowcolor{black!10}\tikz\draw[-latex,dashed,shorten <=0.35em,shorten >=0.35em](0,0)node[protein,fill=mygray,scale=0.55,solid]{}to(0.8,0)node[protein,fill=myred,scale=0.55,solid]{}; & & \footnotesize{12} & & 2 & 0 & 1 & 1 & 0 & 0 & 0 \\
  \rowcolor{black!10}\tikz\draw[-latex,dashed,shorten <=0.35em,shorten >=0.35em](0,0)node[protein,fill=mygray,scale=0.55,solid]{}to(0.8,0)node[protein,fill=mygray,scale=0.55,solid]{}; & & \footnotesize{12} & & 11 & 6 & 1 & 1 & 0 & 1 & 2 \\
  \rowcolor{black!10}\tikz\draw[-latex,dashed,shorten <=0.35em,shorten >=0.5em](0,0)node[protein,fill=mygray,scale=0.55,solid]{}to(0.8,0)node[reaction,fill=mygreen,scale=0.55,solid]{}; & & \footnotesize{13} & & 1 & 0 & 1 & 0 & 0 & 0 & 0 \\
  \rowcolor{black!10}\tikz\draw[-latex,dashed,shorten <=0.35em,shorten >=0.5em](0,0)node[protein,fill=myindigo,scale=0.55,solid]{}to(0.8,0)node[gene,fill=myyellow,scale=0.55,solid]{}; & & \footnotesize{1,2} & & 1160 & 0 & 0 & 0 & 1160 & 0 & 0 \\
  \rowcolor{black!10}\tikz\draw[-latex,dashed,shorten <=0.35em,shorten >=0.5em](0,0)node[protein,fill=myorange,scale=0.55,solid]{}to(0.8,0)node[gene,fill=myyellow,scale=0.55,solid]{}; & & \footnotesize{9} & & 2 & 0 & 0 & 0 & 0 & 2 & 0 \\
  \rowcolor{black!10}\tikz\draw[-latex,dashed,shorten <=0.35em,shorten >=0.35em](0,0)node[protein,fill=myorange,scale=0.55,solid]{}to(0.8,0)node[protein,fill=myred,scale=0.55,solid]{}; & & \footnotesize{12} & & 2 & 2 & 0 & 0 & 0 & 0 & 0 \\ 
  \rowcolor{black!10}\tikz\draw[-latex,dashed,shorten >=0.5em](0,0)node[compound,fill=myblue,scale=0.55,solid]{}to(0.8,0)node[gene,fill=myyellow,scale=0.55,solid]{}; & & \footnotesize{2,5,6,7,10,11} & & 98 & 0 & 0 & 0 & 0 & 0 & 98 \\
  \rowcolor{black!10}\tikz\draw[-latex,dashed,shorten >=0.35em](0,0)node[compound,fill=myblue,scale=0.55,solid]{}to(0.8,0)node[protein,fill=myred,scale=0.55,solid]{}; & & \footnotesize{12} & & 701 & 650 & 50 & 1 & 0 & 0 & 0 \\
  \rowcolor{black!10}\tikz\draw[-latex,dashed,shorten >=0.35em](0,0)node[compound,fill=myblue,scale=0.55,solid]{}to(0.8,0)node[protein,fill=mygray,scale=0.55,solid]{}; & & \footnotesize{12} & & 1728 & 1667 & 59 & 0 & 0 & 0 & 2 \\  
  \rowcolor{black!10}\tikz\draw[-latex,dashed,shorten >=0.5em](0,0)node[compound,fill=myblue,scale=0.55,solid]{}to(0.8,0)node[reaction,fill=mygreen,scale=0.55,solid]{}; & & \footnotesize{13} & & 10 & 0 & 8 & 2 & 0 & 0 & 0 \\
  & & & & 31880 & 22086 & 1703 & 1854 & 2642 & 3210 & 385 \\
  \bottomrule
  \multicolumn{11}{l}{\footnotesize C -- Conjunct encoding; D -- Disjunct encoding; R -- Regulation}
 \end{tabular}
\end{table}

\begin{table}[htb]
 \tabcolsep1.25mm
 \caption{List of the 14 different kinds of regulatory processes subsumed in the edge type \textit{regulation} of the integrative \textit{E.~coli} network (EcoCyc, release 20.0). Each of the 7296 regulatory processes comprises the regulator source ('Regulator') and target ('Regulated entity') as well as the regulatory mode, namely activation ($+$) and inhibition ($-$).}\label{tab:Supp-RegulationTypes}
 \begin{tabular}{rlrll}
  \toprule
  \multicolumn{2}{l}{Regulation type} & \# & Regulator & Regulated entity \\
  \midrule
  \footnotesize{1} & Transcription-Factor-Binding & 4302 & Protein & Transunit, Promoter \\ 
  \footnotesize{2} & Allosteric-Regulation-of-RNAP & 219 & Protein & Promoter \\ 
  \footnotesize{3} & Ribosome-Mediated-Attenuation & 12 & RNA & Terminator \\ 
  \footnotesize{4} & Protein-Mediated-Attenuation & 5 & Protein & Transunit, Terminator \\ 
  \footnotesize{5} & Transcriptional-Attenuation & 3 & Compound & Transunit, Terminator \\ 
  \footnotesize{6} & Rho-Blocking-Antitermination & 3 & Compound & Terminator \\ 
  \footnotesize{7} & Small-Molecule-Mediated-Attenuation & 2 & Compound & Transunit, Terminator \\[0.5em] 
  \footnotesize{8} & RNA-Mediated-Translation-Regulation & 195 & RNA & Transunit, Gene \\ 
  \footnotesize{9} & Protein-Mediated-Translation-Regulation & 56 & Protein & Transunit, Gene \\ 
  \footnotesize{10} & Compound-Mediated-Translation-Regulation & 22 & Protein & Transunit, Gene \\ 
  \footnotesize{11} & Regulation-of-Translation & 4 & Compound & Transunit, Gene \\[0.5em] 
  \footnotesize{12} & Regulation-of-Enzyme-Activity & 2456 & Compound, Protein & Enzyme \\ 
  \footnotesize{13} & Regulation-of-Reactions & 15 & Compound, Protein & Reaction \\ 
  \footnotesize{14} & Regulation & 2 & Protein & Protein \\ 
  \bottomrule
 \end{tabular}
\end{table}

\begin{table}[p]
 \centering
 \caption{Comparison of vertex composition of the integrative \textit{E.~coli} network and the coverage to the integrative model from \citet{Covert2004}, the metabolic model from \citet{Feist2007}, and the transcriptional regulatory network based on the RegulonDB \cite{Gama-Castro2015}.}\label{tab:Supp-ComparisonVertexComposition}
 \begin{tabular}{C{1.25em}l*{4}{r}}
  \toprule
  \multicolumn{3}{l}{Vertices of the integrative \textit{E.~coli} network} & \textit{i}MC1010* & \textit{i}AF1260** & RegulonDB \\
  \midrule 
  \tikz\node[reaction,fill=mygreen,scale=0.55]{}; & Reaction & 4693 & 569/\hphantom{0}767 & 665/1436 & 0/\hphantom{000}0 \\
  \tikz\node[compound,fill=myblue,scale=0.55]{}; & Compound & 2681 & 557/\hphantom{0}615 & 607/\hphantom{0}963 & 0/\hphantom{000}0 \\
  \tikz\node[gene,fill=myyellow,scale=0.55]{}; & Gene & 2545 & 971/1010 & 1168/1260 & 1764/1788 \\
  \tikz\node[protein,fill=myred,scale=0.55]{}; & Protein monomer & 1917 & \multirow{4}{*}{771/\hphantom{0}817} & \multirow{4}{*}{0/\hphantom{000}0} & 185/\hphantom{0}190 \\
  \tikz\node[protein,fill=mygray,scale=0.55]{}; & Protein-protein complex & 929 & & & 11/\hphantom{00}13 \\
  \tikz\node[protein,fill=myindigo,scale=0.55]{}; & Protein-compound complex & 100 & & & 0/\hphantom{000}0 \\
  \tikz\node[protein,fill=myorange,scale=0.55]{}; & Protein-RNA \mbox{complex} & 3 & & & 0/\hphantom{000}0 \\
  \midrule
  \multicolumn{2}{l}{Total (model coverage)} & 12868 & 2868/3209 & 2440/3636 & 1960/1991 \\
  & & & \textcolor{black!50}{89.4\%} & \textcolor{black!50}{67.1\%} & \textcolor{black!50}{98.4} \\
  \multicolumn{2}{l}{Total (EcoCyc coverage)} & & 3156/3209 & 3636/3636 & 1991/1991 \\
  & & & \textcolor{black!50}{98.3\%} & \textcolor{black!50}{100.0\%} & \textcolor{black!50}{100.0\%} \\
  \bottomrule
  \multicolumn{6}{L{35em}}{\footnotesize *~accounted only for intracellular reactions and unique metabolites, in total 1076 and 762} \\
  \multicolumn{6}{L{35em}}{\footnotesize **~accounted only for intracellular reactions and unique metabolites, in total 2382 and 1668}
 \end{tabular}
\end{table}

\begin{landscape}
\begin{table}[htb]
 \centering
 \caption{Most abundant vertices of the downwards (RD~\raise0.35em\hbox{\protect\tikz\protect\draw[-latex](0,0)to(0.45,0);}~MD) and upwards traversing paths (MD~\raise0.35em\hbox{\protect\tikz\protect\draw[-latex](0,0)to(0.45,0);}~RD), their quantity and the respective total degree and the corresponding degree rank.}\label{tab:Supp-TruePathVertices}
 \begin{tabular}{crllrr}
  \toprule
  \multicolumn{2}{l}{Traversing paths} & Vertex ID & Vertex name & Degree & Rank \\
  \midrule
  \multirow{3}{*}{\tikz[-latex]\draw(0,0)node[inner sep=0.2em,outer sep=0em,anchor=south]{RD}to(0,-0.47)node[inner sep=0.2em,outer sep=0em,anchor=north]{MD};} & 7304 & \texttt{PTSH-MONOMER} & HPr (histidine protein) & 37 & 74 \\
  & 7276 & \texttt{PTSH-PHOSPHORYLATED} & HPr-P (phosphorylated HPr) & 35 & 79 \\
  & 3642 & \texttt{RXN0-6718} & EI-P + HPr $\rightarrow$ HPr-P + EI & 4 & 4651 \\
  \cmidrule{2-2}
  & 2878 \\
  \cmidrule{2-6}
  & 7304 & \texttt{PTSH-MONOMER} & HPr (histidine protein) & 37 & 74 \\
  & 7276 & \texttt{PTSH-PHOSPHORYLATED} & HPr-P (phosphorylated HPr) & 35 & 79 \\
  & 2496 & \texttt{RXN0-7166} & PEP + HPr $\leftrightarrow$ HPr-P + Pyr & 4 & 4651 \\
  \cmidrule{2-2}
  & 2430 \\
  \cmidrule{2-6}
  & 4464 & \texttt{RIBONUCLEOSIDE-DIP-REDUCTI-CPLX} & RDPR1 (ribonucleoside-diphosphate reductase) & 29 & 114 \\
  & 4151 & \texttt{RED-THIOREDOXIN-MONOMER} & (reduced) thioredoxin 1 & 34 & 82 \\
  & 4050 & \texttt{OX-THIOREDOXIN-MONOMER} & oxidized thioredoxin 1 & 33 & 89 \\
  \cmidrule{2-2}
  & 2046 \\
  \cmidrule{2-6}
  & 4464 & \texttt{RIBONUCLEOSIDE-DIP-REDUCTI-CPLX} & RDPR1 (ribonucleoside-diphosphate reductase) & 29 & 114 \\
  & 4151 & \texttt{RED-THIOREDOXIN2-MONOMER} & (reduced) thioredoxin 2 & 34 & 82 \\
  & 4050 & \texttt{OX-THIOREDOXIN2-MONOMER} & oxidized thioredoxin 2 & 33 & 89 \\
  \cmidrule{2-2}
  & 2046 \\
  \cmidrule{2-2}\morecmidrules\cmidrule{2-2}
  & 8952 & of overall 18892 \\
  \midrule
  \multirow{3}{*}{\tikz\draw[-latex](0,0)node[inner sep=0.2em,outer sep=0em,anchor=south]{MD}to(0,-0.47)node[inner sep=0.2em,outer sep=0em,anchor=north]{RD};} & 1822 & \texttt{PROTEIN-NRIP} & NtrC-P (phosphorylated NtrC) & 47 & 57 \\
  & 1572 & \texttt{PROTEIN-NRIIP} & NtrB-P (phosphorylated NtrB) & 5 & 2879 \\
  & 1349 & \texttt{NRIPHOS-RXN} & NtrB-P + NtrC $\rightarrow$ NtrB + NtrC-P & 4 & 4651 \\
  \cmidrule{2-2}\morecmidrules\cmidrule{2-2}
  & 1212 & of overall 4070 \\
  \bottomrule
  & \textbf{10164} & \textbf{of overall 22963} \\
 \end{tabular}
\end{table}
\end{landscape}

\begin{table}
 \caption{Compounds that serve as reactants of the phosphoenolpyruvate--carbohydrate phosphotransferase system, so-called PTS-sugars.}\label{tab:Supp-PTS-sugars}
 \begin{tabular}{ll}
  \toprule
  Vertex ID & Vertex name \\
  \midrule
  \texttt{N-ACETYL-D-GLUCOSAMINE\_p} & \textit{N}-acetylglucosamine \\ 
  \texttt{N-ACETYL-D-MANNOSAMINE\_p} & \textit{N}-acetylmannosamine \\ 
  \texttt{NACMUR\_p} & \textit{N}-acetylmuramate \\
  \texttt{ASCORBATE\_p} & Ascorbate \\ 
  \texttt{CELLOBIOSE\_p} & Cellobiose \\ 
  \texttt{DIHYDROXYACETONE} & Dihydroxyacetone \\
  \texttt{GALACTITOL\_p} & Galactitol \\
  \texttt{CPD-12538\_p} & Glucosamine \\ 
  \texttt{2-DEOXY-D-GLUCOSE\_p} & 2-Deoxyglucose \\ 
  \texttt{CPD-15382\_p} & \textit{keto}-Fructose \\ 
  \texttt{GLC\_p} & Glucose \\ 
  \texttt{CPD-3570\_p} & Methylglucoside \\ 
  \texttt{HYDROQUINONE-O-BETA-D-GLUCOPYRANOSIDE\_p} & Hydroquinone-O-glucopyranoside (arbutin) \\ 
  \texttt{MANNITOL\_p} & Mannitol \\ 
  \texttt{CPD-12601\_p} & Mannose \\ 
  \texttt{2-O-ALPHA-MANNOSYL-D-GLYCERATE\_p} & 2-\textit{O}-Mannosylglycerate \\ 
  \texttt{CPD-1142\_p} & Salicin \\
  \texttt{SORBITOL\_p} & Sorbitol \\ 
  \texttt{TREHALOSE\_p} & Trehalose \\ 
  \bottomrule
 \end{tabular}
\end{table}

\begin{table}
 \caption{Regulated entities of the global transcriptional response regulator of the NtrBC system, the phosphorylated NtrC.}\label{tab:Supp-NtrBC-regulatedentities}
 \centering
 \tabcolsep1.2mm
 \begin{tabular}{llp{35em}}
  \toprule
  & \multirow{2}{3em}{Vertex name} & \\
  Vertex ID & & Function of the encoded protein \\
  \midrule
  \texttt{EG10385} & glnG & NtrC (inhibition) \\ 
  \texttt{EG10387} & glnL & NtrB (inhibition) \\ 
  \texttt{EG10383} & glnA & Glutamine synthetase (as 12-fold oligomer; inhibition) \\ 
  \texttt{EG12191} & glnK & PII-2 (as trimer) can activate the adenylylation of glutamine synthetase \\ 
  \texttt{EG10386} & glnH & glutamine ABC transporter - periplasmic binding protein \\ 
  \texttt{EG10388} & glnP & glutamine ABC transporter - membrane subunit \\ 
  \texttt{EG10389} & glnQ & glutamine ABC transporter - ATP binding subunit \\ 
  \texttt{EG11629} & potF & putrescine ABC transporter - periplasmic binding protein \\ 
  \texttt{EG11630} & potG & putrescine ABC transporter - ATP binding subunit \\ 
  \texttt{EG11631} & potH & putrescine ABC transporter - membrane subunit \\ 
  \texttt{EG11632} & potI & putrescine ABC transporter - membrane subunit \\ 
  \texttt{EG12124} & hisJ & histidine ABC transporter - periplasmic binding protein \\ 
  \texttt{EG10007} & hisM & arginine/histidine/lysine/ornithine ABC transporter - membrane subunit \\ 
  \texttt{EG10452} & hisP & arginine/histidine/lysine/ornithine ABC transporter - ATP binding subunit \\ 
  \texttt{EG12125} & hisQ & arginine/histidine/lysine/ornithine ABC transporter - membrane subunit \\ 
  \texttt{EG10072} & argT & arginine/lysine/ornithine ABC transporter - periplasmic binding protein \\ 
  \texttt{EG11821} & amtB & member of NH$_3$/NH$_4^+$ transporters, necessary for growth only at low NH$_3$ levels \\ 
  \texttt{G7071} & cbl & Cbl DNA-binding transcriptional activator \\ 
  \texttt{G7072} & nac & Nac DNA-binding transcriptional dual regulator \\ 
  \texttt{G6943} & astA & Arginine succinyltransferase - 1\textsuperscript{st} step in arginine degradation II (AST pathway) \\ 
  \texttt{G6941} & astB & Succinylarginine dihydrolase (as dimer) - 2\textsuperscript{nd} step in AST pathway \\ 
  \texttt{G6944} & astC & Succinylornithine transaminase - 3\textsuperscript{rd} step in AST pathway \\ 
  \texttt{G6942} & astD & Succinylglutamate semialdehyde dehydrogenase - 4\textsuperscript{th} step in AST pathway \\ 
  \texttt{G6940} & astE & Succinylglutamate desuccinylase - 5\textsuperscript{th} and final reaction in AST pathway \\ 
  \texttt{G6523} & rutA & Uracil oxygenase - 1\textsuperscript{st} step in uracil degradation III \\ 
  \texttt{G6522} & rutB & peroxyureidoacrylate/ureidoacrylate amido hydrolase - 2\textsuperscript{nd} step in uracil degradation III \\ 
  \texttt{G6521} & rutC & (predicted aminoacrylate peracid reductase - 3\textsuperscript{rd} step in uracil degradation III) \\ 
  \texttt{G6520} & rutD & predicted aminoacrylate hydrolase - 4\textsuperscript{th} step in uracil degradation III \\ 
  \texttt{G6519} & rutE & predicted malonic semialdehyde reductase - 5\textsuperscript{th} step in uracil degradation III \\ 
  \texttt{G6518} & rutF & (flavin reductase - activity required for 1\textsuperscript{st} step in uracil degradation (RutA)) \\ 
  \texttt{G6517} & rutG & member of the nucleobase:cation symporter-2 (NCS2) family of transporters (probably for uracil) \\ 
  \texttt{G6782} & ddpX & D-Ala-D-Ala dipeptidase required for wild-type peptidoglycan biosynthesis \\ 
  \texttt{G6781} & ddpA & (predicted peptide ABC transporter - periplasmic binding component) \\ 
  \texttt{G6780} & ddpB & (predicted peptide ABC transporter - membrane component) \\ 
  \texttt{G6779} & ddpC & (predicted peptide ABC transporter - membrane component) \\ 
  \texttt{G6778} & ddpD & (predicted peptide ABC transporter - ATP-binding component) \\ 
  \texttt{G6777} & ddpF & (predicted peptide ABC transporter - ATP-binding component) \\ 
  \texttt{G6969} & yeaG & (impact in adaptation to sustained N starvation, member of Ser protein kinases) \\ 
  \texttt{G6970} & yeaH & (impact in adaptation to sustained N starvation) \\ 
  \texttt{EG12834} & yhdW & (predicted amino acid ABC transporter - membrane component) \\ 
  \texttt{EG12835} & yhdX & (predicted amino acid ABC transporter - ATP-binding component) \\ 
  \texttt{EG12836} & yhdY & (predicted amino acid ABC transporter - membrane component) \\ 
  \texttt{EG12837} & yhdZ & (predicted amino acid ABC transporter - periplasmic binding component) \\ 
  \bottomrule
 \end{tabular}
\end{table}

\begin{landscape}
\tabcolsep0.55mm
\begin{longtable}{rll*{3}{c}*{2}{r*{6}{r}}D{.}{.}{1}}
 \caption{Hubs of the integrative \textit{E.~coli} network with a total degree of at least 50 (DC), their module affiliation, and the differentiation in in-degree (In) and out-degree (Out) including the affiliation and linkage assignments. The last column denotes the intra-domain degree fraction, $\xi$, here given as percental fraction.}\label{tab:Supp-DegreeCentrality} \\
  \toprule
  & & & \multicolumn{3}{l}{Affiliation} & & \multicolumn{3}{l}{in-affiliation} & \multicolumn{3}{l}{in-linkage} & & \multicolumn{3}{l}{out-affiliation} & \multicolumn{3}{l}{out-linkage} & \multicolumn{1}{c}{\multirow{2}{2.2em}{\centering $\xi$ [\%]}} \\
  \cmidrule{4-6}\cmidrule(r){8-10}\cmidrule(l){11-13}\cmidrule(r){15-17}\cmidrule(l){18-20}
  \multicolumn{1}{l}{DC} & Vertex ID & Vertex name & RD & PI & MD & In & RD & PI & MD & C & D & R & Out & RD & PI & MD & C & D & R & \\
  \midrule
  \endfirsthead
  \multicolumn{21}{r}{Continued} \\
  \toprule
  & & & \multicolumn{3}{l}{Affiliation} & & \multicolumn{3}{l}{in-affiliation} & \multicolumn{3}{l}{in-linkage} & & \multicolumn{3}{l}{out-affiliation} & \multicolumn{3}{l}{out-linkage} & \multicolumn{1}{c}{\multirow{2}{2.2em}{\centering $\xi$ [\%]}} \\
  \cmidrule{4-6}\cmidrule(r){8-10}\cmidrule(l){11-13}\cmidrule(r){15-17}\cmidrule(l){18-20}
  \multicolumn{1}{l}{DC} & Vertex ID & Vertex name & RD & PI & MD & In & RD & PI & MD & C & D & R & Out & RD & PI & MD & C & D & R & \\ 
  \midrule
  \endhead
  \midrule
  \multicolumn{21}{r}{Continued on next page} \\
  \endfoot
  \bottomrule
  \endlastfoot
  1412 & \texttt{PROTON} & Proton & & & $\checkmark$ & 1027 & & 15 & 1012 & & 1027 & & 385 & 1 & 9 & 375 & 384 & & 1 & 98.2 \\
  \rowcolor{black!10}930 & \texttt{WATER} & H$_2$O & & & $\checkmark$ & 226 & & 74 & 152 & & 226 & & 704 & & 23 & 681 & 704 & & & 89.6 \\
  \rowcolor{chocolate!100!black!15}515 & \texttt{CPLX0-226} & Crp-cAMP, transcriptional dual regulator & & $\checkmark$ & & 1 & & 1 & & & 1 & & 514 & 513 & 1 & & 1 & & 513 & 0.4 \\
  \rowcolor{black!10}489 & \texttt{Pi} & Phosphate (P) & & & $\checkmark$ & 410 & & 8 & 402 & 1 & 409 & & 79 & & 2 & 77 & 40 & & 39 & 98.0 \\
  439 & \texttt{ATP} & ATP & & & $\checkmark$ & 17 & & 3 & 14 & & 17 & & 422 & & 47 & 375 & 383 & & 39 & 88.6 \\
  \rowcolor{black!10}419 & \texttt{PROTON\_p} & Proton (periplasmic) & & & $\checkmark$ & 96 & & & 96 & & 96 & & 323 & & & 323 & 323 & & & 100.0 \\
  401 & \texttt{CPLX0-7534\_o} & [OmpF]$_3$, outer membran porin F complex & & & $\checkmark$ & 1 & & 1 & & 1 & & & 400 & & & 400 & & 400 & & 99.8 \\
  \rowcolor{black!10}401 & \texttt{CPLX0-7533\_}o & [OmpC]$_3$, outer membran porin C complex & & & $\checkmark$ & 1 & & 1 & & 1 & & & 400 & & & 400 & & 400 & & 99.8 \\
  399 & \texttt{CPLX0-7530\_o} & [OmpE]$_3$, outer membran porin E complex & & & $\checkmark$ & 1 & & 1 & & 1 & & & 398 & & & 398 & & 398 & & 99.7 \\
  \rowcolor{black!10}350 & \texttt{ADP} & ADP & & & $\checkmark$ & 299 & & 35 & 264 & & 299 & & 51 & & 9 & 42 & 22 & & 29 & 87.4 \\
  293 & \texttt{CPLX0-7797} & FNR, transcriptional dual regulator & $\checkmark$ & & & 1 & 1 & & & 1 & & & 292 & 292 & & & & & 292 & 100.0 \\ 
  \rowcolor{black!10}247 & \texttt{WATER\_p} & H$_2$O (periplasmic) & & & $\checkmark$ & 21 & & & 21 & & 21 & & 226 & & & 226 & 226 & & & 100.0 \\
  241 & \texttt{NAD} & NAD$^+$ & & & $\checkmark$ & 112 & & 1 & 111 & 1 & 111 & & 129 & & 1 & 128 & 118 & & 11 & 99.2 \\
  \rowcolor{black!10}235 & \texttt{NADH} & NADH/H$^+$ & & & $\checkmark$ & 115 & & & 115 & & 115 & & 120 & & 1 & 119 & 110 & & 10 & 99.6 \\
  222 & \texttt{PC00027} & IHF, transcriptional dual regulator & $\checkmark$ & & & 2 & & 2 & & 2 & & & 220 & 220 & & & & & 220 & 99.1 \\
  \rowcolor{black!10}221 & \texttt{CPLX0-7705} & Fis, transcriptional dual regulator & $\checkmark$ & & & 1 & 1 & & & 1 & & & 220 & 220 & & & & & 220 & 100.0 \\ 
  187 & \texttt{PD00288} & H-NS, transcriptional dual regulator & $\checkmark$ & & & 1 & 1 & & & & 1 & & 186 & 186 & & & & & 186 & 100.0 \\ 
  \rowcolor{black!10}178 & \texttt{PPI} & Pyrophosphate & & & $\checkmark$ & 147 & & 8 & 139 & & 147 & & 31 & & & 31 & 9 & & 22 & 95.5 \\
  177 & \texttt{Pi\_p} & Phosphate (periplasmic) & & & $\checkmark$ & 168 & & & 168 & & 168 & & 9 & & & 9 & 3 & & 6 & 100.0 \\
  \rowcolor{black!10}176 & \texttt{PHOSPHO-ARCA} & ArcA-P, transcriptional dual regulator & $\checkmark$ & & & 3 & & 3 & & 1 & 2 & & 173 & 173 & & & & & 173 & 98.3 \\
  159 & \texttt{NADP} & NADP$^+$ & & & $\checkmark$ & 114 & & 4 & 110 & & 114 & & 45 & & 1 & 44 & 34 & & 11 & 96.9 \\
  \rowcolor{black!10}151 & \texttt{NADPH} & NADPH/H$^+$ & & & $\checkmark$ & 33 & & & 33 & & 33 & & 118 & & 5 & 113 & 114 & & 4 & 96.7 \\
  144 & \texttt{CO-A} & Coenzyme A & & & $\checkmark$ & 50 & & 2 & 48 & 1 & 49 & & 94 & & 6 & 88 & 77 & & 17 & 94.4 \\
  \rowcolor{black!10}138 & \texttt{GLT} & Glutamate & & & $\checkmark$ & 72 & & & 72 & & 72 & & 66 & & & 66 & 61 & & 5 & 100.0 \\
  131 & \texttt{CPLX0-7639} & [Fur-Fe$^{2+}$]$_2$, transcriptional dual regulator & $\checkmark$ & & & 1 & & 1 & & 1 & & & 130 & 130 & & & & & 130 & 99.2 \\
  \rowcolor{black!10}125 & \texttt{2-KETOGLUTARATE} & 2-Ketoglutarate & & & $\checkmark$ & 58 & & & 58 & & 58 & & 67 & & 2 & 65 & 59 & & 8 & 98.4 \\
  124 & \texttt{PHOSPHO-NARL} & NarL-P, transcriptional dual regulator & $\checkmark$ & & & 3 & & 3 & & 1 & 2 & & 121 & 121 & & & & & 121 & 97.6 \\
  \rowcolor{chocolate!100!black!15}121 & \texttt{CPLX0-8070} & DksA-ppGpp & & $\checkmark$ & & 1 & & 1 & & & 1 & & 120 & 119 & 1 & & 1 & & 119 & 1.7 \\
  115 & \texttt{AMP} & AMP & & & $\checkmark$ & 80 & & 8 & 72 & & 80 & & 35 & & 1 & 34 & 10 & & 25 & 92.2 \\
  \rowcolor{black!10}111 & \texttt{PYRUVATE} & Pyruvate & & & $\checkmark$ & 59 & & 5 & 54 & & 59 & & 52 & & 5 & 47 & 36 & & 16 & 91.0 \\
  98 & \texttt{AMMONIUM} & NH$_4^+$ & & & $\checkmark$ & 70 & & & 70 & 1 & 69 & & 28 & & & 28 & 13 & & 15 & 100.0 \\
  \rowcolor{black!10}87 & \texttt{CARBON-DIOXIDE} & CO$_2$ & & & $\checkmark$ & 81 & & 3 & 78 & & 81 & & 6 & & & 6 & 6 & & & 96.6 \\
  \rowcolor{chocolate!100!black!15}84 & \texttt{PC00061} & Cra, transcriptional dual regulator & & $\checkmark$ & & 3 & & 3 & & 1 & 2 & & 81 & 79 & 2 & & 2 & & 79 & 6.0 \\
  \rowcolor{black!10}82 & \texttt{CPLX0-3930} & FlhDC, transcriptional dual regulator & $\checkmark$ & & & 2 & & 2 & & 2 & & & 80 & 80 & & & & & 80 & 97.6 \\
  81 & \texttt{S-ADENOSYLMETHIONINE} & Adenosylmethionine & & & $\checkmark$ & 3 & & 1 & 2 & & 3 & & 78 & 1 & 7 & 70 & 75 & & 3 & 88.9 \\
  \rowcolor{chocolate!100!black!15}81 & \texttt{GUANOSINE\_TETRAPHOSPHATE} & Guanosine 5'-diphosphate 3'-diphosphate (ppGpp) & & & $\checkmark$ & 4 & & 1 & 3 & & 4 & & 77 & 64 & 1 & 12 & 2 & & 75 & 18.5 \\
  \rowcolor{chocolate!100!black!15}78 & \texttt{CPLX0-8047} & NsrR-nitric oxide & & $\checkmark$ & & 1 & & 1 & & & 1 & & 77 & 76 & 1 & & 1 & & 76 & 2.6 \\
  \rowcolor{black!10}76 & \texttt{ADENOSYL-HOMO-CYS} & Adenosylhomocysteine & & & $\checkmark$ & 65 & & 4 & 61 & & 65 & & 11 & & 3 & 8 & 1 & & 10 & 90.8 \\
  71 & \texttt{ZN+2} & Zinc (Zn$^{2+}$) & & & $\checkmark$ & 5 & & 3 & 2 & & 5 & & 66 & & 5 & 61 & 7 & & 59 & 88.7 \\
  \rowcolor{black!10}69 & \texttt{ACETYL-COA} & acetyl-CoA & & & $\checkmark$ & 33 & & 1 & 32 & & 33 & & 36 & & 3 & 33 & 28 & & 8 & 94.2 \\
  68 & \texttt{OXYGEN-MOLECULE} & O$_2$ & & & $\checkmark$ & 7 & & & 7 & & 7 & & 61 & & 1 & 60 & 57 & & 4 & 98.5 \\
  \rowcolor{black!10}68 & \texttt{PHOSPHO-CPXR} & CpxR, transcriptional dual regulator & $\checkmark$ & & & 1 & & 1 & & 1 & & & 67 & 67 & & & 1 & & 66 & 98.5 \\
  63 & \texttt{GTP} & GTP & & & $\checkmark$ & 2 & & & 2 & & 2 & & 61 & & 2 & 59 & 41 & & 20 & 96.8 \\
  \rowcolor{chocolate!100!black!15}63 & \texttt{MONOMER0-155} & Lrp-Leucine, transcriptional dual regulator & & $\checkmark$ & & 1 & & 1 & & & 1 & & 62 & 61 & 1 & & 1 & & 61 & 3.2 \\
  62 & \texttt{PHOSPHO-PHOB} & PhoB-P, transcriptional dual regulator & $\checkmark$ & & & 3 & & 3 & & 1 & 2 & & 59 & 59 & & & & & 59 & 95.2 \\
  \rowcolor{black!10}59 & \texttt{PC00010} & LexA, transcriptional repressor & $\checkmark$ & & & 1 & & 1 & & 1 & & & 58 & 58 & & & & & 58 & 98.3 \\
  \rowcolor{chocolate!100!black!15}58 & \texttt{PD00353} & Lrp, transcriptional dual regulator& & $\checkmark$ & & 2 & 1 & 1 & & & 2 & & 56 & 55 & 1 & & 1 & & 55 & 3.4 \\
  \rowcolor{black!10}58 & \texttt{CFA-CPLX} & Cyclopropane fatty acyl phospholipid synthase & & & $\checkmark$ & 9 & & 1 & 8 & 1 & & 8 & 49 & & & 49 & & 49 & & 98.3 \\
  57 & \texttt{PHOSPHO-PHOP} & PhoP-P, transcriptional dual regulator & $\checkmark$ & & & 2 & & 2 & & 1 & 1 & & 55 & 55 & & & & & 55 & 96.5 \\
  \rowcolor{black!10}54 & \texttt{AAS-MONOMER} & Acyltransferase & & & $\checkmark$ & 54 & & & 54 & & 1 & 53 & & & & & & & & 100.0 \\
  53 & \texttt{ALKAPHOSPHA-CPLX\_p} & Alkaline phosphatase (periplasmic) & & & $\checkmark$ & 8 & & 1 & 7 & 1 & & 7 & 45 & & & 45 & & 45 & & 98.1 \\
  \rowcolor{black!10}52 & \texttt{PHOSPHO-NARP} & NarP-P, transcriptional dual regulator & $\checkmark$ & & & 3 & & 3 & & 1 & 2 & & 49 & 49 & & & & & 49 & 94.2 \\
  51 & \texttt{D-ALANINE} & Alanine & & & $\checkmark$ & 25 & & 1 & 24 & & 25 & & 26 & & 1 & 25 & 21 & & 5 & 96.1 \\
  \rowcolor{black!10}50 & \texttt{SUC} & Sucrose & & & $\checkmark$ & 23 & & & 23 & & 23 & & 27 & & & 27 & 22 & & 5 & 100.0 \\
 \end{longtable}
\end{landscape}

\begin{landscape}
\tabcolsep1.65mm
\begin{longtable}{cll*{3}{c}*{11}{C{0.975em}}}
 \caption{Top betweenness central vertices of the integrative \textit{E.~coli} network and involved systems. The central reactions of each system are shaded in the corresponding system's color.}\label{tab:Supp-BetweennessCentrality} \\
  \toprule
  Betweenness & Vertex ID & Vertex name & GD & PI & MD & \multicolumn{11}{l}{System} \\
  \midrule
  \endfirsthead
  \multicolumn{17}{r}{Continued} \\
  \toprule
  Betweenness & Vertex ID & Vertex name & GD & PI & MD & \multicolumn{11}{l}{System} \\
  \midrule
  \endhead
  \midrule
  \multicolumn{17}{r}{Continued on next page} \\
  \endfoot
  \midrule
  \multicolumn{17}{l}{\colsquare{dred} \texttt{ATPSYN-RXN}: \texttt{ADP} + \texttt{Pi} + 4 \texttt{PROTON}$_\texttt{p}$ $\leftrightarrow$ \texttt{ATP} + \texttt{WATER} + 3 \texttt{PROTON}} \\
  \multicolumn{17}{l}{\colsquare{dblue} \texttt{ADENYLATECYC-RXN}: \texttt{ADENYLATECYC-MONOMER} + \texttt{ATP} $\rightarrow$ \texttt{CAMP} + \texttt{PPI}; \texttt{RXN0-269}: \textcolor{black!50}{\texttt{PC00004}} + \texttt{CAMP} $\rightarrow$ \texttt{CPLX0-226}} \\
  \multicolumn{17}{l}{\colsquare{gold} \texttt{PHOR-RXN}: \textcolor{black!50}{\texttt{PHOR-MONOMER}$_\texttt{i}$} + \texttt{ATP} + \texttt{Pi} $\rightarrow$ \texttt{PHOSPHO-PHOR}$_\texttt{i}$ + \texttt{ADP}; \texttt{PHOBR-RXN}: \textcolor{black!50}{\texttt{PHOB-MONOMER}} + \texttt{PHOSPHO-PHOR}$_\texttt{i}$ $\rightarrow$ \texttt{PHOSPHO-PHOB} + \textcolor{black!50}{\texttt{PHOR-MONOMER}$_\texttt{i}$}} \\
  \multicolumn{17}{l}{\colsquare{teal} \texttt{RXN0-261}: \textcolor{black!50}{\texttt{PD00353}} + \texttt{LEU} $\leftrightarrow$ \texttt{MONOMER0-155}} \\
  \multicolumn{17}{l}{\colsquare{purple} \texttt{RXN0-5252}: \textcolor{black!50}{\texttt{PD00260}} + \texttt{FE+2} $\leftrightarrow$ \texttt{CPLX0-7620} $\dashrightarrow$ \texttt{CPLX0-7639}} \\
  \multicolumn{17}{l}{\colsquare{black} \texttt{EG30063} $\dashv$ \texttt{EG10671} $\rightarrow$ \texttt{EG10671-MONOMER} $\dashrightarrow$ \texttt{CPLX0-7534}$_\texttt{o}$; \texttt{EG10670} $\rightarrow$ \texttt{EG10670-MONOMER} $\dashrightarrow$ \texttt{CPLX0-7533}$_\texttt{o}$; \texttt{EG10729} $\rightarrow$ \texttt{MONOMER0-282} $\dashrightarrow$ \texttt{CPLX0-7530}$_\texttt{o}$} \\
  \multicolumn{17}{l}{\colsquare{dgreen} \texttt{PPPGPPHYDRO-RXN}: \textcolor{black!50}{\texttt{GDP-TP}} + \texttt{WATER} $\rightarrow$ \texttt{GUANOSINE\_TETRAPHOSPHATE} + \texttt{Pi} + \texttt{PROTON}} \\
  \multicolumn{17}{l}{\colsquare{steelblue} \texttt{PEPDEPHOS-RXN}: \textcolor{black!50}{\texttt{PHOSPHO-ENOL-PYRUVATE}} + \texttt{ADP} + \texttt{PROTON} $\leftrightarrow$ \texttt{PYRUVATE} + \texttt{ATP}} \\
  \multicolumn{17}{l}{\colsquare{chocolate} \texttt{CARBAMATE-KINASE-RXN}: \textcolor{black!50}{\texttt{CARBAMOYL-P}} + \texttt{ADP} + \texttt{PROTON} $\rightarrow$ \texttt{CARBON-DIOXIDE} + \texttt{AMMONIUM} + \texttt{ATP}} \\
  \multicolumn{17}{l}{\colsquare{olive} \texttt{ADENYLYLSULFKIN-RXN}: \textcolor{black!50}{\texttt{PAPS}} + \texttt{ADP} + \texttt{PROTON} $\leftrightarrow$ \textcolor{black!50}{\texttt{APS}} + \texttt{ATP}} \\
  \bottomrule
  \endlastfoot
  0.29906 & \texttt{PROTON} & Proton & & & $\checkmark$ & \colsquare{dred} & & \colsquare{gold} & & & & \colsquare{dgreen} & \colsquare{steelblue} & \colsquare{chocolate} & \colsquare{olive} \\
  0.19894 & \texttt{WATER} & H$_2$O & & & $\checkmark$ & \colsquare{dred} & & & & & & \colsquare{dgreen} \\
  0.09670 & \texttt{ATP} & ATP & & & $\checkmark$ & \colsquare{dred} & \colsquare{dblue} & \colsquare{gold} & & & & & \colsquare{steelblue} & \colsquare{chocolate} & \colsquare{olive} \\
  0.08320 & \texttt{Pi} & Phosphate (P) & & & $\checkmark$ & \colsquare{dred} & & & & & & \colsquare{dgreen} \\ 
  0.06466 & \texttt{PROTON\_p} & Proton (periplasmic) & & & $\checkmark$ & \colsquare{dred} \\
  0.06186 & \texttt{ADP} & ADP & & & $\checkmark$ & \colsquare{dred} & & \colsquare{gold} & & & & & \colsquare{steelblue} & \colsquare{chocolate} & \colsquare{olive} \\
  0.05369 & \texttt{CAMP} & cyclic-AMP (cAMP) & & & $\checkmark$ & & \colsquare{dblue} \\
  \rowcolor{dblue!10}0.05362 & \texttt{RXN0-269\_f} & & & $\checkmark$ & & & \colsquare{dblue} & & & & & & & & & \\
  0.05362 & \texttt{CPLX0-226} & Crp-cAMP, transcriptional dual regulator & & $\checkmark$ & & & \colsquare{dblue} \\
  \rowcolor{dblue!10}0.05361 & \texttt{ADENYLATECYC-RXN} & & & & $\checkmark$ & & \colsquare{dblue} & & & & & & & & & \\
  \rowcolor{gold!10}0.04082 & \texttt{PHOR-RXN} & & & $\checkmark$ & & & & \colsquare{gold} & & & & & & & & \\
  0.04066 & \texttt{PHOSPHO-PHOB} & PhoB-P, transcriptional dual regulator & $\checkmark$ & & & & & \colsquare{gold} & & & & & & & & \\
  \rowcolor{gold!10}0.04056 & \texttt{PHOBR-RXN} & & & $\checkmark$ & & & & \colsquare{gold} & & & & & & & & \\
  0.04052 & \texttt{PHOSPHO-PHOR\_i} & PhoR, sensory histidine kinase (inner membrane) & & $\checkmark$ & & & & \colsquare{gold} & & & & & & & & \\
  0.03880 & \texttt{LEU} & Leucine & & & $\checkmark$ & & & & \colsquare{teal} & & & & & & & \\
  \rowcolor{teal!10}0.03840 & \texttt{RXN0-261\_f} & & & $\checkmark$ & & & & & \colsquare{teal} & & & & & & & \\
  0.03839 & \texttt{MONOMER0-155} & Lrp-Leucine, transcriptional dual regulator & & $\checkmark$ & & & & & \colsquare{teal} & & & & & & & \\
  0.03722 & \texttt{FE+2} & Ferrous (Fe$^{2+}$) & & & $\checkmark$ & & & & & \colsquare{purple} & & & & & \\
  \rowcolor{purple!10}0.03567 & \texttt{RXN0-5252\_f} & & & $\checkmark$ & & & & & & \colsquare{purple} & & & & & & \\
  0.03567 & \texttt{CPLX0-7620} & Fur-Fe$^{2+}$ & & $\checkmark$ & & & & & & \colsquare{purple} & & & & & \\
  0.03540 & \texttt{CPLX0-7639} & [Fur-Fe$^{2+}$]$_2$, transcriptional dual regulator & $\checkmark$ & & & & & & & \colsquare{purple} & & & & & \\
  0.03396 & \texttt{EG10671} & ompF, outer membrane porin F gene & $\checkmark$ & & & & & & & & \colsquare{black} & & & & & \\
  0.03395 & \texttt{EG10671-MONOMER} & OmpF, outer membrane porin F & & $\checkmark$ & & & & & & & \colsquare{black} & & & & & \\
  0.03395 & \texttt{CPLX0-7534\_o} & [OmpF]$_3$, outer membran porin F complex & & & $\checkmark$ & & & & & & \colsquare{black} & & & & & \\
  0.03242 & \texttt{ADENYLATECYC-MONOMER} & CyaA, adenylate cyclase & & & $\checkmark$ & & \colsquare{dblue} \\
  0.02743 & \texttt{GUANOSINE\_TETRAPHOSPHATE} & Guanosine 5'-diphosphate 3'-diphosphate (ppGpp) & & & $\checkmark$ & & & & & & & \colsquare{dgreen} \\
  0.02663 & \texttt{EG10670} & ompC, outer membrane porin C gene & $\checkmark$ & & & & & & & & \colsquare{black} & & & & & \\
  0.02663 & \texttt{EG10670-MONOMER} & OmpC, outer membrane porin C & & $\checkmark$ & & & & & & & \colsquare{black} & & & & & \\
  0.02663 & \texttt{CPLX0-7533\_o} & [OmpC]$_3$, outer membrane porin C complex & & & $\checkmark$ & & & & & & \colsquare{black} & & & & & \\
  0.02560 & \texttt{PPI} & Pyrophosphate & & & $\checkmark$ & & \colsquare{dblue} \\
  0.02489 & \texttt{WATER\_p} & H$_2$O (periplasmic) & & & $\checkmark$ & & & & & & & & & & & \opensquare{black} \\
  0.02432 & \texttt{EG10729} & ompE, outer membrane porin E gene & $\checkmark$ & & & & & & & & \colsquare{black} & & & & & \\
  0.02432 & \texttt{MONOMER0-282} & OmpE, outer membrane porin E & & $\checkmark$ & & & & & & & \colsquare{black} & & & & & \\
  0.02432 & \texttt{CPLX0-7530\_o} & [OmpE]$_3$, outer membran porin E complex & & & $\checkmark$ & & & & & & \colsquare{black} & & & & & \\
  0.02319 & \texttt{NAD} & NAD$^+$ & & & $\checkmark$ & & & & & & & & & & & \opensquare{black} \\
  0.02214 & \texttt{Pi\_p} & Phosphate (periplasmic) & & & $\checkmark$ & & & & & & & & & & & \opensquare{black} \\
  \rowcolor{dgreen!10}0.02119 & \texttt{PPPGPPHYDRO-RXN} & & & & $\checkmark$ & & & & & & & \colsquare{dgreen} & & & & \\
  \rowcolor{dred!10}0.01890 & \texttt{ATPSYN-RXN\_f} & & & & $\checkmark$ & \colsquare{dred} & & & & & & & & & & \\
  0.01803 & \texttt{PYRUVATE} & Pyruvate & & & $\checkmark$ & & & & & & & & \colsquare{steelblue} \\
  0.01786 & \texttt{GLT} & Glutamate & & & $\checkmark$ & & & & & & & & & & & \opensquare{black} \\
  0.01607 & \texttt{CARBON-DIOXIDE} & CO$-2$ & & & $\checkmark$ & & & & & & & & & \colsquare{chocolate} \\
  0.01603 & \texttt{NADP} & NADP$^+$ & & & $\checkmark$ & & & & & & & & & & & \opensquare{black} \\
  \rowcolor{steelblue!10}0.01353 & \texttt{PEPDEPHOS-RXN\_2} & & & & $\checkmark$ & & & & & & & & \colsquare{steelblue} & & & \\
  \rowcolor{steelblue!10}0.01252 & \texttt{PEPDEPHOS-RXN\_1} & & & & $\checkmark$ & & & & & & & & \colsquare{steelblue} & & & \\
  \rowcolor{olive!10}0.01173 & \texttt{ADENYLYLSULFKIN-RXN\_r} & & & & $\checkmark$ & & & & & & & & & & \colsquare{olive} & \\
  0.01171 & \texttt{AMMONIUM} & NH$_4^+$ & & & $\checkmark$ & & & & & & & & & \colsquare{chocolate} \\
  \rowcolor{chocolate!10}0.01161 & \texttt{CARBAMATE-KINASE-RXN} & & & & $\checkmark$ & & & & & & & & & \colsquare{chocolate} & & \\
  0.01144 & \texttt{CO-A} & Coenzyme A & & & $\checkmark$ & & & & & & & & & & & \opensquare{black} \\
  \rowcolor{teal!10}0.01117 & \texttt{ABC-35-RXN\_2} & & & & $\checkmark$ & & & & \colsquare{teal} & & & & & & & \\
  \rowcolor{teal!10}0.01107 & \texttt{ABC-35-RXN\_1} & & & & $\checkmark$ & & & & \colsquare{teal} & & & & & & & \\
  0.01093 & \texttt{EG30063} & micF, mRNA-interfering complementary RNA gene & $\checkmark$ & & & & & & & & \colsquare{black} & & & & & \\ 
 \end{longtable}
\end{landscape}

\begin{sidewaystable}[htb]
 \caption{Top ten key elements of the integrative \textit{E.~coli} network with respect to protein interface-specific degree (piDC) and betweenness centrality (piBC), respectively. The traversing path systems depict the embedding of system components in the protein interface (Figures~\ref{fig:TruePathsHistogram} and \ref{fig:PTS-RNR-NtrBC}, Table~\ref{tab:Supp-TruePathVertices}).}\label{tab:Supp-KeyPIElements}
 \centering
 \begin{tabular}{rllrr*{3}{C{2.37em}}}
  \toprule
  Rank & Vertex ID & Vertex name & piDC & piBC & \multicolumn{3}{l}{Traverse path system} \\
  \midrule
  1 & \texttt{PTSH-MONOMER} & HPr (histidine protein) & 1 & 1 & \colsquare{black!70} & & \\
  \rowcolor{black!5}2 & \texttt{PTSH-PHOSPHORYLATED} & HPr-P (phosphorylated HPr) & 1 & 2 & \colsquare{black!70} & & \\
  3 & \texttt{RED-THIOREDOXIN-MONOMER} & $_{\text{red}}$Trx1 (reduced thioredoxin 1) & 3 & 3 & & \colsquare{goldenrod} & \\
  \rowcolor{black!5}3 & \texttt{RED-THIOREDOXIN2-MONOMER} & $_{\text{red}}$Trx2 (reduced thioredoxin 2) & 3 & 3 & & \colsquare{goldenrod} & \\
  5 & \texttt{OX-THIOREDOXIN-MONOMER} & $_{\text{ox}}$Trx1 (oxidized thioredoxin 1) & 3 & 5 & & \colsquare{goldenrod} & \\
  \rowcolor{black!5}5 & \texttt{OX-THIOREDOXIN2-MONOMER} & $_{\text{ox}}$Trx2 (oxidized thioredoxin 2) & 3 & 5 & & \colsquare{goldenrod} & \\
  7 & \texttt{EG50003-MONOMER} & acyl carrier protein (ACP) & 8 & 14 & & & \opensquare{black} \\
  \rowcolor{black!5}8 & \texttt{FLAVODOXIN1-MONOMER} & flavodoxin 1 & 10 & 17 & & & \opensquare{black} \\
  9 & \texttt{OX-FLAVODOXIN1} & oxidized flavodoxin 1 & 10 & 18 & & & \opensquare{black} \\
  \rowcolor{black!5}9 & \texttt{PROTEIN-CHEA} & chemotaxis protein CheA & 9 & 19 & & & \opensquare{black} \\
  \vdots \\
  \rowcolor{goldenrod!10}12 & \texttt{RIBONUCLEOSIDE-DIP-REDUCTI-CPLX} & RDPR1 (ribonucleoside-diphosphate reductase) & 7 & 24 & & \colsquare{goldenrod} & \\
  \vdots \\
  \rowcolor{black!20}20 & \texttt{RXN0-6718} & EI-P + HPr $\rightarrow$ HPr-P + EI & 63 & 7 & \colsquare{black!70} & & \\
  \vdots \\
  \rowcolor{black!5}69 & \texttt{ADPREDUCT-RXN\_f\_1} & NDP + $_{\text{red}}$Trx1 $\rightarrow$ dNDP + $_{\text{ox}}$Trx1 + H$_2$O & 169 & 8 & & & \opensquare{black} \\
  69 & \texttt{ADPREDUCT-RXN\_f\_2} & NDP + $_{\text{red}}$Trx2 $\rightarrow$ dNDP + $_{\text{ox}}$Trx2 + H$_2$O & 169 & 8 & & & \opensquare{black} \\
  \rowcolor{black!5}\vdots & & & & & & & \\
  72 & \texttt{THIOREDOXIN-REDUCT-NADPH-RXN\_1} & $_{\text{ox}}$Trx1 + NADPH/H$^+$ $\rightarrow$ $_{\text{red}}$Trx1 + NADP$^+$ & 169 & 9 & & & \opensquare{black} \\
  \rowcolor{black!5}72 & \texttt{THIOREDOXIN-REDUCT-NADPH-RXN\_2} & $_{\text{ox}}$Trx2 + NADPH/H$^+$ $\rightarrow$ $_{\text{red}}$Trx2 + NADP$^+$ & 169 & 9 & & & \opensquare{black} \\
  \bottomrule
  \multicolumn{6}{l}{\scriptsize\mylittlesquare{black!70} \texttt{RXN0-6718}: \texttt{PTSH-MONOMER} + \textcolor{black!50}{\texttt{PTSI-PHOSPHORYLATED}} $\rightarrow$ \texttt{PTSH-PHOSPHORYLATED} + \textcolor{black!50}{\texttt{PTSI-MONOMER}}} \\[-0.4em]
  \multicolumn{6}{l}{\scriptsize\mylittlesquare{goldenrod} \texttt{RIBONUCLEOSIDE-DIP-REDUCTI-CPLX} $\dashrightarrow$ \textcolor{black!50}{\texttt{RIBONUCLEOSIDE-DIP-REDUCTI-RXN}}:} \\[-0.4em]
 \end{tabular}
\end{sidewaystable}

\clearpage
\begin{figure}[htb]
 \centering
 \includegraphics[width=0.6\textwidth]{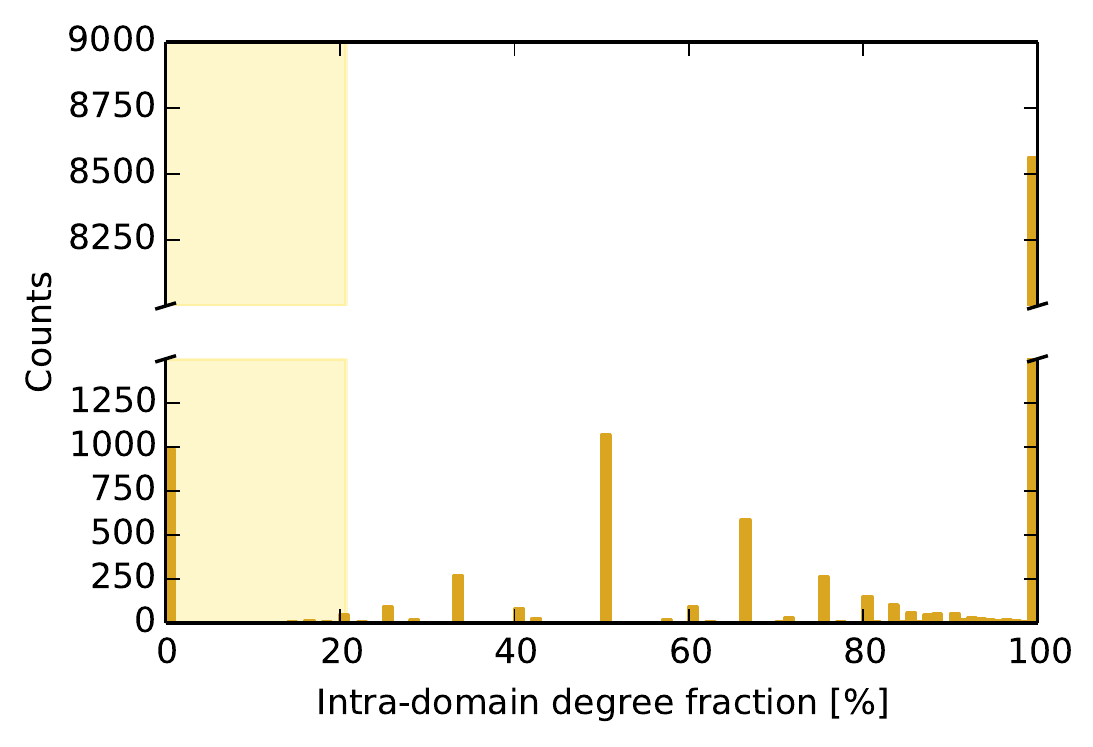}
 \caption{Distribution of intra-domain degree fraction of the integrative \textit{E.~coli} network. The yellow shaded area represents the significant low intra-domain degree fractions tested via z-score.}\label{fig:Supp-IntraDegree}
\end{figure}

\begin{table}[htb]
 \centering
 \caption{Hubs and non-hubs of the integrative \textit{E.~coli} network with a significant low intra-domain degree fraction ($\xi$, tested via z-score) and a total degree (DC) larger than 12, and their domain affiliation (RD, PI and MD).}\label{tab:Supp-IntraDegree}
 \tabcolsep1.25mm
 \begin{tabular}{C{0.65em}l*{3}{c}D{.}{.}{2}D{.}{.}{4}r}
  \toprule
  \multicolumn{2}{l}{Vertex name} & RD & PI & MD & \multicolumn{1}{c}{$\xi$} & \multicolumn{1}{l}{$p$-value} & DC \\
  \midrule
  \multirow{7}{*}{\rotatebox{90}{Hubs}} & CRP-cAMP DNA-binding transcriptional dual regulator & & $\checkmark$ & & 0.39 & 0.0042 & 515 \\ 
  & DksA-ppGpp & & $\checkmark$ & & 1.65 & 0.0049 & 121 \\ 
  & Cra, transcriptional dual regulator & & $\checkmark$ & & 5.95 & 0.0071 & 84 \\ 
  & Guanosine 5'-diphosphate 3'-diphosphate (ppGpp) & & & $\checkmark$ & 18.52 & 0.0208 & 81 \\ 
  & NsrR-nitric oxide & & $\checkmark$ & & 2.56 & 0.0052 & 78 \\ 
  & Lrp-Leucine, transcriptional dual regulator & & $\checkmark$ & & 3.17 & 0.0055 & 63 \\ 
  & Lrp, transcriptional dual regulator & & $\checkmark$ & & 3.45 & 0.0056 & 58 \\ 
  \midrule
  \multirow{18}{*}{\rotatebox{90}{Non-hubs}} & ModE-MoO\textsubscript{4}\textsuperscript{2-} DNA-binding transcriptional dual regulator & & $\checkmark$ & & 4.17 & 0.006 & 48 \\ 
  & NtrC-P, transcriptional dual regulator & & $\checkmark$ & & 8.51 & 0.0089 & 47 \\ 
  & NagC, transcriptional dual regulator & & $\checkmark$ & & 4.35 & 0.0061 & 46 \\ 
  & PdhR, transcriptional dual regulator & & $\checkmark$ & & 4.55 & 0.0062 & 44 \\ 
  & ArgR-arginine, transcriptional dual regulator & & $\checkmark$ & & 5 & 0.0065 & 40 \\ 
  & DksA RNA polymerase-binding transcription factor & & $\checkmark$ & & 5.13 & 0.0068 & 39 \\ 
  & PurR-Hypoxanthine, transcriptional repressor & & $\checkmark$ & & 6.06 & 0.0071 & 33 \\ 
  & CysB-acetylserine, transcriptional dual regulator & & $\checkmark$ & & 6.25 & 0.0072 & 32 \\ 
  & FhlA-Formate, transcriptional activator & & $\checkmark$ & & 6.25 & 0.0072 & 32 \\ 
  & FadR, transcriptional dual regulator & & $\checkmark$ & & 20.69 & 0.0245 & 29 \\ 
  & RutR, transcriptional dual regulator & & $\checkmark$ & & 18.18 & 0.0202 & 22 \\ 
  & NsrR, transcriptional repressor & & $\checkmark$ & & 14.29 & 0.0148 & 21 \\ 
  & CytR, transcriptional repressor & & $\checkmark$ & & 18.75 & 0.0207 & 16 \\ 
  & AraC-arabinose, transcriptional activator & & $\checkmark$ & & 12.5 & 0.0125 & 16 \\ 
  & GntR, transcriptional repressor & & $\checkmark$ & & 20 & 0.0228 & 15 \\ 
  & DnaA-ATP, transcriptional dual regulator & & $\checkmark$ & & 14.29 & 0.0148 & 14 \\ 
  & TrpR-Tryptophan, transcriptional repressor & & $\checkmark$ & & 14.29 & 0.0148 & 14 \\ 
  & Mlc, transcriptional repressor & & $\checkmark$ & & 15.38 & 0.0162 & 13 \\ 
  \bottomrule
 \end{tabular}
\end{table}
\end{document}